\documentclass[aps,prd,twocolumn,amsmath,amsfonts,amssymb,nofootinbib,eqsecnum,%
  secnumarabic,notitlepage,superscriptaddress]{revtex4-1}
\usepackage[utf8]{inputenc}
\usepackage{amsmath,amsfonts,amssymb,graphicx,subfigure}
\usepackage{slashed}
\usepackage{multirow}
\usepackage{lineno}
\usepackage{array}
\usepackage[dvipsnames]{xcolor}
\usepackage[colorlinks=true, linkcolor = blue]{hyperref} 
\hypersetup{colorlinks=true, linkcolor=ForestGreen, citecolor=ForestGreen,
filecolor=ForestGreen, urlcolor=ForestGreen}
\graphicspath{{Plots/}}
\def\ie{{\it i.e.}}

\newcommand{\be}{\begin{equation}}
\newcommand{\ee}{\end{equation}}

\newcommand{\met}{{E_T^{\rm miss}}}
\newcommand{\eV}{{\rm ~eV}}

\newcommand{\GeV}{{\rm ~GeV}}
\newcommand{\TeV}{{\rm ~TeV}}
\newcommand{\pb}{{\rm ~pb}}
\newcommand{\fb}{{\rm ~fb}}
\newcommand{\ab}{{\rm ~ab}}
\newcommand{\invfb}{{\rm ~fb^{-1}}}
\newcommand{\invab}{{\rm ~ab^{-1}}}

\newcommand{\mgamc}{{\sc\small MG5aMC}}
\newcommand{\mgFull}{{\sc\small MadGraph5\_aMC@NLO}}

\newcommand{\confirm}[1]{{\color{black}#1}}

\begin{document}
\leftline{}
\rightline{CP3-20-50, DESY 20-186, MCNet-20-24, VBSCAN-PUB-11-20, IFJPAN-IV-2021-2}

\title{Majorana Neutrinos in Same-Sign $W^\pm W^\pm$ Scattering at the LHC:\\ Breaking the TeV Barrier}

\author{Benjamin~Fuks}
\email{fuks@lpthe.jussieu.fr}
\affiliation{Sorbonne Universit\'e, CNRS, Laboratoire de Physique Th\'eorique et
  Hautes \'Energies, LPTHE, F-75005 Paris, France}
\affiliation{Institut Universitaire de France, 103 boulevard Saint-Michel,
  75005 Paris, France}

\author{Jonas Neundorf}
\email{jonas.neundorf@desy.de}
\affiliation{Deutsches Elektronen-Synchrotron, Notkestraße 85, 22607 Hamburg, Germany}

\author{Krisztian Peters}
\email{krisztian.peters@desy.de}
\affiliation{Deutsches Elektronen-Synchrotron, Notkestraße 85, 22607 Hamburg, Germany}

\author{Richard Ruiz}
\email{rruiz@ifj.edu.pl}
\affiliation{Institute of Nuclear Physics, Polish Academy of Sciences, ul. Radzikowskiego, Cracow 31-342, Poland}
\affiliation{Centre for Cosmology, Particle Physics and Phenomenology {\rm (CP3)},\\
Universit\'e Catholique de Louvain, Chemin du Cyclotron, Louvain la Neuve, B-1348, Belgium}

\author{Matthias Saimpert}
\email{matthias.saimpert@cern.ch}
\affiliation{CERN - 1211 Geneva 23 - Switzerland}

\begin{abstract}
We revisit the sensitivity to non-resonant, heavy Majorana neutrinos $N$ in same-sign $W^\pm W^\pm$ scattering at the $\sqrt{s}=13$ TeV LHC and its high-luminosity upgrade. As a benchmark scenario, we work in the context of the Phenomenological Type I Seesaw model, relying on a simulation up to next-to-leading order in QCD with parton shower matching. After extensively studying the phenomenology of the  $pp\to\mu^\pm\mu^\pm j j$ process at the amplitude and differential levels,  we design a  simple collider analysis with remarkable signal-background separation power. At 95\% confidence level we find that the squared muon-heavy neutrino mixing element $\vert V_{\mu N} \vert^{2}$ can be probed down to about $0.06-0.3 ~ (0.03-0.1)$ for $m_N = 1-10~{\rm TeV}$ with $\mathcal{L}=300$ fb$^{-1}~(3$ ab$^{-1})$. For heavier masses of $m_N = 20~{\rm TeV}$, we report sensitivity for $\vert V_{\mu N} \vert^{2}\gtrsim 0.5~(0.3)$. The $W^\pm W^\pm$ scattering channel can greatly extend the mass range covered by current LHC searches for heavy Majorana neutrinos and particularly adds invaluable sensitivity above a few hundred GeV. We comment on areas where the analysis can be improved as well as on the applicability to other tests of neutrino mass models.
\end{abstract}

\date{\today}

\maketitle

\section{Introduction}\label{sec:intro}

Following the discovery of neutrino oscillations \cite{Ahmad:2002jz, Ashie:2005ik}, uncovering the origin of neutrinos' tiny masses and their large mixing angles are among the most pressing questions in particle physics today \cite{Strategy:2019vxc,EuropeanStrategyGroup:2020pow}.
To address these mysteries, neutrino mass models, collectively known as Seesaw models, do so by hypothesizing a variety of states that couple to the Standard Model's (SM) lepton and Higgs sectors \cite{Ma:1998dn}. 
These states include new charged or gauge-singlet (or sterile) fermions, scalars with exotic gauge quantum numbers, as well as gauge bosons of new symmetries, and mediate the non-conservation of lepton number (LN) and/or charged lepton flavor number over a range of mass scales and coupling strengths. For reviews of Seesaw models and their tests, see Refs.~\cite{Atre:2009rg,Deppisch:2015qwa,Cai:2017jrq,Cai:2017mow}.
Despite these viable solutions, there remains a lack of clear guidance from both experiment and theory as to what is realized by nature. It is therefore necessary to broadly approach this challenge in complementary aspects.

To this extent, tests of neutrino mass models at the Large Hadron Collider (LHC) are
supported by a number of signatures,
including searches for dijet resonances~\cite{Aad:2019hjw,Sirunyan:2019vgj}, 
many-lepton final states~\cite{Chatrchyan:2012ya,Sirunyan:2018mtv,Aaboud:2018qcu,Aad:2019kiz,Sirunyan:2019bgz,Aad:2020fzq} 
and LN-violating lepton pairs~\cite{Sirunyan:2018mtv,Sirunyan:2018xiv,Sirunyan:2018pom,Sirunyan:2018xiv,Aaboud:2018spl}, 
but rely mostly on mechanisms mediated by quark-antiquark $(q\overline{q})$ annihilation~\cite{Keung:1983uu}. However, due to its high center-of-mass energy $(\sqrt{s})$, the LHC is also effectively an electroweak (EW) boson collider \cite{Dawson:1984gx,Kane:1984bb,Kunszt:1987tk}. This in turn opens a multitude of complementary channels. For example, in the context of the Phenomenological Type I Seesaw model~\cite{delAguila:2008cj,Atre:2009rg}, the $W\gamma$ fusion channel \cite{Datta:1993nm,Dev:2013wba,Alva:2014gxa,Degrande:2016aje} has already helped direct searches for heavy neutrinos $N$ with  masses beyond a few hundred GeV improve sensitivity to  active-sterile neutrino mixing matrix elements $V_{\ell N}$. In fact, with an  integrated luminosity of $\mathcal{L}\approx36\invfb$ of proton-proton collisions at
$\sqrt{s}=13\TeV$,  $\vert V_{\ell N}\vert^2\gtrsim  \mathcal{O}(0.01-1)$ are excluded for lepton flavors $\ell\in\{e,\mu\}$ and sterile neutrino masses in the range $m_N=100\GeV-1\TeV$~\cite{
Sirunyan:2018mtv,Sirunyan:2018xiv}. With the full LHC data set, this degree of sensitivity is
anticipated to reach masses up to
$m_N=3-4\TeV$~\cite{Pascoli:2018heg}.

\begin{figure}
\includegraphics[width=\columnwidth]{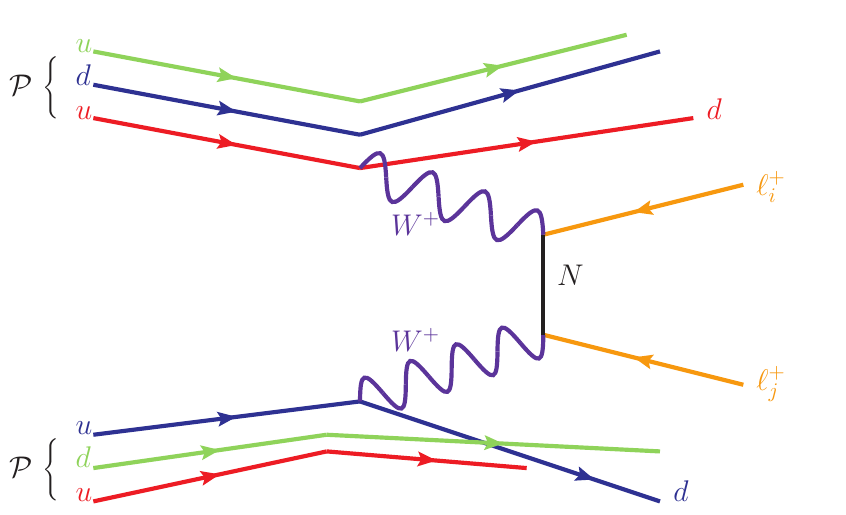}
\caption{
  Diagrammatic representation of same-sign $\ell_i^+\ell_j^+$ production  in same-sign $W^+ W^+$ scattering in $pp$ collisions mediated by  a Majorana neutrino $N$. Drawn with \texttt{JaxoDraw}~\cite{Binosi:2008ig}.
}
\label{fig:wwScattHeavyN_diagram}
\end{figure}

Motivated by the recent experimental observations of  EW vector boson fusion / scattering (VBF) at the LHC \cite{Chatrchyan:2013foa,Aad:2014zda,Sirunyan:2017ret,Aaboud:2018ddq,Sirunyan:2020gyx},
we revisit the sensitivity of same-sign $W^\pm W^\pm$ scattering to TeV-scale Majorana neutrinos at $\sqrt{s}=13\TeV$. As a benchmark scenario, we work in the 
framework of the Phenomenological Type I Seesaw and 
focus on the production of same-sign muon pairs $(\mu^\pm \mu^\pm)$ without substantial transverse momentum imbalance via the spacelike exchange of an $N$  in $W^\pm W^\pm$ scattering
 \cite{Dicus:1991fk},
 \begin{equation}
 W^\pm W^\pm \to \mu^\pm \mu^\pm.
 \label{eq:wwScattProc}
 \end{equation}
As shown in figure~\ref{fig:wwScattHeavyN_diagram}, this is essentially a realization of neutrinoless $\beta\beta$ $(0\nu\beta\beta)$ decay at large momentum transfer when mediated at dimension $d=7$~\cite{delAguila:2012nu,Lehman:2014jma,Aoki:2020til,Fuks:2020zbm}.
While past studies have investigated the importance of this channel
\cite{Dicus:1991fk,Datta:1993nm,Ali:2001gsa,Panella:2001wq,Chen:2008qb,Atre:2009rg}, most works were restricted to sub-TeV $m_N$, and therefore subject to  signal processes with more dominant cross sections. As {the heavy neutrino exchange} in equation~\eqref{eq:wwScattProc} is always non-resonant, the channel is complementary to other processes, such as the $q\overline{q'}\to N\ell$ annihilation and $W\gamma\to N\ell$ fusion mechanisms, which become kinematically inaccessible 
for sterile neutrinos that are too heavy.
For similar reasons,
the channel is robust against the impact of long $N$ lifetimes.

To conduct this study, we employ a state-of-the-art simulation tool chain that allows us to model SM backgrounds and, for the first time, the $W^\pm W^\pm  \to \ell_i^\pm \ell_j^\pm$ signal process at next-to-leading order (NLO) in QCD with parton shower (PS) matching. We report remarkable signal-background separation power and attribute this to the signal process exhibiting both VBF and LN-violating topologies. For the $pp \to \mu^\pm \mu^\pm j j$ collider signature with forward jet-tagging and simple selection cuts, we find that \confirm{$\vert V_{\mu N} \vert^{2}\gtrsim 0.06-0.3~(0.03-0.1)$ can be probed at 95\% confidence level (CL)  for $m_N = 1-10~{\rm TeV}$ with $\mathcal{L}=300$ fb$^{-1}$ (3 ab$^{-1}$). For  masses of  $m_N = 20\TeV$ we find  sensitivity for $\vert V_{\mu N} \vert^{2}$ down to $0.5~(0.3)$.} 

The remainder of this work is organized in the following manner: First, we describe in section \ref{sec:theory} our theoretical framework and give an overview of current experimental constraints. Next, we summarize our computational setup (section~\ref{sec:setup}) and  simulation prescriptions (section~\ref{sec:modeling}). In section \ref{sec:wwScattLHC} we explore extensively the phenomenology of the $W^\pm W^\pm \to \ell_i^\pm \ell_j^\pm$ signal process at the amplitude and differential levels.
This is then used in section~\ref{sec:analysis} to 
design our collider analysis and estimate the LHC's sensitivity to LN violation (LNV) in $W^\pm W^\pm$ scattering.
We provide an outlook in section \ref{sec:outlook} on areas where our analysis can be improved as well as its applicability to other tests of neutrino mass models at colliders.
Finally, we conclude in section~\ref{sec:conclusions}. Where relevant, technical derivations and details on software modifications are reported in  appendices \ref{app:0vBBLimits}, \ref{app:fxfxCuts}, and \ref{app:helicity}.

\section{Theoretical Framework}\label{sec:theory}
We describe in this section the theoretical framework in which we work (section~\ref{sec:theory_model}) and summarize current experimental constraints on the model (section~\ref{sec:theory_constraints}).

\subsection{The Phenomenological Type I Seesaw Model}\label{sec:theory_model}
To study the sensitivity of the LHC to the LN-violating $W^\pm W^\pm \to \ell_i^\pm \ell_j^\pm$ process,
we work in the context of the Phenomenological Type I Seesaw {model}~\cite{delAguila:2008cj,Atre:2009rg}.
Like the eponymous mechanism~\cite{Minkowski:1977sc, Yanagida:1979as,GellMann:1980vs,Glashow:1979nm,Mohapatra:1979ia,Shrock:1980ct,Schechter:1980gr} or its low-scale variants~\cite{Mohapatra:1986aw,Mohapatra:1986bd,Bernabeu:1987gr,Akhmedov:1995ip,Akhmedov:1995vm,Gavela:2009cd}, the model hypothesizes the existence of Majorana neutrinos $(N_{k'})$ that {couple} to SM particles through mass mixing with light neutrinos $(\nu_{k})$. More precisely, the renormalizable Lagrangian of the theory,
\begin{equation}
\mathcal{L}_{\rm Type~I} = \mathcal{L}_{\rm SM} + \mathcal{L}_{\rm Kin.} +\mathcal{L}_{\rm Y},  
\end{equation} 
extends the SM Lagrangian $(\mathcal{L}_{\rm SM})$ by kinetic and Majorana mass terms $(\mathcal{L}_{\rm Kin.} )$
for $n_R\geq2$ right-handed (RH) neutrinos $(\nu_R^i)$, as well as 
by Yukawa couplings $(\mathcal{L}_{\rm Y})$ between the SM Higgs field, the  SM lepton doublets $(L^j)$ and {the gauge-singlet fermions} $\nu_R^i$.

After EW symmetry breaking, 
flavor eigenstates of active, left-handed (LH) neutrinos $(\nu_{L\ell})$ can be generically~\cite{Atre:2009rg}  decomposed into light $(\nu_{k})$ and heavy $(N_{k'})$ mass eigenstates with mass eigenvalues $m_{\nu_k}$ and $m_{N_{k'}}$:
\begin{equation}
\nu_{L\ell} = \sum_{k=1}^3 U_{\ell k} \nu_k + \sum_{k'=4}^{n_R+3} V_{\ell k'} N_{k'}.
\label{eq:nuMix}
\end{equation}
Here, the complex-valued matrix elements $U_{\ell k}~(V_{\ell k'})$ parameterize the mixing between the LH interaction state $\nu_{L\ell}$ and the light (heavy) mass eigenstate $\nu_k~(N_{k'})$.
Formally, the matrix elements satisfy the relationship $U U^\dagger + VV^\dagger  = \mathbb{I}$. For clarity and without loss of generality,
we order states such that $m_{N_{k'}} < m_{N_{k'+1}}$.

To leading order in active-sterile mixing $\vert V_{\ell k'}\vert$, equation~\eqref{eq:nuMix} gives rise to the following effective, charged current component of the theory's interaction Lagrangian,
\begin{eqnarray}
\Delta\mathcal{L} &=& -\frac{g_W}{\sqrt{2}}W^+_\mu 
\sum_{k=1}^3
\sum_{\ell}^\tau\left[\overline{\nu}_k U^*_{\ell k}\gamma^\mu P_L \ell\right]
\nonumber\\
& & 
 -\frac{g_W}{\sqrt{2}}W^+_\mu 
\sum_{k'=4}^{n_R+3}
\sum_{\ell}^\tau\left[\overline{N}_{k'}V^*_{\ell k'}\gamma^\mu P_L \ell\right]
+\text{H.c.}  \quad
\label{eq:nuLag}
\end{eqnarray}
In the above {expression}, $g_W\approx0.65$ is the SM  weak gauge coupling constant
and $P_{L/R}=(1/2)(1\mp\gamma^5)$ are the standard chiral projection operators in four-component notation.
Similar terms can be derived for the neutral current $(Z)$ and Higgs interactions for both heavy Dirac and Majorana neutrinos~\cite{Atre:2009rg}. For simplicity we {solely investigate the phenomenology of the} lightest heavy neutrino mass eigenstate, which we relabel as $N\equiv N_{k'=4}$ {(so that} $V_{\ell N} \equiv V_{\ell k'=4}$), and decouple all other {heavy eigenstates}.

In short-distance scattering and decay processes involving only a single heavy neutrino,
the mixing factors that appear in equation~\eqref{eq:nuLag}
act as effective couplings and factor out of   amplitudes.
For resonant production of a {heavy neutrino}, this allows one to define a ``bare'' cross section $\sigma_0$ such that~\cite{Han:2006ip}
\begin{eqnarray}
\sigma(pp \to N\ell^\pm+X) \equiv
 \vert V_{\ell N}\vert^2 \times
\sigma_0(pp \to N\ell^\pm+X).
\label{eq:bareXSecS}
\qquad
\end{eqnarray}
When the $W^\pm W^\pm \to \ell_i^\pm \ell_j^\pm$ process involves only a single $t$-channel {exchange of a heavy neutrino},
such a factorization is also possible.
More specifically, one can define a ``bare'' $t$-channel cross section given by
\begin{eqnarray}
\sigma(pp &\to& \ell_i^\pm \ell_j^\pm+X) \equiv \nonumber \\
& & \vert V_{\ell_i N}V_{\ell_j N}\vert^2 \times \sigma_0(pp \to \ell_i^\pm \ell_j^\pm+X) .
\label{eq:bareXSecT}
\qquad
\end{eqnarray}
For our purposes, it suffices to add that these expressions hold at NLO in QCD~\cite{Ruiz:2015zca,Degrande:2016aje}.
Under certain assumptions, the above decomposition or a similar one can hold for processes involving multiple interfering heavy neutrino mass eigenstates. Such expressions, however, may not always be tractable due to large interference effects. 
For further discussions,  see Refs.~\cite{Pilaftsis:1991ug,Kersten:2007vk,Moffat:2017feq,Chao:2009ef,Godbole:2020doo}.

\subsection{Model Constraints}\label{sec:theory_constraints}

In its most general construction, the free parameters of the Phenomenological Type I Seesaw {model} (beyond those of the SM) consist of the {neutrino} masses $m_{\nu_k}$, $m_{N_k'}$, and the neutrino mixing elements $U_{\ell  k}$, $V_{\ell k'}$. Imposing flavor symmetries on the lepton sector, however, can reduce the number of independent degrees of freedom,
as discussed in Refs.~\cite{Pilaftsis:1991ug,Kersten:2007vk,Moffat:2017feq,King:2017guk,Xing:2019vks} and references therein. It is also possible to constrain {these} parameters by tying the lightness of the $\nu_k$ neutrinos to beyond the SM physics,
such as to dark matter and the baryon asymmetry of the universe, as done for example in Refs.~\cite{Asaka:2005pn,Asaka:2005an}.
For our purposes, we take {mass} and mixing {parameters} to be phenomenologically independent. We do this to develop a collider analysis in a flavor-model-independent fashion and is motivated by the desire to broaden sensitivity to a range of ultraviolet completions.

Beyond theoretical considerations are the following recent experimental constraints on the model:
\begin{itemize}
\item \textbf{Direct constraints from $0\nu\beta\beta$ searches:}
After an exposure of 127.2~kg-yr, the GERDA experiment reports {at 90\% CL} the following lower limit on the $0\nu\beta\beta$ decay half-life in  $^{76}$Ge~\cite{Agostini:2020xta}:
\begin{equation}
    T_{1/2}^{0\nu} >  1.8\cdot 10^{26}~{\rm yr}.
\end{equation}
Assuming that the nuclear $0\nu\beta\beta$ process is {only} mediated by heavy neutrinos, 
this translates into {an} upper limit on {their} masses and mixing of
\begin{equation}
     \left\vert \sum_{k'=4}^{n_R+3}  \frac{V_{ek'}^2}{m_k'} \right\vert <
     (2.33-4.12)\cdot 10^{-6} \TeV^{-1},
     \label{eq:0vBBLimit}
\end{equation}
where the variation stems from the uncertainties in the nuclear matrix element. For further details on the derivation of this constraint, see appendix~\ref{app:0vBBLimits}.

\item \textbf{Direct constraints from collider searches:} At $\sqrt{s}=13\TeV$ and with  $\mathcal{L}\approx36\invfb$ of data, searches for the $pp\to \ell_i\ell_j\ell_k+\met$ signature with $\ell\in\{e,\mu\},$ by the CMS experiment constrain active-sterile neutrino mixing at  95\% CL to be \cite{Sirunyan:2018mtv},
\begin{align}
    &\quad \vert V_{\ell N}\vert^2  \lesssim 10^{-5}-10^{-2} ~ \text{for}~ 1\GeV \!<\!  m_N \!<\! m_W,  \nonumber\\
    & \quad \vert V_{\ell N}\vert^2  \lesssim 10^{-2}-1 ~  \text{for}~ m_W \!<\! m_N  \!<\!1.2\TeV.
\end{align} 
Constraints from the ATLAS experiment with the same integrated luminosity  are comparable for $m_N<m_W$ but weaker for $m_N >  m_W$ due to the  absence of the $W\gamma$ channel in their signal modeling \cite{Aad:2019kiz}. Searches for the $pp\to \ell_i^\pm\ell_j^\pm+nj$ signature by CMS yield only slightly more stringent constraints due to a larger signal-over-background ratio~\cite{Sirunyan:2018xiv}.

  \item \textbf{Indirect constraints on  $V_{\ell k'}$:} For $n_R = 3$ sterile neutrinos
 with masses above the EW scale,  
 a global study of  precision EW data,
searches for non-unitarity in quark mixing, and
 searches  for lepton flavor violation and non-universality of weak decays
constrain active-sterile neutrino mixing to be~\cite{Fernandez-Martinez:2016lgt},
\begin{align}
\sqrt{2\vert \eta_{ee}\vert} 		< 0.050,	&\quad \sqrt{2\vert \eta_{e\mu}\vert} 		< 0.026,	\nonumber\\
\sqrt{2\vert \eta_{\mu\mu}\vert} 	< 0.021,	&\quad \sqrt{2\vert \eta_{e\tau}\vert} 		< 0.052,	\nonumber\\
\sqrt{2\vert \eta_{\tau\tau}\vert} 	< 0.075,	&\quad \sqrt{2\vert \eta_{\mu\tau}\vert} 	< 0.035,
\end{align}
at   95\% CL. The parameter $\eta_{\ell\ell'}$ is related to the heavy neutrino mixing matrix $V_{\ell k'}$ by 
$\sqrt{2\vert \eta_{\ell\ell'}\vert} = \sum_{k'=4}^{6} \sqrt{V_{\ell k'}V^*_{\ell' k'}}$.
{For the scenario that we consider}, {\it i.e.,} $\ell=\ell'=\mu$ with only one heavy neutrino species, this translates to {an} upper limit of
\be
  \vert V_{\mu N}\vert^2  <  4.41\cdot10^{-4}\quad
{\rm at~95\%~CL}\ .
\ee

\item \textbf{Direct constraints on the absolute mass scale of light neutrinos:} Attempts to measure  the light neutrino mass scale directly from the kinematic end point in $\beta$ decay with the KATRIN experiment~\cite{Osipowicz:2001sq} {constrain} the  light neutrino masses to {satisfy}~\cite{Aker:2019uuj}
\be
  m(\nu_e) < 1.1\eV\quad
{\rm at~90\%~CL}\ .
\ee

 \item \textbf{Constraints on neutrino masses from cosmology:} An analysis of neutrinos' impact on the cosmic microwave background, supernovae, large scale structure,  and big bang nucleosynthesis  constrains the sum of {light neutrino masses} to be~\cite{Loureiro:2018pdz}
\be
  \sum m_\nu \lesssim 0.26\eV\quad
  {\rm at~95\%~CL}\ .
\ee

 \item \textbf{Neutrino oscillation measurements of  $U_{\ell k}$:} In the absence of sterile neutrinos or additional new physics, the {elements of the light neutrino mixing matrix} $U_{\ell k}$ have been fit to or constrained by long and short baseline neutrino oscillation data under the condition that {$U$} is unitary~\cite{Esteban:2018azc}. Relaxing this constraint, however, greatly weakens the goodness of fit, particularly in the $\tau$ flavor sector~\cite{Parke:2015goa,Ellis:2020ehi}.

\end{itemize}

\section{Computational Setup}\label{sec:setup}
We now summarize  the computational setup used  in this  study. We start with section~\ref{sec:setup_mc} where we document our Monte Carlo (MC) simulation chain.
In section~\ref{sec:setup_sm}  we list the numerical values used for SM inputs,
and similarly in section~\ref{sec:setup_seesaw} the numerical values used for non-SM inputs. 
A  description on how we model the $W^\pm W^\pm \to  \ell_i^\pm \ell_j^\pm$ signal process
and leading background processes
in $pp$ collisions is deferred to section~\ref{sec:modeling}.

\subsection{Monte Carlo Setup}\label{sec:setup_mc}

To investigate same-sign $W^\pm W^\pm$ scattering when mediated by a heavy Majorana neutrino at the $\sqrt{s}=13\TeV$ LHC, we employ a state-of-the-art simulation tool chain.
For hard, parton-level scattering processes, we use the general-purpose MC event generator~\textsc{MadGraph5\_aMC@NLO} (\mgamc)~(version 2.7.1.2)~\cite{Stelzer:1994ta,Alwall:2014hca},
which enables us to simulate tree-induced processes in the SM up to NLO in QCD~\cite{Frixione:2002ik,Frederix:2009yq,Hirschi:2011pa,Hirschi:2015iia}.
Processes involving heavy neutrinos are simulated up to NLO in QCD by importing into \mgamc~the default variant of the \textsc{HeavyN}~\cite{Degrande:2016aje}
  \textsc{FeynRules} UFO libraries~\cite{Christensen:2008py,Christensen:2009jx,Degrande:2011ua,Alloul:2013bka,Degrande:2014vpa}. 
For select backgrounds we perform jet matching up to one additional parton (at the Born level) at NLO  in QCD precision using the FxFx matching procedure \cite{Frederix:2012ps}, as further detailed in section~\ref{sec:modeling}. To simulate the decay of $W$ bosons, we impose the spin-correlated narrow width approximation as implemented in \textsc{MadSpin}~\cite{Artoisenet:2012st,Alwall:2014bza}.

All signal and background events are passed through \textsc{Pythia8} (version 243)~\cite{Sjostrand:2014zea} for QCD and QED parton showering up to leading logarithmic (LL) accuracy, hadronization, and multiparton interaction / underlying event modeling. Decays of heavy-flavored hadrons and $\tau$ leptons are handled internally by the PS. Following Refs.~\cite{Fuks:2019iaj,Jager:2020hkz}, we apply the improved color reconnection~\cite{Christiansen:2015yqa} and dipole recoil~\cite{Cabouat:2017rzi} models available in  \textsc{Pythia8}. For simplicity, activity from additional proton-proton interactions occurring during the same bunch crossing, \ie, pileup, is assumed to be subtracted from experimental data by dedicated algorithms and, therefore, are not included in our simulations.

In our analysis, hadron-level events are passed to the fast detector simulator \textsc{Delphes} (version 3.4.2)~\cite{deFavereau:2013fsa}. There, particle-level clustering of hadrons is handled according to the anti-$k_T$ algorithm~\cite{Catani:1993hr,Ellis:1993tq,Cacciari:2008gp} as implemented in \textsc{FastJet} \cite{Cacciari:2005hq,Cacciari:2011ma}, with a radius parameter $R=0.4$. To emulate experimental reconstruction with realistic detector resolution and particle identification, detector responses are tuned using the \textsc{ATLAS} configuration card available in the \textsc{Delphes} repository. Particle-level distributions at LO+PS and NLO+PS were checked using \textsc{MadAnalysis5} (version 1.8)~\cite{Conte:2012fm,Conte:2014zja,Conte:2018vmg}.

\subsection{Standard Model Inputs}\label{sec:setup_sm}
For SM inputs we work in the $n_f=4$ active quark flavor scheme with a Cabibbo-Kobayashi-Maskawa mixing matrix that is diagonal with unit entries. Unless specified, we assume the following mass and coupling values:
\begin{eqnarray}
m_t(m_t) &= 172.9\GeV, 	& ~m_b(m_b)=4.7\GeV,  \nonumber\\
m_\tau   &= 1.777\GeV, 	& ~m_h = 125.1\GeV,  \nonumber\\
M_Z 	 &= 91.188\GeV, & 
~\alpha_{\rm QED}^{-1}(M_Z)=132.5070, \nonumber\\
 G_F &= 1.166390\cdot & \hspace*{-.2cm}10^{-5}\GeV^{2}.
\label{eq:sminputs}
\end{eqnarray}
For scattering computations at both LO and NLO in QCD we employ the NNPDF3.1 NLO+LUXqed parton distribution function (PDF) set with $\alpha_s(M_Z)=0.118$ \texttt{(lhaid=324900)} \cite{Manohar:2016nzj,Manohar:2017eqh,Bertone:2017bme}. For non-perturbative dynamics we tune the shower  with the ATLAS A14 central tune \texttt{(Tune:pp=21)} as paired with the NNPDF2.3 LO+QED PDF set with $\alpha_s(M_Z)=0.130$ \texttt{(pdfcode=247000)}~\cite{Ball:2013hta}.
PDF and $\alpha_s$ scale evolution are handled by \textsc{LHAPDF} (version 6.2.3) \cite{Buckley:2014ana}. PDF uncertainties are extracted using replicas PDFs~\cite{Buckley:2014ana,Bertone:2017bme}.

For signal and background processes, we set the nominal $(\zeta=1.0)$ collinear factorization $(\mu_f)$ and QCD renormalization $(\mu_r)$
scales to be half the sum over each visible, final-state  particle's transverse energy:
\begin{equation}
\mu_f, \mu_r = \zeta \times \mu_0, ~\text{with}~ \mu_0 = \frac{1}{2} \sum_{f} \sqrt{m_f^2 + p_{T,f}^2}.
\end{equation}
Here $m_f$ and $p_{T,f}$ stand for the mass and transverse momentum of the final-state particle $f$ respectively. The shower factorization scale $(\mu_s=\zeta\times\hat{\mu}_s)$ is kept at its default value prescribed in Ref.~\cite{Alwall:2014hca}. To estimate the size of  higher-order QCD corrections, we vary discretely and independently $\mu_f, \mu_r, \mu_s,$ over the set $\zeta=\{0.5,1.0,2.0\}$ to obtain a nine- or 27-point uncertainty band.

\subsection{Heavy Neutrino Inputs}\label{sec:setup_seesaw}
For simulations involving the heavy neutrino $N$ we use the default inputs of the Majorana neutrino variant of the \textsc{HeavyN} NLO UFO libraries~\cite{Degrande:2016aje}. SM particle masses are updated according to equation~\eqref{eq:sminputs}. As we are interested in the benchmark scenario featuring only one heavy mass eigenstate, additional heavy mass eigenstates are decoupled by setting $m_{N_5},m_{N_6} = 10^{10}\GeV$. To deactivate  $e$ and $\tau$ flavor mixing, we set
\begin{equation}
\vert V_{e N}\vert^2, \vert V_{\tau N}\vert^2 =0, \quad \vert V_{\mu N}\vert^2 = 1.0.
\label{eq:mixingInputs}
\end{equation}
Sensitivity to smaller values of $\vert V_{\mu N}\vert $ are obtained by a na\"ive rescaling of cross sections, which is permissible by equation \eqref{eq:bareXSecT}.
Notably, as $N$ is never resonantly produced, its total width $(\Gamma_N)$ and lifetime can be ignored.

\section{Signal and Background Modeling in MadGraph5\_aMC@NLO}\label{sec:modeling}

Generically speaking, collider processes that feature either LNV or VBF  exhibit characteristic kinematical and topological properties that enable remarkable background rejection capabilities. Consequentially, background modeling for the $W^\pm W^\pm\to\ell^\pm_i\ell^\pm_j$ channel is hindered by high background rejection rates, and hence by poor MC efficiencies. In this context, we report the development of efficient MC modeling prescriptions that overcome such difficulties for 
our signal (section~\ref{sec:modeling_sig}) and 
leading backgrounds (section~\ref{sec:modeling_bkg})  at NLO+PS within the \mgamc~simulation framework.

\subsection{Signal Modeling}\label{sec:modeling_sig}
To model the  $W^\pm W^\pm \to \ell^\pm_i \ell^\pm_j$ process when mediated  by the exchange of a $t$-channel Majorana neutrino $N$, as shown in figure~\ref{fig:wwScattHeavyN_diagram}, 
we consider the gauge-invariant set of   $W^\pm W^\pm$ scattering diagrams  contributing to the following $2\to4$, Born-level process at $\mathcal{O}(\alpha^4)$
\begin{equation}
q_1 ~q_2 ~ \to ~ q'_1 ~q'_2 ~\ell^\pm_i ~\ell^\pm_j.
\label{eq:modeling_signal}
\end{equation}
Here $q$ {denotes} any light quark or antiquark. We neglect interference with the $s$-channel process, $q\overline{q}\to W^{\pm*} \to N\ell^\pm_i \to \ell^\pm_i \ell^\pm_j q \overline{q'}$. To justify this, we require that the leading dijet system at the analysis level carries a large invariant mass, which suppresses the $N^{(*)}\to  W^{(*)}{\ell} \to  q \overline{q'}{\ell}$ splitting chain. For $m_N$ below the TeV scale, neglecting the $q\overline{q'}$ annihilation mechanism is also justified by the narrow width approximation. Under this the leading contributions to the $s$-channel process are factorizable and non-interfering with  $t$-channel  diagrams; interfering contributions  are $\mathcal{O}(\Gamma_N/m_N)\ll1$, and hence insignificant.

To carry out this modeling in \mgamc~at NLO in QCD using the \textsc{HeavyN} libraries, we employ the 
syntax\footnote{
For further details on syntax and usage,
see Refs.~\cite{Alwall:2014hca,Degrande:2016aje}.
}
\begin{verbatim}
import model SM_HeavyN_NLO
define p = g u c d s u~ c~ d~ s~
define j = p
generate    p p > mu+ mu+ j j QED=4 QCD=0 $$ 
        w+ w- / n2 n3 [QCD]
add process p p > mu- mu- j j QED=4 QCD=0 $$ 
        w+ w- / n2 n3 [QCD]
\end{verbatim}
Formally, the Born-level matrix element for equation~\eqref{eq:modeling_signal} is finite in the absence of phase space cuts. At $\mathcal{O}(\alpha_s)$, however, an infrared-safe definition for external states is needed. We therefore require at the generator-level that QCD partons are sequentially clustered according to the anti-$k_T$ algorithm with $R=0.4$ and that the transverse momentum $(p_T)$ and pseudorapidity $(\eta)$ of these  clusters satisfy the following demands:
\begin{equation}
    p_T^j > 20\GeV \quad\text{and}\quad \vert\eta^j\vert < 5.5.
    \label{cuts:gen_jets}
\end{equation}

As a technical remark, we relax checks on infrared pole cancellation in {\sc MadFKS}. Such checks are automatically raised in \mgamc~for VBF processes due to the possible omission of virtual diagrams that are mixed NLO QCD-EW corrections and not pure QCD contributions at $\mathcal{O}(\alpha_s)$. In our case such diagrams do not exist and therefore bypassing the check only impacts the MC efficiency. To do this we set  \texttt{\#IRPoleCheckThreshold=-1.0d0} in the file \texttt{Cards/FKS\_params.dat}.

\subsection{Background Modeling}\label{sec:modeling_bkg}

Due to the presence of forward, high-$p_T$ jets, a high-energy, same-sign lepton pair, and an absence of (light) neutrinos in our signal process,  several backgrounds processes that are traditionally present in collider searches for LNV can be readily suppressed through simple kinematic requirements. Among many examples are: demanding that the leading dijet system carries a large invariant mass, stringent $p_T$ cuts on the same-sign leptons, and vetoing events with three or more charged leptons. As a result, background categories such as associated top quark production, $t\overline{t}B$ with $B\in\{W^\pm/\gamma^*/Z,h\}$; single top quark channels, $tB$; and triboson production $WWV$ can be neglected for the purposes of our study.

The leading background channels that remain after such baseline selection criteria include: 
the mixed EW-QCD channel $W^\pm W^\pm jj$ (section \ref{sec:modeling_bkg_qcd}), 
the pure EW channel $W^\pm W^\pm jj$ (section \ref{sec:modeling_bkg_ew}), 
and the mixed EW-QCD diboson+jets process $W^\pm V + nj$ 
with $V\in\{\gamma^*/Z\}$ (section \ref{sec:modeling_bkg_diboson}). To model these processes in \mgamc, we employ the  prescriptions described below.

\subsubsection{QCD Production of Same-Sign $W^\pm W^\pm jj$}\label{sec:modeling_bkg_qcd}
Due to its similar topology and kinematic scales, the mixed EW-QCD production (henceforth labeled QCD production) of  $pp\to W^\pm W^\pm jj$  is a prominent background  for the $pp\to\ell^\pm\ell^\pm j j$ signal process. However, a defining characteristic of this mode, which at the Born level occurs at $\mathcal{O}(\alpha^2\alpha_s^2)$, is the $t$-channel exchange of a gluon. This indicates an intermediate flow of color. As such, the presence of a third, high-$p_T$ jet is significantly more likely in this process than in the signal process. Such radiation can induce recoils in the $(W^\pm W^\pm)$-system, siphon energy from the two forward jets, or give rise to excess central hadronic activity~\cite{Barger:1990py,Barger:1991ar,Bjorken:1992er,Fletcher:1993ij,Barger:1994zq}. One is thus motivated to consider the process at NLO in QCD. 

We do this in \mgamc~ by using the syntax:
\begin{verbatim}
import model loop_sm
generate    p p > w+ w+ j j QED=2 QCD=2 [QCD]
add process p p > w- w- j j QED=2 QCD=2 [QCD]
\end{verbatim}
As in the signal process, we impose the criteria of equation~\eqref{cuts:gen_jets} on outgoing QCD partons. We report in the first line of table~\ref{tab:mcxsec_bkg} the corresponding generator-level cross section at $\sqrt{s}=13\TeV$, which reaches $\sigma\sim385\fb$, along with residual $\mu_f, \mu_r$ scale and PDF uncertainties. Before parton showering, resonant $W$ bosons are decayed to muons (see section~\ref{sec:setup_mc} for related details). 

For simplicity, we neglect contributions from leptonic tau decays. We do so because the presence of additional light neutrinos results in events with characteristically softer muons and larger momentum imbalances. Such features can be tamed by tuning the pre-selection and signal region cuts used in section \ref{sec:analysis}.

\begin{table}[!t]
 \setlength\tabcolsep{2pt}
 \renewcommand{\arraystretch}{1.6}
\begin{center}
\caption{
  Generator-level cross sections [fb] and cuts, $\mu_f,\mu_r$ scale uncertainty [\%], {PDF  uncertainties} [\%], and perturbative order  for leading backgrounds at $\sqrt{s}=13\TeV$.
} \resizebox{\columnwidth}{!}{
\begin{tabular}{c c c c c c}
\hline\hline
Process   & Order & Cuts & $\sigma^{\rm Gen.}$ [fb] & $\pm\delta_{\mu_f,\mu_r}$ & $\pm\delta_{\rm PDF}$ \\
\hline
$W^\pm W^\pm jj$ (QCD) & NLO in QCD & Eq.~\eqref{cuts:gen_jets} &
$385$ & $^{+10\%}_{-10\%}$ & $^{+1\%}_{-1\%}$ \\
\hline
\multirow{2}{*}{$W^\pm W^\pm jj$ (EW)} & \multirow{2}{*}{NLO in QCD} 
& Eq.~\eqref{cuts:gen_jets} + & \multirow{2}{*}{$254$} &
\multirow{2}{*}{$^{+1\%}_{-1\%}$} & \multirow{2}{*}{$^{+1\%}_{-1\%}$}\\
    & & diagram removal & & & \\
\hline
Inclusive $W^\pm V~(3\ell\nu)$  & FxFx (1j) &  Eqs.~\eqref{cuts:gen_3lv_lep}, \eqref{cuts:gen_3lv_FxFx} &  $2,520$ 
& $^{+5\%}_{-6\%}$ & $^{+1\%}_{-1\%}$
\\
\hline\hline
\end{tabular}
} 
\label{tab:mcxsec_bkg}
\end{center}
\end{table}

\subsubsection{EW Production of Same-Sign $W^\pm W^\pm jj$ }\label{sec:modeling_bkg_ew}
Like the previous case, pure EW production of  $pp\to W^\pm W^\pm jj$, which at the Born level occurs at $\mathcal{O}(\alpha^4)$, is a leading background to the $pp\to\ell^\pm\ell^\pm j j$ signal process. Similarities to the signal process's topology, kinematic characteristics, and, importantly, color flow are identifiable in the $W^\pm W^\pm \to W^\pm W^\pm$, VBF sub-process. While featuring a  large cross section at the inclusive level, other sub-processes, such as resonant triboson production, are essentially removed through basic selection criteria. 

To efficiently model the EW same-sign $W^\pm W^\pm$ channel at NLO in QCD, we make a variant of the so-called ``vector boson fusion approximation'' and neglect resonant triboson production in a gauge-invariant manner. We do this under the presumption that relevant analysis-level selection cuts are applied. Other non-resonant, interfering diagrams are kept. The   syntax we employ is 
\begin{verbatim}
generate    p p > w+ w+ j j $$
        a z w+ w- QCD=0 QED=4 [QCD]
add process p p > w- w- j j $$ 
        a z w+ w- QCD=0 QED=4 [QCD]
\end{verbatim}

We apply the same loose, generator-level cuts as listed in equation~\eqref{cuts:gen_jets} and report good numerical stability. In the second line of table~\ref{tab:mcxsec_bkg},
we report the generator-level cross section  at $\sqrt{s}=13\TeV$, which reaches $\sigma\sim254\fb$, and associated uncertainties. Resonant $W$ bosons are decayed to muons before parton showering.

\subsubsection{Inclusive Diboson Spectrum}\label{sec:modeling_bkg_diboson}

Due to its picobarn-scale rate, the inclusive $pp\to 3\ell\nu+X$ spectrum (which we imprecisely label ``diboson spectrum'' even though the process includes interference with all non-resonant diagrams) contributes to the same-sign dilepton signature $pp\to  \ell^\pm \ell^\pm j j$ through pathological configurations of the final-state kinematics. Such configurations include, for example, when two or more initial-state QCD emissions both possess large $p_T$ but the odd-sign charged lepton is not successfully identified because it is too soft in $p_T$ or too forward in $\vert\eta\vert$.

While a bulk of these phase space configurations are captured by the fixed-order matrix element for the $pp\to 3\ell\nu jj$ process,  which at the Born level occurs at $\mathcal{O}(\alpha^4 \alpha_s^2)$, convergence in perturbative QCD requires that outgoing QCD partons are hard (high $p_T$) and central (low $\vert\eta\vert$). In light of the anticipated scales of the $(3\ell\nu)$-system, the component of phase space where one QCD parton is central and one is  forward is better described  by the fixed-order matrix element for the $pp\to 3\ell\nu j$ sub-process, with the second $j$ being populated by the PS. Likewise, for two forward emissions, the phase space is better described best by the    $pp\to 3\ell\nu $ sub-process with two PS emissions.

To model these complications  we extend the brute-force prescription of Ref.~\cite{Pascoli:2018heg}. This entails starting with the  $2\to4$ process, $pp\to 3\ell\nu$ at NLO in QCD, with all interfering, non-resonant diagrams and without invoking the narrow width approximation for intermediate $W ,Z$ bosons. Instead, loose, generator-level cuts on leptons are imposed to regulate $s$- and $t$-channel divergences, thereby keeping matrix elements finite and perturbative. Specifically, we require that  charged leptons satisfy
\begin{equation}
    m_{\rm os}^{\ell\ell}>8\GeV \quad\text{and}\quad \vert \eta^\ell \vert < 4.0,
    \label{cuts:gen_3lv_lep}
\end{equation}
where $m_{\rm os}^{\ell\ell}$ is the invariant mass of any opposite-sign, charged lepton pair, independent of flavor. While smaller $m_{\rm os}^{\ell\ell}$ thresholds can still  regulate $\gamma^*\to\ell\ell$ splittings, we refrain from doing so to avoid contributions from vector meson resonances. Such states are not modeled with the perturbative event generator~\mgamc.

To account for {additional} central jet multiplicities, we {match} the  inclusive $pp\to 3\ell\nu$ spectrum at NLO+PS up to its first jet multiplicity (relative to the Born level) at NLO in QCD using the FxFx prescription~\cite{Frederix:2012ps}. The relevant \mgamc~syntax in this case is
\begin{verbatim}
define ell  = e+ mu+ ta+ e- mu- ta-
define vv   = ve vm vt ve~ vm~ vt~
generate    p p > ell ell ell vv   
        QED=4 QCD=0 [QCD] @0
add process p p > ell ell ell vv j 
        QED=4 QCD=1 [QCD] @1
\end{verbatim}
Explicitly, we use the following jet-matching inputs:
\begin{equation}
    p_T^j > 30\GeV, \quad \vert\eta^j\vert < 5.5, \quad Q_{\rm cut}^{\rm FxFx} = 65\GeV,
        \label{cuts:gen_3lv_FxFx}
\end{equation}
where $Q_{\rm cut}^{\rm FxFx}$ is the FxFx matching scale.
With this setup, hadronic observables are accurate to at least LO+PS(LL), with the two leading jets being defined at all momenta and rapidities. We report in the third line of table~\ref{tab:mcxsec_bkg}
 the generator-level cross section of the diboson spectrum  at NLO in QCD with FxFx matching to the first jet multiplicity (FxFx1j) and its associated uncertainties. At $\sqrt{s}=13\TeV$ the rate is about $\sigma\sim2.5\pb$, which is just slightly larger than the rate at NLO, which we compute to be $\sigma\sim2.3\pb$.

With this setup, we capture configurations where the odd-sign charged lepton in the $3\ell\nu2j+X$ final state is too forward or too soft to be  identified as an analysis-quality charged lepton. A disadvantage of this setup, however, is the limited MC statistics when the two same-sign charged leptons carry $p_T^\ell\gtrsim 100-150\GeV$ but the odd-sign lepton is much softer. To enrich MC statistics for this region of phase space, we introduce tailored generator-level cuts into the \mgamc~phase space integration routines. Enriched samples are combined with the baseline FxFx1j sample. Overlap is removed through cuts on $p_T^{\ell_2}$. For technical details of this modeling, see appendix~\ref{app:fxfxCuts}.

\begin{figure}
\includegraphics[width=\columnwidth]{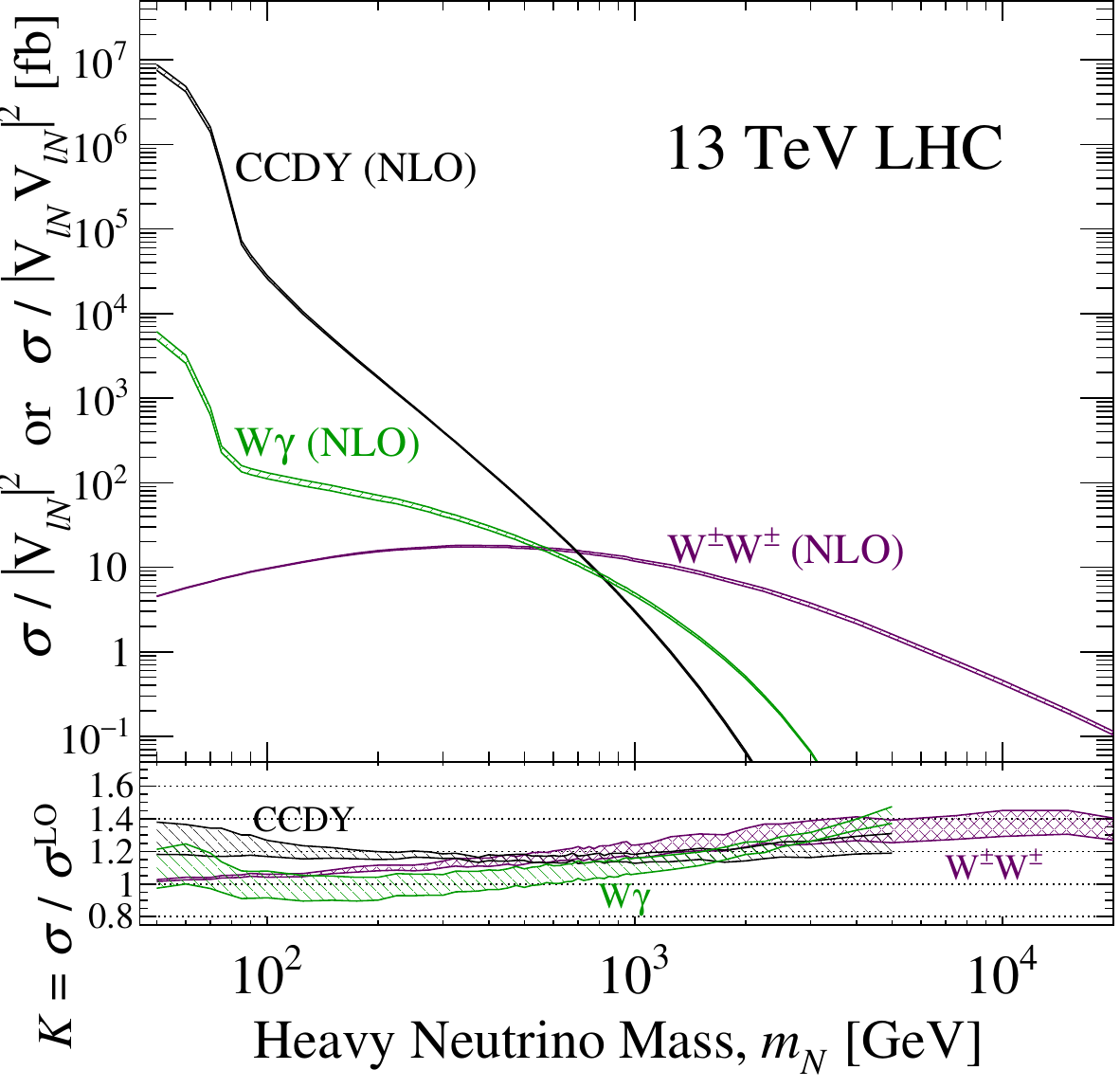}
\caption{
Upper:
As a function of heavy neutrino mass $m_N$ [GeV], the bare cross section $\sigma/\vert V_{\ell N}V_{\ell N}\vert^2$ [fb] for the $W^\pm W^\pm$ signal process at NLO in QCD (purple band), as well as the bare cross sections $\sigma/\vert V_{\ell N}\vert^2$ for the CCDY (black band) and $W\gamma$ fusion (green band) processes at NLO in QCD. Band thickness corresponds to the residual scale uncertainty.
Lower: The QCD $K$-factor for each channel.
}
\label{fig:wwScatt_XSec_vs_mN_LHCX13}
\end{figure}

\section{Heavy Neutrinos in $W^\pm W^\pm$ Scattering at the LHC}\label{sec:wwScattLHC}

In this section we investigate the phenomenology of the $W^\pm W^\pm \to \ell^\pm_i \ell^\pm_j$ process when mediated by a heavy Majorana neutrino at the LHC. To do this, we examine the integrated (section~\ref{sec:wwScattLHC_total}) and differential (section~\ref{sec:wwScattLHC_diff}) cross sections of the $W^\pm W^\pm$ channel, and place special focus on the low- (section \ref{sec:wwScattLHC_total_loMass}) and high-mass (section \ref{sec:wwScattLHC_total_hiMass}) limits of the intermediate  neutrino, on the impact of QCD corrections (section \ref{sec:wwScattLHC_total_qcd}), and on potential violations of partial-wave unitarity (section \ref{sec:wwScattLHC_total_unitarity}).

\subsection{Total Production Rate}\label{sec:wwScattLHC_total}

As a first step, we present in the upper panel of figure~\ref{fig:wwScatt_XSec_vs_mN_LHCX13} and as a function of heavy neutrino mass $m_N$, the total cross section for the full $2\to 4$, hadron-level process
\begin{equation}
    p p \to \ell^\pm \ell^\pm j j +X.
\label{eq:2to4proc}\end{equation}
More precisely, we evaluate the bare cross section, as defined in equation~\eqref{eq:bareXSecT}, at NLO in QCD and for LHC collisions at $\sqrt{s}=13\TeV$, assuming the exchange of a single heavy neutrino that couples to a single charged lepton flavor. In equation~\eqref{eq:2to4proc}, $X$ denotes any additional hadronic and photonic activity present in the inclusive process. The band thickness corresponds to the residual renormalization and collinear factorization scale dependence at NLO, as quantified in section~\ref{sec:setup_sm}. We assume the generator-level cuts of equation \eqref{cuts:gen_jets}. To quantify the size of $\mathcal{O}(\alpha_s)$ corrections, we  show in the lower panel of figure~\ref{fig:wwScatt_XSec_vs_mN_LHCX13} the NLO in QCD $K$-factor, defined as the  ratio  of the NLO and LO cross sections:
\begin{equation}
    K^{\rm NLO} = \sigma^{\rm NLO} ~/~ \sigma^{\rm LO}.
    \label{eq:kFacto}
\end{equation} 

For the mass range $m_N = 40\GeV-20\TeV$, we report that bare cross sections at NLO in QCD, QCD $K$-factors, as well as scale and PDF uncertainties roughly span
\begin{eqnarray}
    \sigma^{\rm NLO}        &:& 0.1-20\fb,\\
    K^{\rm NLO}             &:& 1.05-1.4, \\
    \delta\sigma_{\mu_r,\mu_f} / \sigma &:& \pm1\% - \pm5\%, \\
    \delta\sigma_{\rm PDF} / \sigma     &:&  \pm1\% - \pm2\%.
\end{eqnarray}
A summary of bare cross sections and uncertainties for the $W^{\pm}W^{\pm}\to \ell^{\pm} \ell^{\pm}$ process at representative heavy neutrino masses is listed in table~\ref{tab:heavyNxsec}. 

To compare to other heavy neutrino processes,\footnote{For these additional channels, we follow the prescription of Ref.~\cite{Degrande:2016aje} with updated inputs as listed in section~\ref{sec:modeling}. To regulate the $W\gamma$ fusion matrix element, we use the phase space cuts of equation~\eqref{cuts:gen_jets} as well as require $p_T^\ell > 10\GeV$ and $\vert\eta^\ell\vert<4.0$. Total widths of SM particles are kept at their SM values.
}
we also present in figure~\ref{fig:wwScatt_XSec_vs_mN_LHCX13} the bare cross sections at NLO in QCD, as  defined in  equation~\eqref{eq:bareXSecS}, the associated scale uncertainties, and QCD $K$-factor  for the $2\to2$, charged current Drell-Yan (CCDY) process (black band),
\begin{eqnarray}
p p \to W^{\pm (*)} + X \to N \ell^\pm + X,
\end{eqnarray}
and the $2\to 3$, $W\gamma$ fusion process (green band),
\begin{eqnarray}
p p \to N \ell^\pm j + X.
\end{eqnarray}

\begin{table}[!t]
\renewcommand{\arraystretch}{1.4}
\begin{center}
 \setlength\extrarowheight{3pt}
 \scriptsize
\caption{
For representative heavy neutrino masses $(m_N)$
and active-sterile mixing $V_{\ell N}=1$,
the  $pp\to\ell^\pm\ell^\pm j j+X$ cross section [fb]
at NLO in QCD,
with residual scale uncertainties [\%],
PDF uncertainties [\%],
and NLO $K$-factor.
}\resizebox{.9\columnwidth}{!}{
\begin{tabular}{c c c c c}
\hline\hline
$m_N$   
& $\sigma^{\rm NLO}$ [fb] & $\pm\delta_{\mu_f,\mu_r}$ 
& $\pm\delta_{\rm PDF}$ 
& $K^{\rm NLO}$ \\
\hline
150\GeV & $13.3$ & $^{+1\%}_{-2\%}$ & $^{+1\%}_{-1\%}$ & $1.09$ \\
1.5\TeV & $8.45$ & $^{+4\%}_{-4\%}$ & $^{+1\%}_{-1\%}$ & $1.26$ \\
5.0\TeV & $1.52$ & $^{+5\%}_{-5\%}$ & $^{+2\%}_{-2\%}$ & $1.32$ \\
15\TeV  & $0.190$ & $^{+5\%}_{-5\%}$ & $^{+2\%}_{-2\%}$ & $1.32$ \\
\hline\hline
\end{tabular}
} 
 \label{tab:heavyNxsec}
\end{center}
\end{table}

We find several notable observations:
First is that, quantitatively, the bare, same-sign $WW$ cross section is about \confirm{$4-6~(1-3)$ orders of magnitude smaller} than the CCDY $(W\gamma)$ process for $m_N\sim50-100\GeV$. This is much smaller than the \confirm{$4~(2)$} orders of magnitude that one expects from na\"ive power counting. Second is that while the bare rates of resonant channels fall precipitously  for increasing $m_N$, which is due to suppression in both the matrix element and available phase space, the $W^\pm W^\pm$ rate moderately \textit{increases} before slowly decreasing. 
For an active-sterile mixing of $\vert V_{\ell N}\vert^2 =1$, this leads to the  $W^\pm W^\pm$ rate surpassing the $W\gamma$ rate at \confirm{$m_N\sim500-600\GeV$} and the CCDY rate at \confirm{$m_N\sim700\GeV$}. Due to the different sensitivities of the three channels to active-sterile  mixing, the crossover occurs at higher neutrino masses for smaller values of $\vert V_{\ell N}\vert^2$. \confirm{For example: at $\vert V_{\ell N}\vert^2 =0.1$, the $W^\pm W^\pm$ cross section surpasses the $W\gamma$ (CCDY) rate at $m_N=1.8-1.9~(1.2-1.3)\TeV$.}

While mixing can alter the precise values of these crossovers, the qualitative picture does not change.  For instance: independent of $\vert V_{\ell N}\vert^2$, the $W^\pm W^\pm\to\ell^\pm\ell^\pm$ cross section exhibits a qualitatively different dependence on  $m_N$ than in the CCDY and $W\gamma$ channels. This leads to the  $W^\pm W^\pm$ rate at $\sqrt{s}=13\TeV$ to be the same at both $m_N\sim 40\GeV$ and $m_N\sim2.5\TeV$. Moreover, unlike resonant production of heavy neutrinos via $q\overline{q'}$ annihilation or $W\gamma$ fusion,  heavy  neutrinos in $t$-channel processes like $W^\pm W^\pm \to \ell^\pm_i \ell^\pm_j$ are non-resonant. So while there is a kinematic suppression in the $W^\pm W^\pm$ matrix element at very large $m_N$, there is no corresponding phase space suppression. This manifests in the cross section as a milder dependence on increasing sterile neutrino masses. Interestingly, as heavy neutrinos are never on-shell in $t$-channel exchanges they can never manifest as a long-lived particle. Therefore, search complications associated with displaced vertices are not present.

In comparison to past work, this is the first evaluation of the full $2\to4$, same-sign $W^\pm W^\pm$ scattering process in $pp$ collisions at NLO in QCD. At LO, the literature ~\cite{Dicus:1991fk,Ali:2001gsa,Panella:2001wq,Atre:2009rg,Aoki:2020til} is admittedly in disagreement with itself. Qualitatively, the dependence on collider energy and heavy neutrino masses in all these works are consistent. Quantitatively, large differences exist. In some cases, differences can be traced to the omission of numerical pre-factors in analytic and/or numerical results, theoretical uncertainties associated with the effective $W$ approximation~\cite{Dawson:1984gx,Kane:1984bb,Kunszt:1987tk}, and uncertainties in PDF sets. In other cases, the lack of documented  inputs and possible phase space cuts hinder precise comparisons. Support for our numerical results include agreement with analytical expressions for helicity amplitudes. For further details, we refer to sections \ref{sec:wwScattLHC_total_loMass} and \ref{sec:wwScattLHC_total_hiMass}, and appendix \ref{app:helicity}.

To further understand the dependence on $m_N$ in the $W^\pm W^\pm$ channel, we consider for illustration purposes the matrix element for the $2\to2$, $W^\pm W^\pm \to \ell^\pm \ell^\pm$ sub-process. For the momentum and helicity assignments
\begin{align}
    W^+_\mu(p_1^W,\lambda_1^W) + W^+_\nu(p_2^W,\lambda_2^W) \to 
    \nonumber\\
    \ell^+(p_1^\ell,\lambda_1^\ell) &+ \ell^+(p_2^\ell,\lambda_2^\ell),
\end{align}
and the invariants
$M_{WW}^2  = (p_1^W+p_2^W)^2$, 
$t = (p_1^W-p_1^\ell)^2$, and
$u = (p_1^W-p_2^\ell)^2$,
 the {helicity} amplitudes are given by
\begin{eqnarray}
-i\mathcal{M} &=&
\varepsilon_\mu(p_1^W,\lambda_1^W)\varepsilon_\nu(p_2^W,\lambda_2^W)\mathcal{T}^{\mu\nu}(p_1^\ell,p_2^\ell,\lambda_1^\ell,\lambda_2^\ell)    
\nonumber\\
& & + (t\leftrightarrow u).
\label{eq:meWWScatt}
\end{eqnarray}
Here $\varepsilon$ are the usual helicity polarization vectors for massive gauge bosons in the unitary gauge, the $(t\leftrightarrow u)$ term accounts for final-state lepton exchange, and following the Feynman rules of Refs.~\cite{Denner:1992vza,Denner:1992me}, the LN-violating current $(\mathcal{T})$  in the \textsc{HELAS} convention~\cite{Murayama:1992gi} 
is\footnote{In several instances our  analytic results differ from those  in Ref.~\cite{Dicus:1991fk}. We do not speculate on their specific origin but note that some omissions are obviously typographical.
}~\cite{Dicus:1991fk}
\begin{align}
& \mathcal{T}^{\mu\nu}  
= 
-i\left(\frac{-ig_W}{\sqrt{2}}\right)^2 
\frac{V_{\ell N} V_{\ell N}}{\left(t - m_N^2\right)} \times
\\
& 
\left[
\overline{u}(p_1^\ell,\lambda^\ell_1=R)\gamma^\mu P_R
\left(\not\!p_N + m_N\right) \gamma^\nu P_L  v(p_2^\ell,\lambda_2^\ell=R)\right]
\nonumber\\
& \qquad =
-i\left(\frac{-ig_W}{\sqrt{2}}\right)^2 
\frac{V_{\ell N} V_{\ell N} }{\left(t - m_N^2\right)} \times m_N \times
\label{eq:lnvCurrent}
\\
&  
\left[
\overline{u}(p_1^\ell,\lambda_1^\ell=R)\gamma^\mu
\gamma^\nu P_L  v(p_2^\ell,\lambda_2^\ell=R)\right].
\nonumber
\end{align}
In the above we assume a clockwise fermion flow of leptons~\cite{Denner:1992vza,Denner:1992me}, $p_N = (p_1^W - p_1^\ell)$ is the momentum of the internal  sterile neutrino, and  $u(p,\lambda)$ and $v(p,\lambda)$ are the standard helicity spinors for massless, spin-$1/2$ fermions.

Crucially, differences in Feynman rules for LN-violating fermion currents relative to the standard rules for LN-conserving ones give rise to an effective parity inversion in the $W_1^+-\ell_1^+-N$ vertex and spinor for $\ell_1^+$ \cite{Kayser:1982br,Mohapatra:1991ng}. This implies~\cite{Han:2012vk,Ruiz:2020cjx} that the successive gauge interactions involving massless particles  in $\mathcal{T}$ are \textit{helicity inverting} and not helicity preserving as one usually finds in SM gauge interactions involving massless, external particles. Subsequently, projection operators select for the heavy neutrino's RH helicity state, and hence the factor of $m_N\times \mathbb{I}_4$ in $\mathcal{T}$. This is in contrast to LN-conserving currents, such as in the process $W^+W^- \to \ell^+\ell^-$, where projection operators select for the LH helicity state, and hence the $\not\! p_N$ term in $\mathcal{T}$.

We report that exact, analytic evaluation of equations \eqref{eq:meWWScatt} and \eqref{eq:lnvCurrent} yields somewhat bulky expressions without obvious insights. This is despite being a $2\to2$ process and can be tied to the added algebraic complication of the incoming $W$ bosons being massive and carrying a longitudinal polarization. Instead, we focus on the low- (section \ref{sec:wwScattLHC_total_loMass}) and high-mass (section \ref{sec:wwScattLHC_total_hiMass}) limits of the intermediate heavy neutrino. For technical details and intermediate expressions, see appendix \ref{app:helicity}.

\subsubsection{Low-Mass Limit}\label{sec:wwScattLHC_total_loMass}

We consider first the limit where masses of both $W$ and $N$ are small compared to the $W^{\pm}W^{\pm}$ scattering scale, \ie, when $m_W,m_N\ll M_{WW}$. In this limit, the LN-violating tensor current in equation \eqref{eq:lnvCurrent} scales as
\begin{align}
 \mathcal{T}^{\mu\nu} &\propto g_W^2 V_{\ell N} V_{\ell N} \frac{m_N ~ M_{WW}}{\left(t - m_N^2\right)}  \\
 & \sim  g_W^2  V_{\ell N} V_{\ell N} \frac{ m_N}{M_{WW}}  + \mathcal{O}\left({\frac{m_N^2}{M_{WW}^2}, \frac{m_W^2}{M_{WW}^2}}\right),
 \label{eq:meTScaleSmallMLimit}
\end{align}
where the $M_{WW}$ factor in the numerator originates from the two  lepton spinors, $u(p^\ell),v(p^\ell)\sim \sqrt{M_{WW}}$. In this same limit, the scattering of longitudinally polarized $W$ bosons is enhanced over the {scattering of} the transverse polarizations. This enhancement can be seen in the polarization vectors themselves, which scale as
\begin{align}
\varepsilon_\mu(p^W,\lambda^W=\pm) &\sim \mathcal{O}(1), \\
\varepsilon_\mu(p^W,\lambda^W=0) &\sim \frac{p^W_\mu}{m_W} + \mathcal{O}\left(\frac{m_W}{M_{WW}}\right) \\
&\sim \mathcal{O}\left(\frac{M_{WW}}{m_W}\right).
\end{align}
This shows that in the high-energy limit the $W^\pm W^\pm\to \ell^\pm\ell^\pm$ process  is driven by $W^\pm_0 W^\pm_0$ scattering and that the corresponding matrix element scales as
\begin{align}
    -i\mathcal{M} &= 
    \varepsilon_\mu(\lambda_1^W=0)\varepsilon_\nu(\lambda_2^W=0)\mathcal{T}^{\mu\nu}   + (t\leftrightarrow u) \\
    &\sim  g_W^2 V_{\ell N} V_{\ell N} \frac{m_N}{m_W^2} M_{WW}.
    \label{eq:meWWScatt_scaling_loMass}
\end{align}
Remarkably, after squaring $\mathcal{M}$, the quadratic dependence on $M_{WW}$ is canceled by the flux factor in the definition of the parton-level cross section $\hat{\sigma}$. This renders the rate independent of $M_{WW}$ but quadratic in $m_N$,
\begin{equation}
    \hat{\sigma}(W^+ W^+\to \ell^+\ell^+) \sim g_W^4  \vert V_{\ell N}\vert^4 \frac{m_N^2}{m_W^4}.
\end{equation}
Thus, we  can attribute the growth in the same-sign  $WW$ scattering rate seen in figure~\ref{fig:wwScatt_XSec_vs_mN_LHCX13} for  sub-TeV heavy neutrinos to the cancellation of momentum scales in high-energy $W^\pm_0 W^\pm_0$ scattering in tandem with helicity inversion in the LN-violating lepton current.

After a more careful computation (see appendix \ref{app:helicity}), the $W^+ W^+\to\ell^+_1\ell^+_2$ cross section for $n_R$ heavy neutrinos is
\begin{align}
    \hat{\sigma}(W^+ W^+\to\ell^+_1\ell^+_2) 
    &= 
    \frac{g_W^4 (2-\delta_{\ell_1\ell_2})}{2^5~3^2\pi m_W^4}
 \Bigg\vert \sum_{k=4}^{n_R+3}V_{\ell_1 k} m_{N_{k}} V_{\ell_2 k}\Bigg\vert^2 
    \nonumber\\
    &+ {\mathcal{O}\left(\frac{m_N^2}{M_{WW}^2}, \frac{M_{WW}^2}{m_W^2}\right)}.
    \label{eq:xsecWWScattLowMLimit}
\end{align}

\subsubsection{High-Mass Limit}\label{sec:wwScattLHC_total_hiMass}

We consider now the kinematic limit where the $W$ boson's mass and all momentum-transfer scales are small compared to the sterile neutrino's mass, \ie, the decoupling limit~\cite{Appelquist:1974tg} where $m_W^2, M_{WW}^2, \vert t\vert,\vert u\vert \ll m_N^2$. In this limit, $N$-exchanges can be treated as contact interactions, and the pole structure of its propagator can be systematically expanded. Doing this causes the LN-violating tensor current of equation \eqref{eq:lnvCurrent} to scale as
\begin{align}
 \mathcal{T}^{\mu\nu} &\propto g_W^2 V_{\ell N} V_{\ell N} 
 \frac{m_N M_{WW}}{\left(t - m_N^2\right)} 
 \\
 &\sim -g_W^2 V_{\ell N} V_{\ell N}\frac{M_{WW}}{m_N}  + \mathcal{O}\left(\frac{\vert t \vert}{m_N^2}\right).
 \label{eq:meTScaleLargeMLimit}
\end{align}
An analogous expression holds for the $u$-channel.

A consequence of this expansion is that the angular dependence that encapsulates forward- and backward-scattering enhancements in gauge interactions becomes a sub-leading contribution in the propagator. This implies that for most $W^\pm W^\pm$ polarization combinations the forward ($t$-channel) and backward ($u$-channel) helicity amplitudes are indistinguishable, save for a relative minus sign that triggers an exact destructive interference. As a result, the only non-vanishing amplitudes are those where the incoming $W^\pm W^\pm$ states carry the same helicity. 

Noting once more the enhancement of longitudinal-longitudinal  scattering over other $W^\pm W^\pm$ helicity configurations, one finds that the leading  contribution to the matrix element for the $W^\pm_0 W^\pm_0\to\ell^\pm\ell^\pm$ process  scales as
\begin{align}
    -i\mathcal{M} &= 
    \varepsilon_\mu(\lambda_1^W=0)\varepsilon_\nu(\lambda_2^W=0)\mathcal{T}^{\mu\nu}   + (t\leftrightarrow u) \\
    &\sim  g_W^2 \frac{V_{\ell N} V_{\ell N}}{m_N} \frac{M_{WW}^3}{m_W^2}.
        \label{eq:meWWScatt_scaling_hiMass}
\end{align}
After squaring, the dependence on $M_{WW}$ is partially compensated by the flux factor in the parton-level cross section. The result is a total rate that scales as 
\begin{equation}
    \hat{\sigma}(W^+ W^+\to \ell^+\ell^+) \sim g_W^4  \frac{\vert V_{\ell N}\vert^4}{m_N^2} \frac{M_{WW}^4}{m_W^4}.
\end{equation}

Immediately, we see that the $m_N^{-2}$ factor originating from the heavy neutrino propagator is never fully offset by the LN-violating current or other factors at the cross section level. Ultimately, this is responsible for the drop in cross section that occurs in figure~\ref{fig:wwScatt_XSec_vs_mN_LHCX13} for increasing $m_N$.

After a more careful computation (see appendix \ref{app:helicity}), one finds that the parton-level, $W^+ W^+ \to\ell^+_1\ell^+_2$ 
cross section for the exchange of $n_R$ heavy neutrinos is:
\begin{align}
  \hat{\sigma}& (W^+W^+\to\ell^+_1\ell^+_2)
   =     \frac{g_W^4 (2-\delta_{\ell_1\ell_2})}{2^7 3^2 \pi}\frac{M_{WW}^4}{m_W^4}\\
&    
  \times\Bigg\vert \sum_{k=4}^{n_R+3} \frac{V_{\ell_1 N_k}  V_{\ell_2 N_k}}{m_{N_k}} \Bigg\vert^2
   + \mathcal{O}\left(
   \frac{M_{WW}^2}{m_N^2},
   \frac{M_{WW}^2}{m_W^2}
   \right).
   \nonumber
\end{align}
We stress that the transition rate's dependence on masses and mixing elements of heavy neutrinos mirrors the scaling behavior found in nuclear $0\nu\beta\beta$ decay rates~\cite{Dicus:1991fk,Atre:2009rg}. Notably, the coherent summation over $V$ permits complex phases to trigger potentially large cancellations in analogy to the  ``funnel behavior'' in $0\nu\beta\beta$ decay.

\subsubsection{QCD Corrections to $W^\pm W^\pm\to \ell^\pm\ell^\pm$}\label{sec:wwScattLHC_total_qcd}

Returning to figure~\ref{fig:wwScatt_XSec_vs_mN_LHCX13}, we recall that past investigations into the $W^\pm W^\pm \to  \ell^\pm_i \ell^\pm_j$ process historically \cite{Dicus:1991fk,Ali:2001gsa,Chen:2008qb} employed the effective $W$ approximation~\cite{Dawson:1984gx,Kane:1984bb,Kunszt:1987tk}. In this approximation, $W$ bosons are treated as constituents of the proton and the $2\to 2$, $W^\pm W^\pm \to  \ell^\pm_i \ell^\pm_j$ scattering rate is convolved with $W$ boson PDFs of the proton. Only  later in Ref.~\cite{Atre:2009rg} was the full $2\to 4$ process with forward jets evaluated. In all these cases, however, only LO estimates of cross sections were calculated. Therefore, we are  motivated to comment on the size of NLO in QCD corrections and residual uncertainties in the full $2\to 4$ process, particularly in relation to those of the CCDY and $W\gamma$ modes, which were first reported in Refs.~\cite{Ruiz:2015gsa,Degrande:2016aje}.

In the  lower panel of figure~\ref{fig:wwScatt_XSec_vs_mN_LHCX13}   we show as a function of heavy neutrino mass  the QCD $K$-factors, as defined in equation  \eqref{eq:kFacto}, for the same-sign $W^\pm W^\pm$, CCDY, and $W\gamma$ processes. Band thicknesses correspond to the residual $\mu_r, \mu_f$ dependence at NLO. Over the mass range $m_N \sim 40\GeV - 20\TeV$, we find that QCD corrections gradually and uniformly increase the total $W^\pm W^\pm$ rate by \confirm{$+5\%$ to $+35\%$}. As in deeply inelastic scattering, one-loop QCD corrections to spacelike EW emissions in VBF do not appreciably alter cross section normalizations~\cite{Han:1992hr}. Consequentially, we attribute the increase in cross section at NLO to real, initial-state radiation. This purported reliance on an $\mathcal{O}(\alpha_s(\mu_r))$ splitting is supported by the increased scale dependence at larger $m_N$ and the lack of a scale dependence in the $2\to4$  process. In comparison to the other  channels, the $K$-factor for $W^\pm W^\pm$ sits just below (above) the CCDY ($W\gamma$) curve for $m_N\sim750\GeV$ and  overtakes both at $m_N\gtrsim1\TeV$.

For corrections beyond NLO in QCD, one can consider two complementary directions.  The first pertains to higher-order QCD corrections while the second pertains to EW corrections. Based on results for VBF production of the SM Higgs {boson}~\cite{Bolzoni:2010xr,Bolzoni:2011cu,Dreyer:2018qbw}, we anticipate that improvements at $\mathcal{O}(\alpha_s^2)$ and $\mathcal{O}(\alpha_s^3)$ have only a modest impact on total cross sections and distributions. At NLO in EW, corrections to the LO rate typically scale as
\begin{align}
     \Bigg\vert\frac{\delta\sigma^{\rm NLO-EW}}{\sigma^{\rm LO}} \Bigg\vert
     \sim \frac{g_W^2}{4\pi}\log\frac{M_{WW}^2}{m_W^2}.
\end{align}
As we show in section \ref{sec:wwScattLHC_diff}, scales for $M_{WW}$ at the LHC range {about} $M_{WW}\sim300\GeV-600\GeV$ for a large array of heavy neutrino masses. This translates to a modest uncertainty of \confirm{$\delta\sigma^{\rm NLO-EW}/\sigma^{\rm LO}\sim9\%$ to $13\%$}. We anticipate that in both cases the impact of missing higher-order corrections is negligible for discovery purposes.


\subsubsection{Partial-wave unitarity in $W^\pm_0 W^\pm_0\to \ell^\pm\ell^\pm$}\label{sec:wwScattLHC_total_unitarity}

As a brief remark, we interestingly note that the scattering amplitude for the $W^\pm W^\pm\to \ell^\pm\ell^\pm$ process exhibits poor high-energy behavior when initiated by a pair of longitudinally polarized $W$ bosons. As evident in equations \eqref{eq:meWWScatt_scaling_loMass} and \eqref{eq:meWWScatt_scaling_hiMass}, and more precisely in appendix \ref{app:helicity},  matrix elements for both the low- and high-mass limits scale with some positive power of the $(WW)$-scattering energy, $M_{WW}$. This is distinct from the CCDY channel where no such scaling behavior is present. For fixed heavy neutrino masses and mixing, such a dependence on $M_{WW}$ implies~\cite{Cornwall:1974km,Lee:1977eg,Chanowitz:1985hj,Appelquist:1987cf,Dicus:2004rg,Dicus:2005ku} that the matrix elements violate partial-wave unitarity above some scattering energy threshold $E_U$, unless additional physics cancels this dependence. As both amplitudes depend on the mass of an internal Majorana neutrino, it is possible that the unitarity violation is actually tied to the explicit breaking of LN symmetry in the Type I Seesaw model. If so, then it can potentially be resolved through the spontaneous breaking of LN symmetry via a Higgs-like mechanism.

While a systematic study of partial-wave unitarity in the Phenomenological Type I Seesaw model {lies beyond the scope of this work}, we can nevertheless {provide} a qualitative outlook. Following Ref.~\cite{Appelquist:1974tg}, the $J=0$ partial-wave amplitude of the $W^\pm_0 W^\pm_0\to \ell^\pm\ell^\pm$ process is related to its matrix element $\mathcal{M}$ by the relationship
\begin{align}
    a_{J=0}&=\frac{1}{32\pi}\int_{-1}^1 {\rm d}\cos\theta_1 ~ \mathcal{M}(W^\pm_0 W^\pm_0\to \ell^\pm_1\ell^\pm_2).
\end{align}
Here $\theta_1$ is the polar angle of $\ell_1$ in the frame of the $(WW)$-system. For the low- and high-mass limits, the partial-wave amplitudes are given to lowest order by
\begin{align}
    \text{low-mass} :
    \quad a_0 \approx &
    (2-\delta_{\ell_1\ell_2})
    \frac{g_W^2 M_{WW}}{16\pi m_W^2}
    \nonumber\\
    &  \times
     \Bigg\vert \sum_{k=4}^{n_R+3} V_{\ell_1 N_k} m_{N_k} V_{\ell_2 N_k} \Bigg \vert,  
 \\
    \text{high-mass} :
        \quad a_0 \approx &
    (2-\delta_{\ell_1\ell_2})
    \frac{g_W^2 M_{WW}^3}{32\pi m_W^2}
    \nonumber\\ 
     & \times   
     \Bigg\vert \sum_{k=4}^{n_R+3}\frac{V_{\ell_1 N_k} V_{\ell_2 N_k}}{m_{N_k}} \Bigg \vert. 
\end{align}
Requiring $\vert a_J \vert < 1$ implies that the $W^\pm_0 W^\pm_0$ channel saturates unitarity at $M_{WW}=E_U$, {with}
\begin{align}
     \text{low-mass} &: E_U = 
     \cfrac{16\pi m_W^2/ {\big[(2-\delta_{\ell_1\ell_2})g_W^2\big]}}{\Bigg\vert \sum_{k=4}^{n_R+3} V_{\ell_1 N_k} m_{N_k} V_{\ell_2 N_k} \Bigg \vert},
     \\
     \text{high-mass} &: E_U^3 = 
     \cfrac{32\pi m_W^2/{\big[(2-\delta_{\ell_1\ell_2})g_W^2\big]}}{\Bigg\vert \sum_{k=4}^{n_R+3}\frac{V_{\ell_1 N_k} V_{\ell_2 N_k}}{m_{N_k}} \Bigg \vert}.
\end{align}

In the low-mass limit and assuming a single heavy neutrino of mass $m_N=1\TeV$ with an active-sterile mixing of $\vert V_{\ell N}\vert^{2}=10^{-1}~(10^{-2})~[10^{-3}]$, partial-wave unitarity saturates at about $E_U\sim 7.6\TeV~(76\TeV)~[760\TeV]$. Working instead in the high-mass limit, the mass and mixing upper bound from equation \eqref{eq:0vBBLimit} as derived from $0\nu\beta\beta$ decay searches implies a lower bound on the saturation scale of $E_U\sim72\TeV-87\TeV$. This suggests that if a discovery of $0\nu\beta\beta$ decay is  made by current-generation experiments, and if the decay is mediated by a heavy Majorana neutrino, then a future $pp$ collider with $\sqrt{s}=100\TeV$ may be able to probe this high-energy behavior. However, we caution that in both cases the saturation scale is acutely sensitive to the values of active-sterile mixing elements and heavy neutrino masses. In order to minimize theoretical bias in estimating the LHC's sensitivity to the $W^\pm W^\pm\to \ell^\pm\ell^\pm$ process, we do not impose the above constraints on the Phenomenological Type I Seesaw's parameter space, and in fact  defer further discussion of partial-wave unitary to future work. 

\subsection{Kinematic Properties of $W^\pm W^\pm \to \ell^\pm_i \ell^\pm_j$}\label{sec:wwScattLHC_diff}

\begin{figure*}[!t]
\subfigure[]{\includegraphics[width=0.48\textwidth]{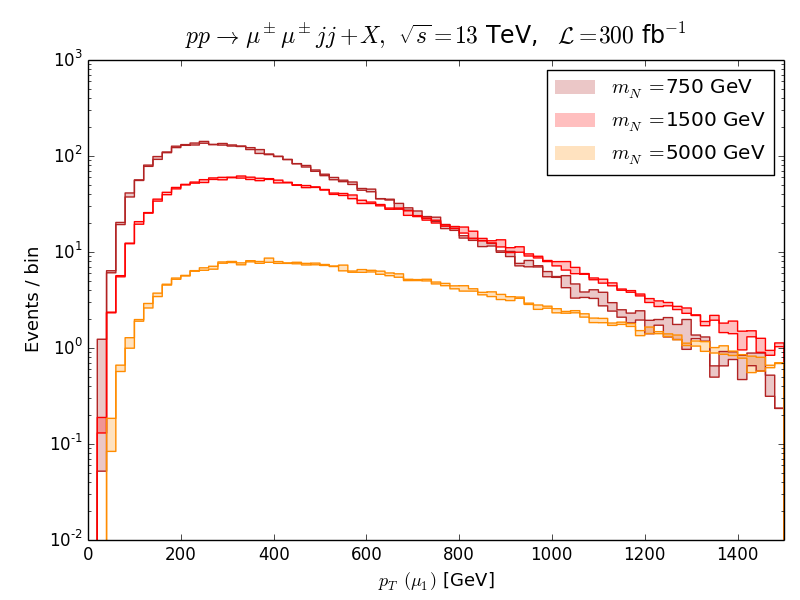} \label{fig:wwScattHeavyN_pTXl1_NLOPS_truth_multiMN}}
\subfigure[]{\includegraphics[width=0.48\textwidth]{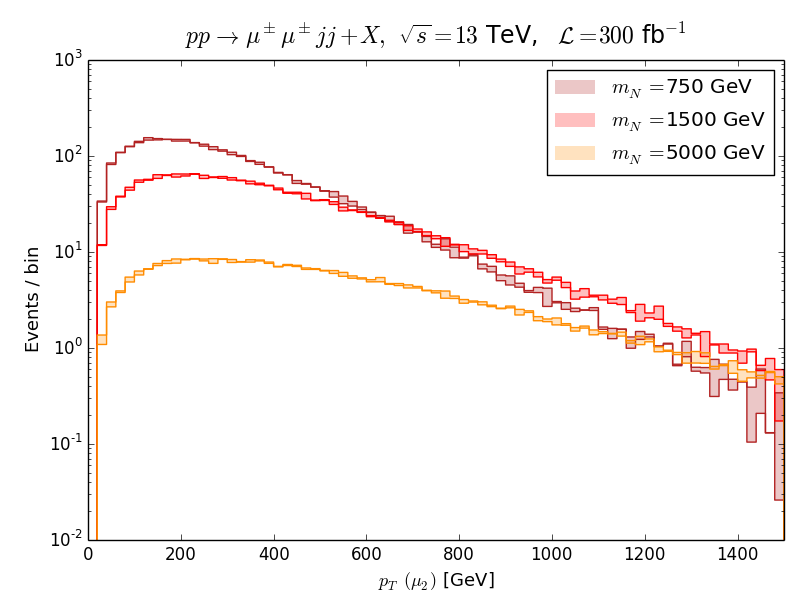} \label{fig:wwScattHeavyN_pTXl2_NLOPS_truth_multiMN}}
\\
\subfigure[]{\includegraphics[width=0.48\textwidth]{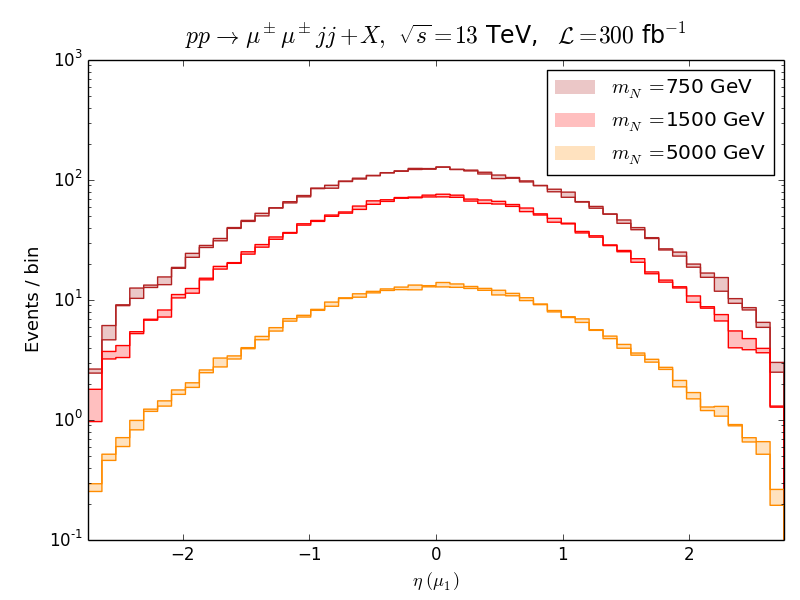} \label{fig:wwScattHeavyN_etal1_NLOPS_truth_multiMN}}
\subfigure[]{\includegraphics[width=0.48\textwidth]{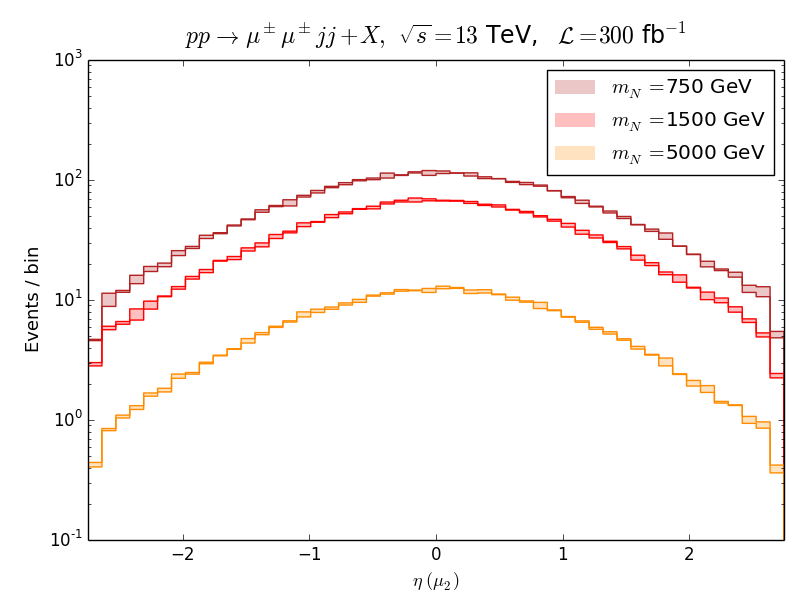} \label{fig:wwScattHeavyN_etal2_NLOPS_truth_multiMN}}
\caption{
Kinematic distributions at $\sqrt{s}=13\TeV$ of the same-sign $W^\pm W^\pm \to \mu^\pm \mu^\pm$ signal process at NLO+PS with residual $\mu_r, \mu_f, \mu_s$ uncertainty envelope (band thickness),  for $m_N=750\GeV$ (darkest), $1.5\TeV$ (dark), and $5\TeV$ (light), of 
the (a,c) leading and (b,d) sub-leading $\mu^\pm$ (a,b) transverse momentum $(p_T)$ and (c,d) pseudorapidity $(\eta)$.
}
\label{fig:wwScattHeavyN_single_lep}
\end{figure*}

\begin{figure*}[!t]
\subfigure[]{\includegraphics[width=0.48\textwidth]{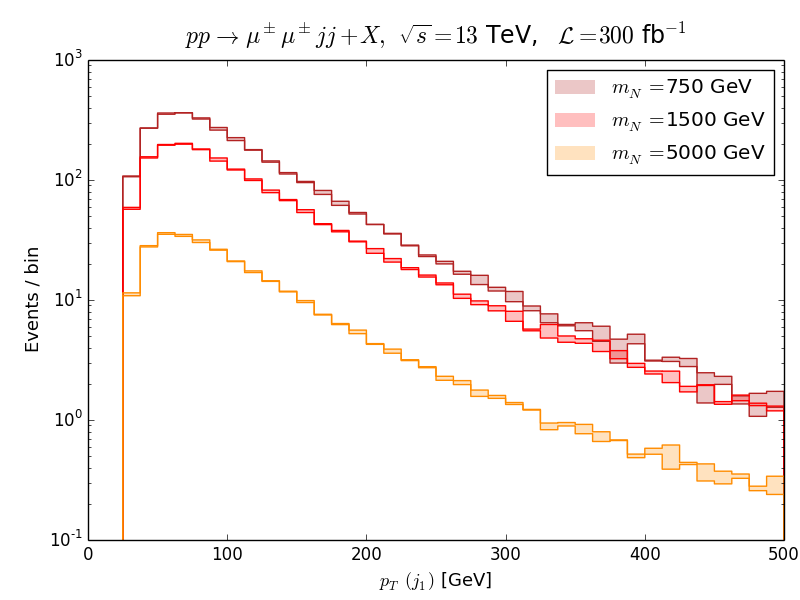} \label{fig:wwScattHeavyN_pTXj1_NLOPS_truth_multiMN}}
\subfigure[]{\includegraphics[width=0.48\textwidth]{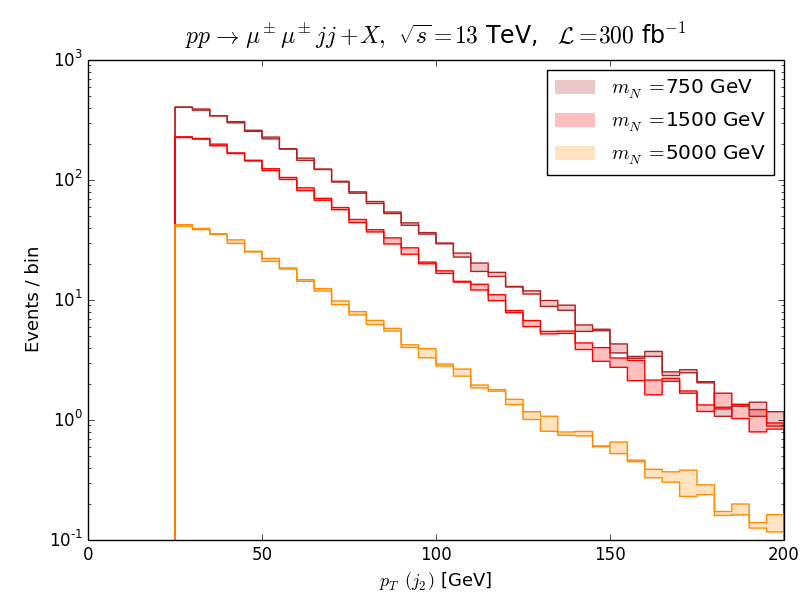} \label{fig:wwScattHeavyN_pTXj2_NLOPS_truth_multiMN}}
\\
\subfigure[]{\includegraphics[width=0.48\textwidth]{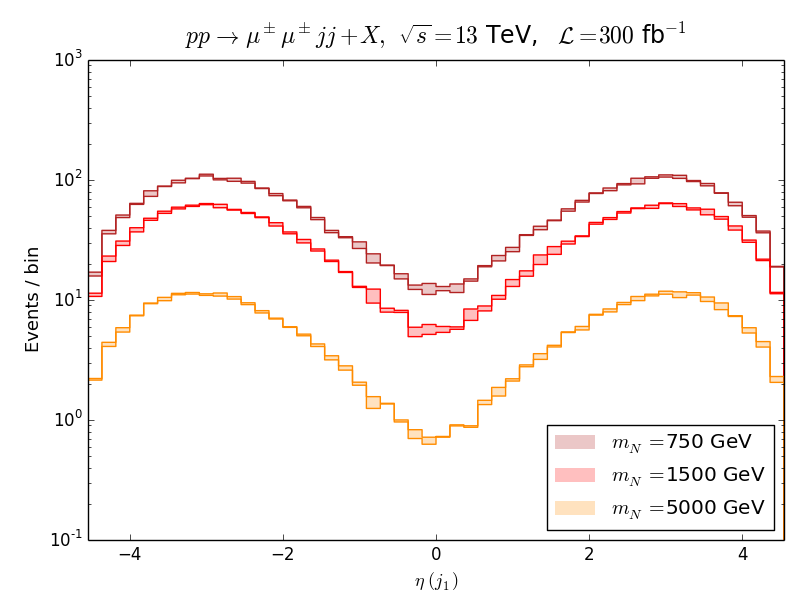} \label{fig:wwScattHeavyN_etaj1_NLOPS_truth_multiMN}}
\subfigure[]{\includegraphics[width=0.48\textwidth]{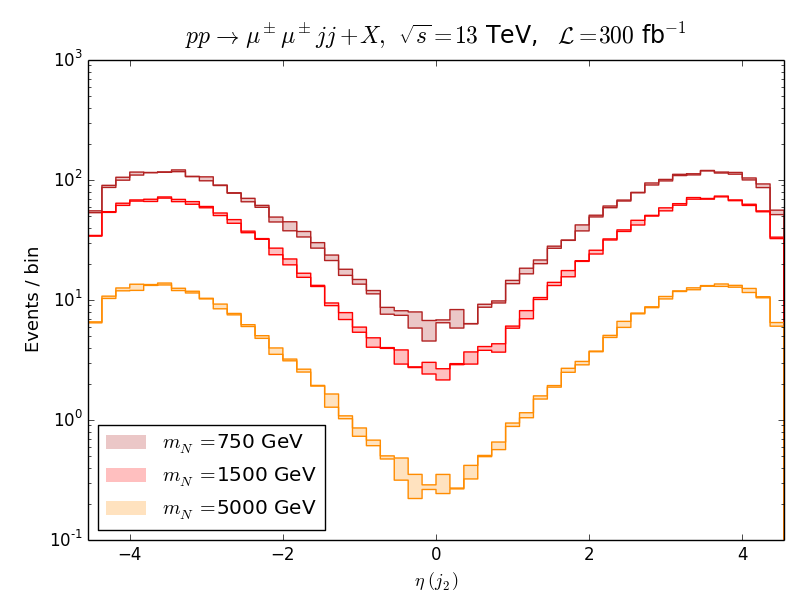} \label{fig:wwScattHeavyN_etaj2_NLOPS_truth_multiMN}}
\caption{
Same as figure \ref{fig:wwScattHeavyN_single_lep} but for the (a,c) leading and (b,d) sub-leading jet.
}
\label{fig:wwScattHeavyN_single_jet}
\end{figure*}

We now turn to exploring the kinematic properties of the $W^\pm W^\pm \to \ell^\pm_i \ell^\pm_j$ signal process at $\sqrt{s}=13\TeV$. As NLO-vs-LO comparisons of VBF kinematics are extensively documented, we restrict ourselves to NLO+PS(LL) distributions where available and neglect comparisons to properties at LO+PS(LL). For concreteness, we fix $\ell_i = \ell_j = \mu$ and set simulation inputs as prescribed in section \ref{sec:setup}. For each of the following observable we assume the representative benchmark masses $m_N=750\GeV$ (darkest), $1.5\TeV$ (dark), and $5\TeV$ (light). Events are normalized to $\mathcal{L}=300\invfb$. Also shown for each distribution is the residual $\mu_r, \mu_f, \mu_s$ uncertainty (band thickness) as obtained from a $27$-point variation envelope. 

 Throughout this section we work with particle-level objects. We do so to emulate detector thresholds (but not detector resolution) according to our analysis in section \ref{sec:event_selection} and to ensure the infrared safety of observable definitions. In practice, this means that after parton showering  we impose anti-$k_T(R=0.4)$ clustering on all hadronic activity. We also require that electron, muon, hadronic tau, and jet candidates satisfy the following requirements
\begin{align}
    p_T^{e~(\mu)~[\tau_h]~\{j\}} &> 10~(27)~[20]~\{25\}\GeV, \\
    \vert\eta^{e~(\mu)~[\tau_h]~\{j\}}\vert &< 2.5~(2.7)~[2.5]~\{4.5\}.
\end{align}
Particle identification efficiencies and mistagging rates are kept at their default values in \textsc{MadAnalysis5}~\cite{Conte:2012fm,Conte:2014zja,Conte:2018vmg}. That is to say, we consider an ideal setup in which identification efficiencies are set to unity and mistagging rates are set to zero. No kinematic smearing is applied.

\begin{figure*}[!t]
\subfigure[]{\includegraphics[width=0.48\textwidth]{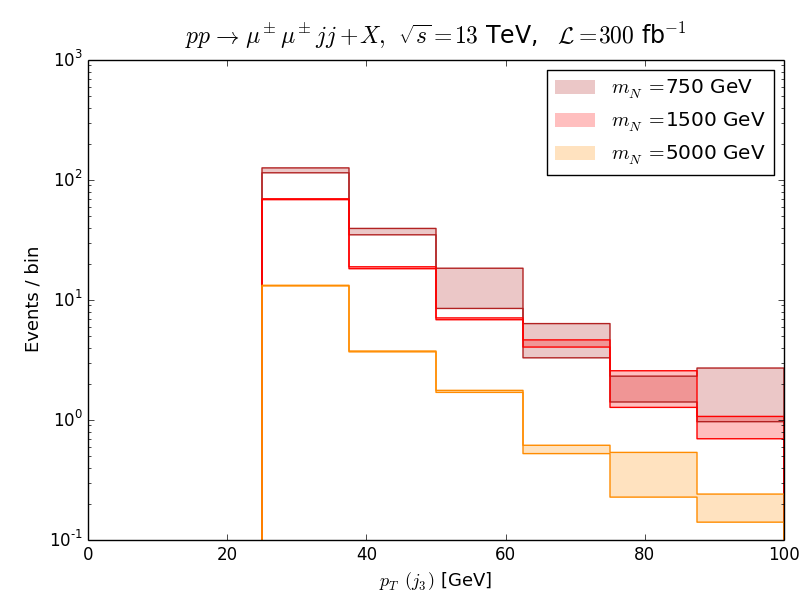} \label{fig:wwScattHeavyN_etaj1_XLOPS_truth_multiMN}}
\subfigure[]{\includegraphics[width=0.48\textwidth]{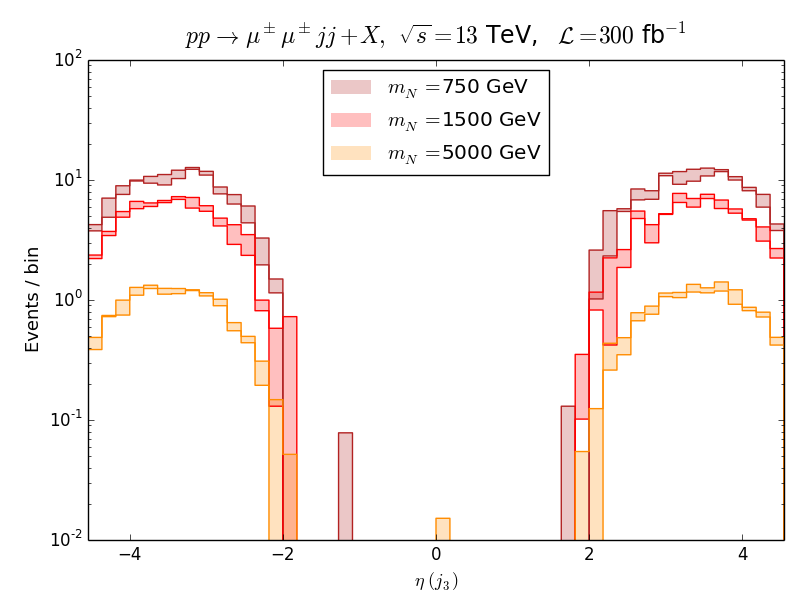} \label{fig:wwScattHeavyN_etaj2_XLOPS_truth_multiMN}}
\caption{
Same as figure \ref{fig:wwScattHeavyN_single_lep} but for the (a) $p_T$ and (b) $\eta$ of the trailing jet $(j_3)$ at LO+PS.
}
\label{fig:wwScattHeavyN_single_jet3}
\end{figure*}

Given these stipulations, we define our signal as the
\begin{equation}
pp \to  \mu^\pm \mu^\pm j j + X
\label{eq:kinematicsProdDef}
\end{equation}
process, where $X$ denotes the possibility of additional hadronic or photonic activity. More precisely, we require events to possess exactly two same-sign $\mu$ candidates and at least two $j$ candidates. Events containing any number of $e$ or $\tau_h$ candidates are rejected. This setup implies that we remain inclusive with respect to soft and forward objects that fail candidacy requirements. For clarity, objects are ranked by their $p_T$, with $p_T^{k}>p_T^{k+1}$.

We start with figure \ref{fig:wwScattHeavyN_single_lep} where we plot the (a,b) $p_T$ and (c,d) $\eta$ distributions of the (a,c) leading $(\mu_1)$ and (b,d) sub-leading $(\mu_2)$ muon in  our signal process at NLO+PS. We  foremost note the lack of any resonant structure in both $p_T^{\mu}$ distributions. This follows from the absence of $s$-channel resonances in the $W^\pm W^\pm \to \mu^\pm \mu^\pm$ sub-process. Instead, we find a $p_T$ behavior reminiscent of open particle production that plateaus for a few hundred GeV and then falls due to kinematic suppression. There  is a steeper falloff for smaller $m_N$. The $p_T$ spectra indicate that the $(\mu^\pm \mu^\pm)$-system, or equivalently the $(W^\pm W^\pm)$-system, possesses a large invariant mass that reaches several hundred GeV. We observe that both muons tend toward smaller values of $\vert\eta\vert$, independent of the heavy neutrino mass, indicating an absence of forward scattering.

Moving onto figure \ref{fig:wwScattHeavyN_single_jet}, we present the same information as in figure \ref{fig:wwScattHeavyN_single_lep} but for the (a,c) leading $(j_1)$ and (b,d) sub-leading  $(j_2)$ jets. In both $p_T$ spectra we observe peaks at $p_T^j \sim m_W/2$, which is characteristic of the VBF process and is due to the recoil against the $t$-channel emission of $W$ bosons. Consistently, we observe in the $\eta$ distributions that the two jets are forward, with maxima in the forward direction near \confirm{$\vert\eta^j\vert \sim 3$} and a suppression of central activity at \confirm{$\vert\eta^j\vert \sim 0$}. We find that the shapes of all observables in figure~\ref{fig:wwScattHeavyN_single_jet} are  insensitive to the  values of $m_N$ under consideration. This follows from the fact that quarks in the $2\to4$ process do not directly couple to the LN-violating current, and therefore act like spectators.

\begin{figure*}[!t]
\subfigure[]{\includegraphics[width=0.48\textwidth]{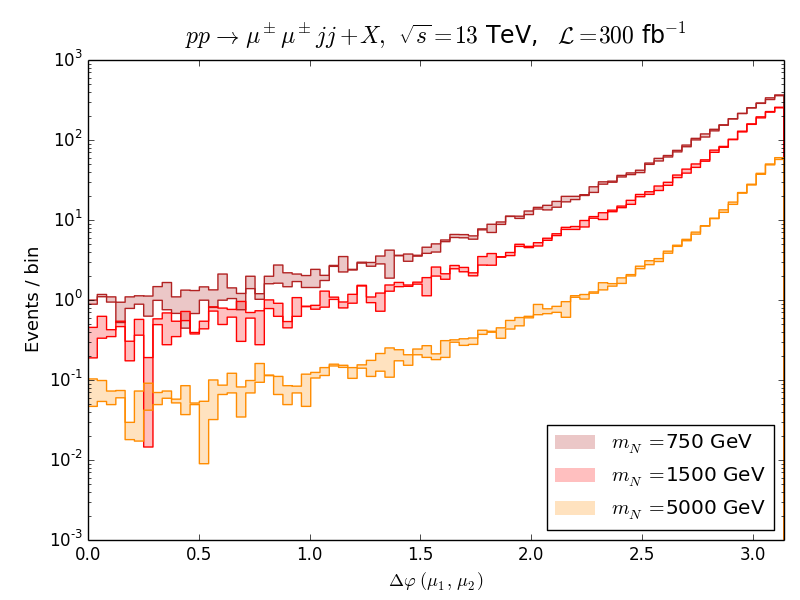} \label{fig:wwScattHeavyN_Phill_NLOPS_truth_multiMN}}
\subfigure[]{\includegraphics[width=0.48\textwidth]{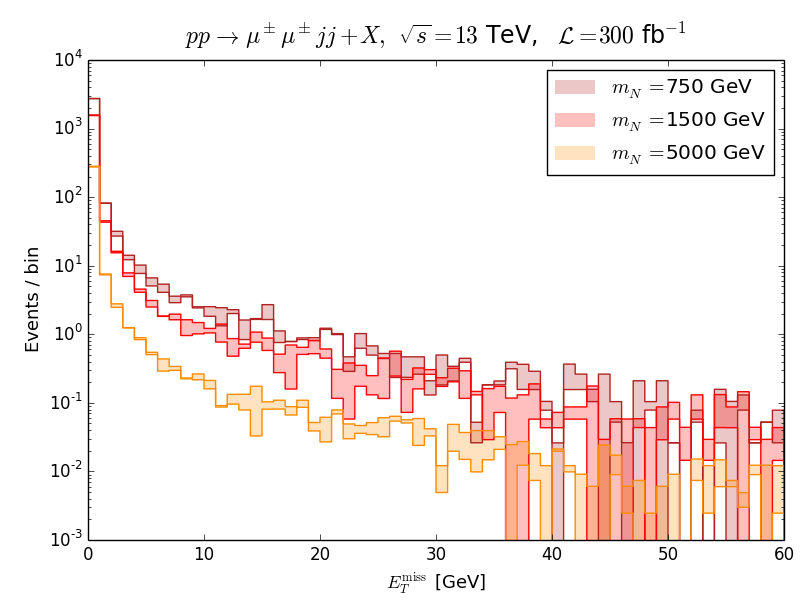} \label{fig:wwScattHeavyN_METXX_NLOPS_truth_multiMN}}
\\
\subfigure[]{\includegraphics[width=0.48\textwidth]{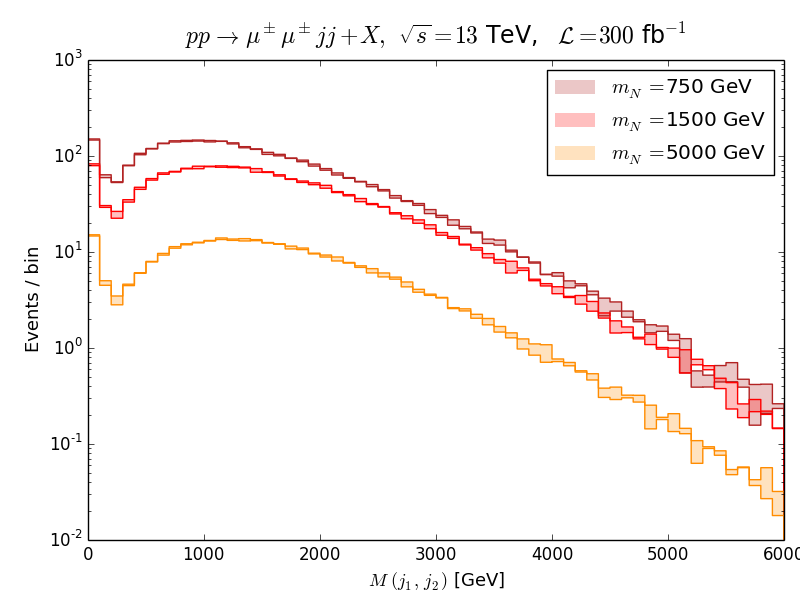} \label{fig:wwScattHeavyN_Mj1j2_NLOPS_truth_multiMN}}
\subfigure[]{\includegraphics[width=0.48\textwidth]{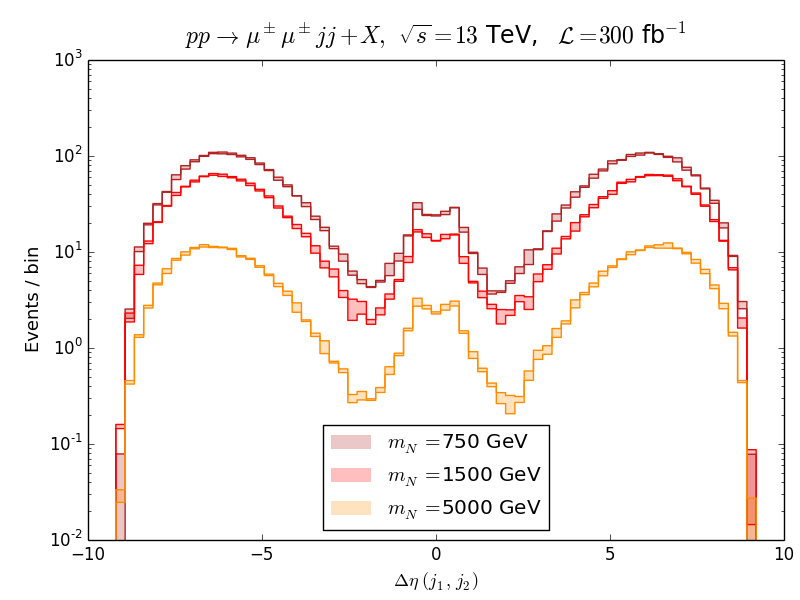} \label{fig:wwScattHeavyN_dEtaj_NLOPS_truth_multiMN}}
\caption{
Same as figure \ref{fig:wwScattHeavyN_single_lep} but for the
(a) azimuthal separation of the leading same-sign $\mu^\pm\mu^\pm$ pair $\Delta\varphi(\mu_1,\mu_2)$,
(b) missing transverse energy $\met$,
(c) invariant mass of the leading dijet system  $M(j_1,j_2)$, and 
(d) pseudorapidity difference of the same system $\Delta\eta(j_1,j_2)$.
}
\label{fig:wwScattHeavyN_composite}
\end{figure*}

\begin{figure*}[!t]
\subfigure[]{\includegraphics[width=0.48\textwidth]{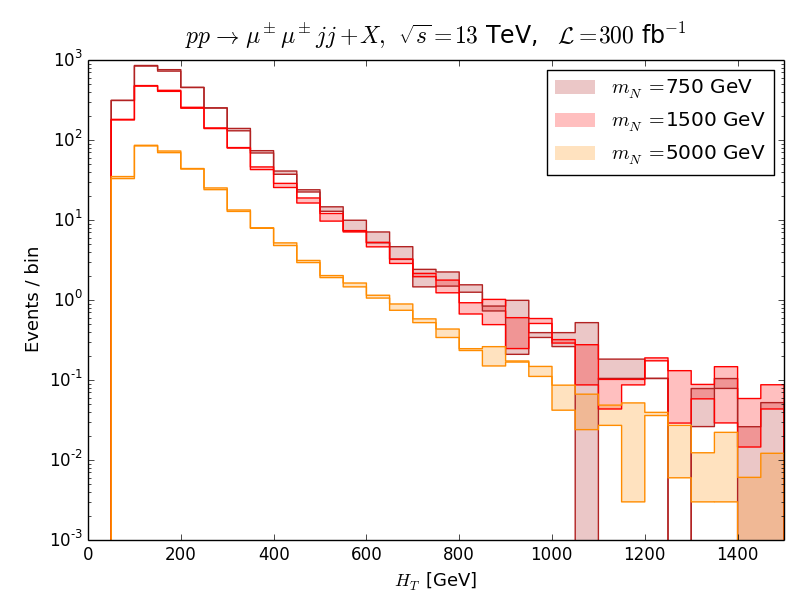} \label{fig:wwScattHeavyN_HTjet_NLOPS_truth_multiMN}}
\subfigure[]{\includegraphics[width=0.48\textwidth]{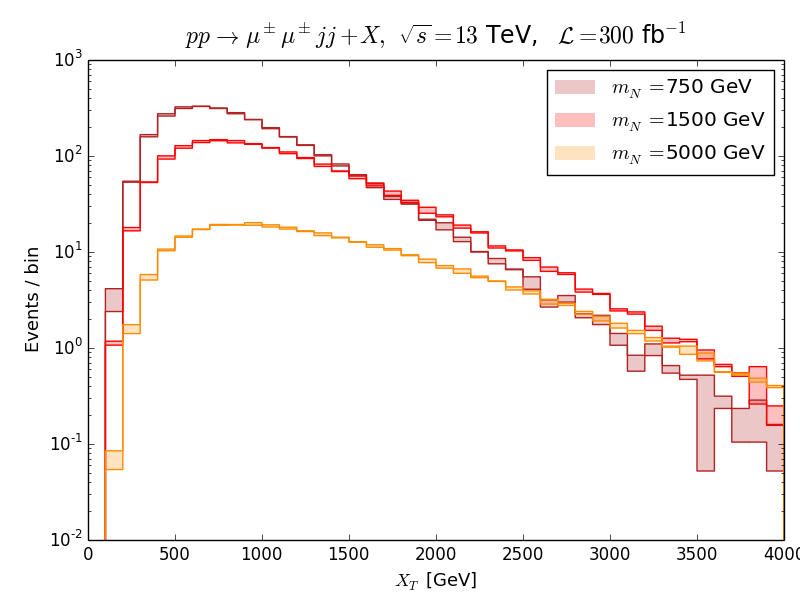} \label{fig:wwScattHeavyN_XTall_NLOPS_truth_multiMN}}
\\
\subfigure[]{\includegraphics[width=0.48\textwidth]{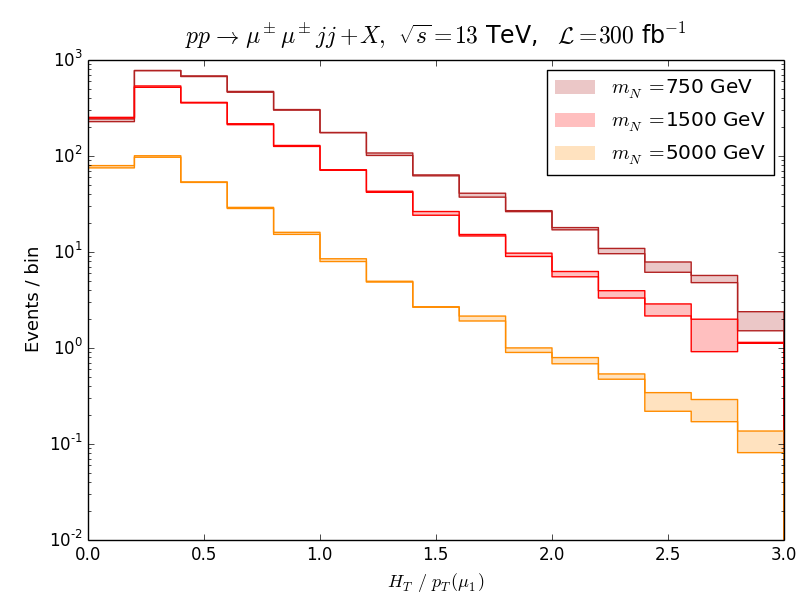} \label{fig:wwScattHeavyN_RHTpT_NLOPS_truth_multiMN}}
\subfigure[]{\includegraphics[width=0.48\textwidth]{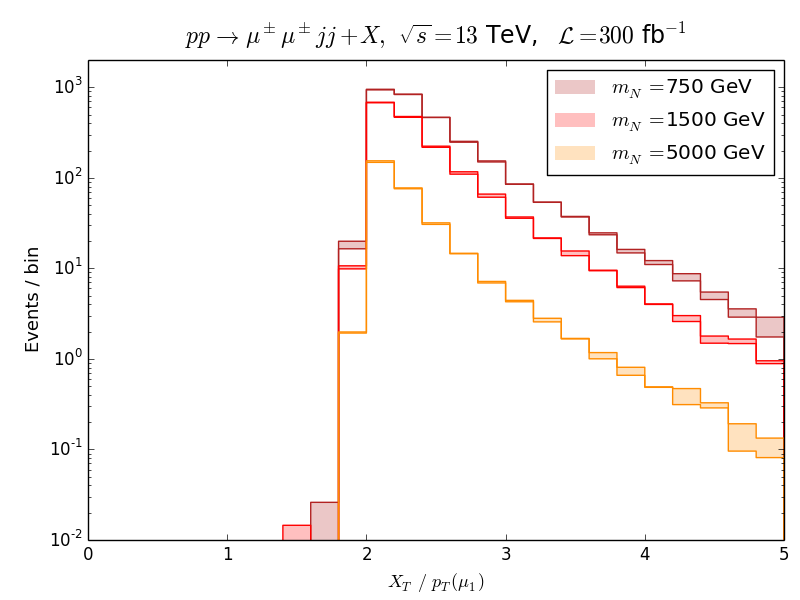} \label{fig:wwScattHeavyN_RXTpT_NLOPS_truth_multiMN}}
\caption{
Same as figure \ref{fig:wwScattHeavyN_single_lep} but for the
(a) scalar sum of all jet $p_T$ $(H_T)$,
(b) scalar sum of all visible $p_T$ $(X_T)$,
(c) ratio $H_T/p_T^{\mu_1}$,
(d) ratio $X_T/p_T^{\mu_1}$.
}
\label{fig:wwScattHeavyN_global}
\end{figure*}

As the $pp \to  \mu^\pm \mu^\pm j j +X$  process is simulated at NLO+PS, one  has access to the $pp \to  \mu^\pm \mu^\pm j j j+X$ process at LO+PS. We are thus able to explore the QCD radiation pattern in the VBF process. In figure \ref{fig:wwScattHeavyN_single_jet3} we show the (a) $p_T$ and (b) $\eta$ distribution of the trailing jet $j_3$ at LO+PS. For all considered values of $m_N$, we observe an inclination toward lower $p_T$, with most of the phase space sitting between the threshold at $p_T^{j_3}\sim p_T=25\GeV$ and $p_T^{j_3}\sim 40\GeV$. This is just below the characteristic $p_T$ of the two leading jets. We note a strong suppression of central $(\vert \eta^{j_3}\vert \lesssim 2)$ tertiary jets. This does not mean an absence of QCD radiation for $\vert \eta^{j_3}\vert \lesssim 2$, only that it is soft. Most of the activity resides in the forward direction, peaking at $\vert \eta\vert\sim 3-4$, again independent of $m_N$. As this is well in the vicinity of the leading jets  it is likely that the $q\to qg$ and $g\to q\overline{q}$ splittings responsible for $j_3$ involve smaller momentum transfers, which results in shallower opening angles between $j_3$ and its companion.

To further investigate the dynamics  of the   $W^{\pm}W^{\pm} \to \mu^\pm \mu^\pm$ sub-process, we consider in figure \ref{fig:wwScattHeavyN_composite} observables that are built from the momenta of two or more particles. We start with figure \ref{fig:wwScattHeavyN_Phill_NLOPS_truth_multiMN}, where we plot the azimuthal separation of the same-sign $\mu^\pm\mu^\pm$ pair,   defined as
\begin{equation}
\Delta\varphi(\mu_1,\mu_2) = \vec{p}_T^{~ \mu_1} \cdot \vec{p}_T^{~ \mu_2} ~/~ \vert \vec{p}_T^{~ \mu_1} \vert \vert ~ \vec{p}_T^{~ \mu_2} \vert.
\end{equation}
We find that the leptons exhibit a strong back-to-back trajectory with the separation peaking (curtailing) at $\Delta\varphi(\mu_1,\mu_2) \approx \pi ~(0)$. This is despite being a $4$-body final state at LO, which would suggest a sizable recoil against the $(jj+X)$-system. For increasing heavy neutrino masses we observe a higher tendency for back-to-back trajectories. The marginal-to-moderate recoil that is found suggests that modeling the $2\to4$ signal process as a $2\to2$ process within the effective $W$ approximation as done in Refs.~\cite{Dicus:1991fk,Ali:2001gsa,Chen:2008qb} is a fair approximation.

In figure \ref{fig:wwScattHeavyN_METXX_NLOPS_truth_multiMN} we focus on the distribution of the missing transverse energy $\met$, defined per event as the magnitude of the two-momentum recoil against all visible (vis) objects, regardless of their energy,
\begin{equation}
\met = \vert \vec{p}_T^{\rm ~miss} \vert, \quad \vec{p}_T^{\rm ~miss} = - \sum_{k\in\{ \rm vis \}} \vec{p}_T^{~k}.
\end{equation}
We find that the distribution strongly peaks at $\met\lesssim10\GeV$, in line with  expectations of a $2\to4$ process without outgoing light neutrinos.  As we are working without any detector resolution effects, the nonzero $\met$  originates from the weak decays to light neutrinos of mesons generated in the parton shower. Aside from differences in the rate normalization,  we observe no substantial  dependence of $\met$ on the heavy neutrino mass.

In figure \ref{fig:wwScattHeavyN_Mj1j2_NLOPS_truth_multiMN} we show the invariant mass distribution of the  two highest $p_T$ (leading) jets, given  by
\begin{equation}
M(j_1,j_2) = \sqrt{(p^{j_1}+p^{j_2})^2}.
\end{equation}
For the heavy neutrino masses under consideration, we see that the peaks of the invariant mass spectra occur at \confirm{$M(j_1,j_2) \sim 1000\GeV-1200\GeV$},  with a peak position at larger $M(j_1,j_2)$ for larger $m_N$.  A narrow collection of events at {$M(j_1,j_2)\ll500\GeV$} is also observed. These low-mass events are attributed to instances of one forward jet possessing relatively low $p_T$ while  another jet undergoes a hard $q^*\to qg$ splitting. In such cases the $(qg)$ pair can be identified as the leading jet pair but still return a small $M(j_1,j_2)$ since this corresponds to the  $(q^*)$-system's virtuality, which is favored to be small in massless QCD. The dependence on $m_N$  indicates that the hadronic activity is not completely decoupled from the $W^\pm W^\pm \to  \mu^\pm \mu^\pm$ sub-process, and therefore can potentially offer a handle on determining the value of $m_N$. This  is  relevant given the  mild scale uncertainty bands. 

The $M(j_1,j_2)$ spectra point to the signal process being driven by valence quark-valence quark scattering involving large momentum fractions, \ie, $x_B > M(j_1,j_2)/\sqrt{s} \sim 0.1$. In comparison to the $p_T^\mu$  distributions of figure \ref{fig:wwScattHeavyN_single_lep}, which show charged lepton momenta reaching a few hundred GeV, we see that comparable momentum fractions are propagated into the $W^\pm W^\pm \to \mu^\pm \mu^\pm$  sub-process. For example: estimating the incoming $W$ boson energies by those of the muons, $E_W \sim E_\mu \sim p_T^\mu \sim 100\GeV-300\GeV$, and the outgoing quark energies from the invariant mass of the two leading jets, which are also back-to-back, \confirm{$E^{out}_q \sim M(j_1,j_2)/2 \gtrsim 500\GeV-1000\GeV$,} then the typical momentum fractions carried by the $W$ reach
\begin{equation}
\confirm{
x_W \equiv \frac{E_W}{E^{in}_q} =  \frac{E_W}{(E_W + E^{out}_q)} \lesssim 0.1-0.4.
}
\end{equation}

Moving onto figure \ref{fig:wwScattHeavyN_dEtaj_NLOPS_truth_multiMN}, we show the pseudorapidity difference between the two leading jets, defined as
\begin{equation}
\Delta\eta(j_1,j_2) = \eta^{j_1} - \eta^{j_2}.
\end{equation}
We report several notable features. First is the symmetric behavior around $\Delta\eta= 0$, which stems from having a symmetric beam configuration. Second is that most of the phase space populates the region where $\vert\Delta\eta\vert\gtrsim2$ and appears independent of heavy neutrino masses. Third is the presence of a modest collection of events with $\vert\Delta\eta\vert\lesssim1$. Such events are  consistent with the low-mass distribution in figure \ref{fig:wwScattHeavyN_Mj1j2_NLOPS_truth_multiMN} originating from $q^*\to qg$ and $g^*\to q\overline{q}$ splittings with relatively small opening angles.

Beyond {one}- and {two}-particle observables are those sensitive to the global activity of the $pp \to \mu^\pm \mu^\pm jj+X$ process. In particular, we consider in figure~\ref{fig:wwScattHeavyN_HTjet_NLOPS_truth_multiMN} the scalar sum of $p_T$ over all jets in an event $(H_T)$,
\begin{equation}
H_T =  \sum_{k\in\{ \rm jets \}} \vert \vec{p}_T^{~k} \vert,
\end{equation}
and in figure~\ref{fig:wwScattHeavyN_XTall_NLOPS_truth_multiMN}, the scalar sum of the $p_T$ of all reconstructed particle candidates (reco) $(X_T)$,
\begin{equation}
X_T =  \sum_{k\in\{ \rm reco \}} \vert \vec{p}_T^{~k} \vert.
\end{equation}

In the first case and for all heavy neutrino masses we observe that $H_T$ peaks at $H_T \sim 100\GeV$ and uniformly decreases for larger values of $H_T$. As the net contribution of the two (three) leading jets in  the signal scales as 
\begin{equation}
p_T^{j_1}+p_T^{j_2} (+p_T^{j_3})\sim  2\times\frac{M_W}{2}  (+p_T^{j \min}) = 80~(105)\GeV,
\end{equation}
the $H_T$ distribution suggests the presence of little high-$p_T$ hadronic activity beyond these leading objects. For the $X_T$ case, we observe a slight dependence on $N$'s mass, with the distributions' maxima occurring at $X_T \sim 600\GeV-750\GeV$ and tending towards larger values for larger $m_N$. We attribute this sensitivity to the dependence of muon $p_T$, as seen in figure \ref{fig:wwScattHeavyN_single_lep},  which also peaks at larger values for increasing $m_N$. In comparing $X_T$ to the $(\mu^\pm \mu^\pm j j(j))$-system itself, we find that the scalar sum of $H_T$  and the same-sign muon pair $p_T$,
\begin{align}
H_T +  p_T^{\mu_1}  + p_T^{\mu_2} \sim 	& ~ 100\GeV + (200\GeV-400\GeV) \nonumber\\
								&  + (100\GeV-200\GeV) \\
							= 	&  ~ 400\GeV - 700\GeV,
\end{align} 
undershoots the peak of $X_T$ by $\Delta X_T \sim 50\GeV - 200\GeV$. This indicates a sizable presence of electromagnetic activity (photons), which one can anticipate from the presence of jets and muons with TeV-scale momenta.

Finally, we consider observables that measure the relative amounts of hadronic and leptonic activity in a given event. Such quantities are   sensitive to the color structure of hard scattering processes \cite{Campanario:2014lza,Pascoli:2018rsg,Pascoli:2018heg,Fuks:2019iaj},  and hence employable in  dynamic jet vetoes for color-singlet signal processes. While a full exploration of jet vetoes in $W^\pm W^\pm \to \mu^\pm \mu^\pm$ is outside our scope, we  consider as representative cases  in figures \ref{fig:wwScattHeavyN_RHTpT_NLOPS_truth_multiMN}
and \ref{fig:wwScattHeavyN_RXTpT_NLOPS_truth_multiMN} respectively, 
the ratio of $H_T$ and the leading charged lepton $p_T$:
\begin{equation}
r^{H_T}_{\mu_1} = H_T / p_T^{\mu_1},
\end{equation}
and the ratio of $X_T$ and the leading charged lepton $p_T$:
\begin{equation}
r^{X_T}_{\mu_1} = X_T / p_T^{\mu_1}.
\end{equation}

In the first case, we observe that a majority of the phase space sits well below $r^{H_T}_{\mu_1} = 1$. This indicates that on an event-by-event basis more  transverse momentum is carried by the leading muon than in all jets combined. This is unlike diboson and top quark processes where the situation is reversed \cite{Pascoli:2018rsg,Pascoli:2018heg,Fuks:2019iaj}. We find that the ratios peak just above $r^{H_T}_{\mu_1} \sim 0.25$ and are largely independent of the heavy neutrino masses that we consider. This is also consistent with  na\"ive estimations from the individual $H_T$ and $p_T^{\mu_1}$  distributions, which suggest 
\begin{equation}
\frac{H_T}{p_T^{\mu_1}}  \sim \frac{100\GeV}{(200\GeV-400\GeV)} \sim 0.25-0.5.
\end{equation} 

In the $r^{X_T}_{\mu_1}$ case, we observe a sharp cut-off  at $r^{X_T}_{\mu_1}\sim 2$ that is largely independent of the heavy neutrino's mass. This can be tied to the disparity of momentum scales between leptons and jets. In particular, since $p_T^{\mu_1}, p_T^{\mu_2}   \gg p_T^{j_k}$, one finds that the ratio scales roughly as 
\begin{equation}
r^{X_T}_{\mu_1} \sim \frac{H_T +  p_T^{\mu_1}  + p_T^{\mu_2}}{p_T^{\mu_1}} \sim  \frac{p_T^{\mu_1}  + p_T^{\mu_2}}{p_T^{\mu_1}} \sim 2.
\end{equation}
The approximation $p_T^{\mu_1}\sim p_T^{\mu_2}$ is again consistent with  $2\to2$ scattering and the back-to-back trajectories found in the azimuthal separation distribution in figure \ref{fig:wwScattHeavyN_Phill_NLOPS_truth_multiMN}.

\section{Sensitivity at the LHC and HL-LHC}\label{sec:analysis}

In this section we estimate the discovery potential of heavy Majorana neutrinos in same-sign $WW$ scattering at the LHC and its high-luminosity upgrade.  After summarizing our simulated detector setup in section~\ref{sec:analysis_pid}, we build our event selection menu in section~\ref{sec:event_selection}, and present our findings in section~\ref{sec:results}. Our analysis includes signal and background processes that are normalized to an integrated luminosity of $\mathcal{L}=300\invfb$ for the $\sqrt{s}=13\TeV$ LHC, and to $\mathcal{L}=3\invab$ for the HL-LHC.

\subsection{Detector Modeling and Particle Identification}\label{sec:analysis_pid}

Particle objects considered throughout the analysis are defined using the ATLAS configuration card available from the \textsc{Delphes} repository. In the results that follow the ATLAS card was modified to construct jet candidates from hadronic activity using the anti-$k_T$ sequential clustering algorithm with a distance parameter $R=0.4$. This value is more widely used in recent ATLAS data analyses than  the default value of $R=0.6$. All the other parameters in the configuration card, which include particle identification and mis-tagging efficiencies as well as the fiducial geometry definition, are left unchanged. 

As summarized in section \ref{sec:setup_mc}, the impact of pileup is assumed to be subtracted from events as one would do with real data. While we neglect the presence of additional low-$p_T$, pileup jets in our samples, the impact  on  particle resolution is at least partly encapsulated in the momentum smearing routines in \textsc{Delphes}.

\subsection{Event Selection}\label{sec:event_selection}

\begin{table}
  \renewcommand{\arraystretch}{1.3}
    \caption{Pre-selection and signal region cuts.}
    \centering
    \resizebox{0.825\columnwidth}{!}{
    \begin{tabular}{c}
    \hline\hline
    Pre-selection Cuts \\ 
    $p_T^{\mu_1~(\mu_2)} > 27~(10)\GeV$,    \quad 
    \quad $\vert\eta^\mu\vert<2.7$,         \quad 
    $n_\mu=2$,  \\
    $p_T^{j}>25\GeV$,       \quad 
    $\vert\eta^j\vert<4.5$, \quad 
    $n_j\geq2$, \\ 
    $Q_{\mu_1}\times Q_{\mu_2}=1$, \quad  
    $M(j_1,j_2) > 700 \GeV$\\
    \hline
    Signal Region Cuts \\    
    $p_T^{\mu_1}, ~ p_T^{\mu_2} > 300\GeV$ \\
    \hline\hline
    \end{tabular}
    }
    \label{tab:cut_summary}
\end{table}

Our analysis is designed to be as simple as pragmatically possible. We do this to establish a baseline sensitivity and discovery potential at the LHC that broadly covers Majorana  masses spanning $m_N=50\GeV-20\TeV$. Investigating improvements that target specific mass regimes is left to future work. As discussed in  section~\ref{sec:outlook_ex}, we do not exploit all the kinematic characteristics reported in section~\ref{sec:wwScattLHC_diff}. We omit, for example,  cuts on $\met$ or hadronic activity that are well-established handles in searches for LNV at colliders~\cite{delAguila:2008cj,Atre:2009rg,Pascoli:2018heg}. Thus, a more tailored analysis by ATLAS or CMS should yield improvements over the outlook presented here. 

To identify our LN-violating collider signature, 
\begin{align}
    p p \to \mu^\pm \mu^\pm j j + X,
    \label{eq:event_sig}
\end{align}
which is characterized by two same-sign muons and  at least two jets, we first apply a loose event selection (called pre-selection in the following). This reduces the number of background processes that must be considered  while keeping a high selection efficiency for the signal process. The pre-selection also includes  requirements needed to ensure the near $100\%$ efficiency of the inclusive, single-muon trigger chains used to record collision data in ATLAS~\cite{Aad:2016jkr} and CMS~\cite{Sirunyan:2018fpa} during Run 2 and the future Run 3 of the LHC. A summary of pre-selection requirements, adapted to Run 2 trigger and acceptance thresholds of ATLAS, is listed in the top of Table~\ref{tab:cut_summary}. 

At pre-selection we require that events have exactly two isolated muon candidates. Muon candidates must have the same electric charge $Q$ and reside within the fiducial volume of $\vert\eta\vert<2.7$. The leading muon must have $p_T > 27\GeV$ in order to ensure high efficiency of the muon trigger, whereas the sub-leading muon $p_T$ just needs to be above the reconstruction threshold of $p_T > 10 \GeV$. Events with additional lepton candidates, including hadronically decaying $\tau$ leptons, are vetoed.

At least two jets with $p_T > 25\GeV$ and $\vert\eta\vert<4.5$  must be present in each event. The invariant mass of the system constituted by  the two highest $p_\mathrm{T}$ jets passing these criteria, $M(j_1,j_2)$,  must also exceed $700\GeV$. This suppresses interfering sub-processes (see section \ref{sec:modeling_sig}) and enriches the signal sample with a topology that corresponds to the scattering of weak vector bosons. 

For the heavy neutrino mass range under consideration, we report that about \confirm{$\mathcal{A}\sim20\%$ to $40\%$} of signal events survive pre-selection cuts. For representative $m_N$ (first column), we list in Table~\ref{tab:signalRates_by_cuts} the generator-level cross section $\sigma^{\rm Gen.}$ (second column) assuming a nominal active-sterile mixing as set in equation \eqref{eq:mixingInputs}, the pre-selection-level cross section $\sigma^{\rm Pre.}$ (third column), and the pre-selection acceptance rate $\mathcal{A}=\sigma^{\rm Pre.}/\sigma^{\rm Gen.}$.

\begin{table}[]
  \renewcommand{\arraystretch}{1.3}
  \setlength\tabcolsep{6pt}
  \centering
    \caption{Visible signal cross sections (and efficiencies) after applying different selections to the simulated events.}
\resizebox{.825\columnwidth}{!}{        
    \begin{tabular}{r|c c l c l}
    \hline
    \hline
    $m_N$\quad~ & 
    $\sigma^\mathrm{Gen.}$ [fb] &
    \multicolumn{2}{c}{$\sigma^\mathrm{Pre.}$ [fb]~$(\mathcal{A})$} &
    \multicolumn{2}{r}{$\sigma^\mathrm{SR}$ [fb]~$(\varepsilon)$}  
     \\ \hline
150\GeV & 13.3     & 3.7  & $(28\%)$     & 0.5 &$(14\%)$\\
1.5\TeV & 8.45     & 3.18 & $(38\%)$     & 1.9 &$(63\%)$ \\
5\TeV   & 1.52     & 0.58 & $(38\%)$     & 0.46 &$(79\%)$\\
15\TeV  & 0.190    & 0.072 & $(38\%)$    & 0.056 &$(78\%)$ \\
         \hline
         \hline
    \end{tabular}}
    \label{tab:signalRates_by_cuts}
\end{table}

\begin{table}[]
  \renewcommand{\arraystretch}{1.3}
     \setlength\extrarowheight{3pt}  
  \centering
    \caption{Expected number of SM background events in the Signal Region at the 
    (HL-)LHC with $\mathcal{L}=300~\invfb$ (3 ab$^{-1}$).
    }
\resizebox{.825\columnwidth}{!}{         
    \begin{tabular}{l|ccc|c}
    \hline    \hline
    Collider & QCD $W^\pm W^\pm j j$  & EW $W^\pm W^\pm j j$ &  $W^\pm V (3\ell\nu)$ & Total \\ \hline
    LHC		& 0.05 	& 0.52	& 0.14	& 0.71	\\
    HL-LHC	& 0.49 	& 5.17	& 1.40	& 7.10	\\
   \hline \hline
    \end{tabular}}
     \label{tab:background}    
\end{table}

\begin{figure*}[!t]
\subfigure[]{\includegraphics[width=0.49\textwidth]{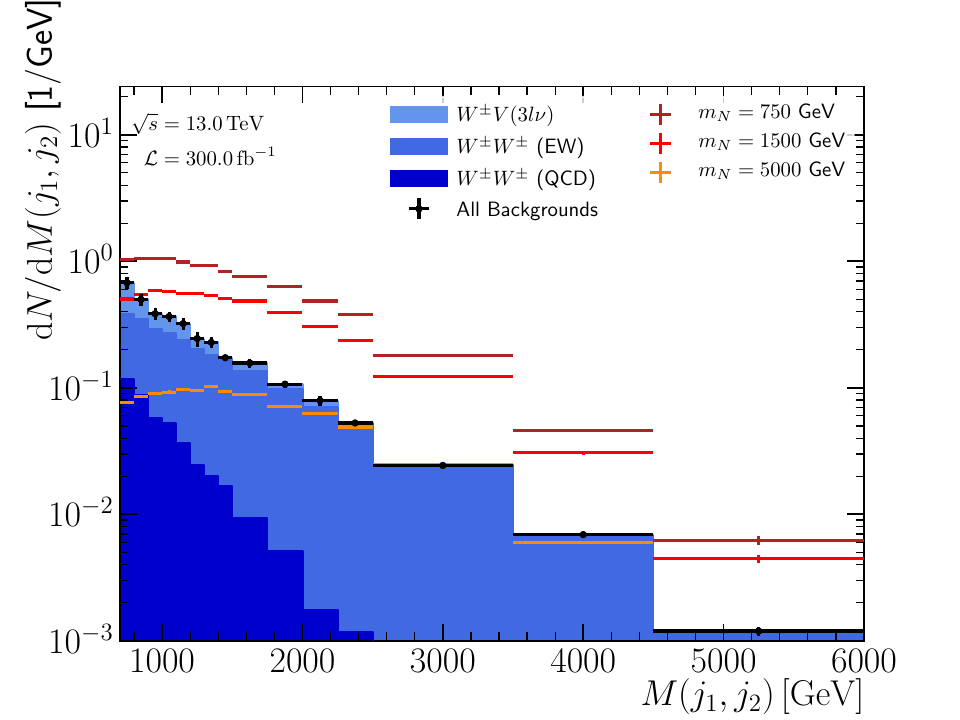} \label{fig:h_mjj_Preselection_0}}
\subfigure[]{\includegraphics[width=0.49\textwidth]{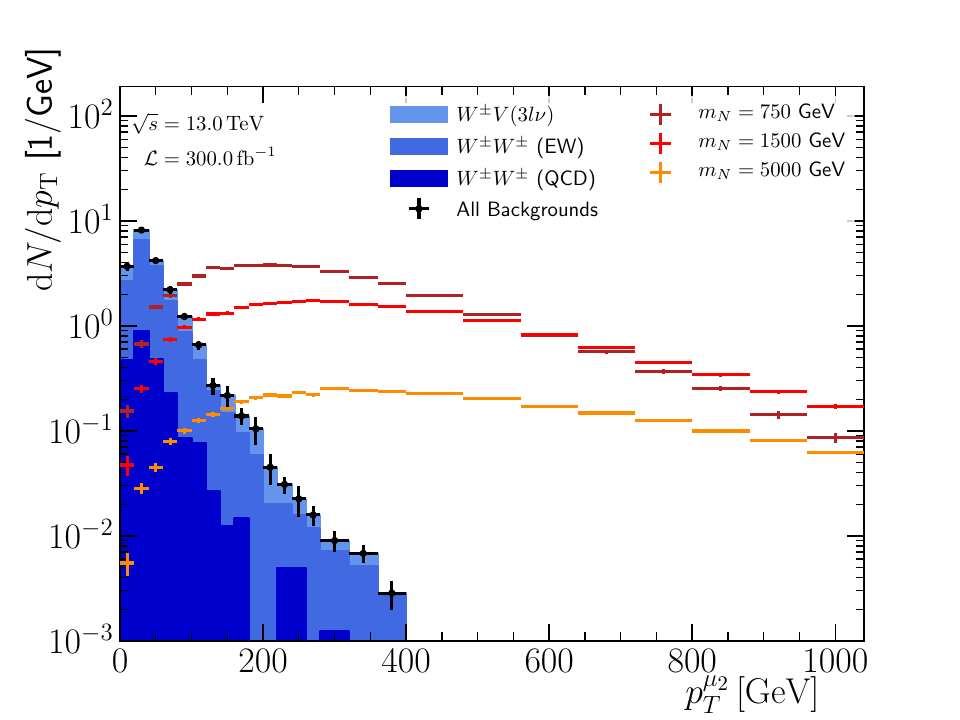} \label{fig:h_ptmu2_Preselection_0}}
\\
\caption{The distribution for (a) $M(j_1,j_2)$ and (b) $p_T^{\mu_2}$  after pre-selection cuts  for the signal process assuming representative heavy neutrino masses $m_N=750\GeV$ (darkest cross), $1.5\TeV$ (dark  cross), and $5\TeV$ (light cross). Also shown are the inclusive $W^\pm V$ (light), EW $W^\pm W^\pm jj$ (dark), and QCD $W^\pm W^\pm jj$ (darkest) backgrounds, as well as their sum (black cross). 
}
    \label{fig:kinematics_SR}
\end{figure*}

After pre-selection cuts, we anticipate that the leading background processes consist of mixed EW-QCD production of $W^\pm W^\pm jj$, pure EW production of $W^\pm W^\pm jj$,  and the inclusive diboson(+jets) spectrum $W^\pm V + nj$ with $V\in\{\gamma^*/Z^{(*)}\}$. For compactness we label these:
\begin{align}
        \text{$W^\pm W^\pm jj$ (QCD)}    &:~ pp\to W^\pm W^\pm jj \to 2\mu^\pm jj+X,\\
        \text{$W^\pm W^\pm jj$ (EW)}     &:~ pp\to W^\pm W^\pm jj \to 2\mu^\pm jj+X,\\
        \text{$W^\pm V(3\ell\nu)$}  &:~ pp\to 3\ell\nu+X.
\end{align}
Our MC modeling of these backgrounds is described in section~\ref{sec:modeling_bkg} and their generator-level rates are summarized in Table \ref{tab:mcxsec_bkg}. We have checked that other processes satisfying the same-sign muon signature do not appreciably survive pre-selection. For example: using the $K$-factors of Ref.~\cite{vonBuddenbrock:2020ter}, we estimate that the rate for the $t\overline{t}W^\pm\to2\mu^\pm+X$ process after pre-selection cuts, but minus the $M(j_1,j_2)$ requirement, is  \confirm{$\sigma_{t\overline{t}W^\pm}\sim1.2\ab$}. The $M(j_1,j_2)$ criterion reduces this an order of magnitude.

We acknowledge that we do not fully account for backgrounds that are difficult to simulate accurately with MC simulations alone. This includes opposite-sign muon pairs in which one muon is reconstructed with the incorrect charge, or events where one of the same-sign muons has a ``fake'' origin, \textit{e.g.}, a muon candidate originating from a jet. While such backgrounds are sub-dominant in the dimuon final state, this is less so for other lepton flavors~\cite{Aad:2016tuk,Alvarez:2016nrz,CMS-DP-2017-036,Pascoli:2018heg}. Whatever the final state, such backgrounds should be investigated carefully by experiments through dedicated studies based on collision data, as usually done in collider searches for LNV~\cite{Aaboud:2018spl,Sirunyan:2018xiv}.

To help define our analysis's signal region, we present in figure~\ref{fig:kinematics_SR} the distribution of (a) $M(j_1,j_2)$ and (b) $p_T^{\mu_2}$  for the signal process after pre-selection cuts, assuming representative heavy neutrino masses $m_N=750\GeV$ (darkest cross), $1.5\TeV$ (dark  cross), and $5\TeV$ (light cross). We also plot after pre-selection cuts the  $W^\pm V$ (light), EW $W^\pm W^\pm jj$ (dark), and QCD $W^\pm W^\pm jj$ (darkest) backgrounds, as well as their sum (black cross). The curves are normalized to $\mathcal{L}=300\invfb$.

Qualitatively, we observe that background processes tend towards smaller values of transverse momentum and invariant mass while the signal process tends towards larger values and exhibit broader, wider distributions. More quantitatively, we observe in figure~\ref{fig:h_mjj_Preselection_0} that background processes peak at $M(j_1,j_2)\lesssim800\GeV$ and taper off for larger invariant masses. This contrasts with the signal samples, which peak at $M(j_1,j_2)\gtrsim900\GeV$, plateau for a couple hundred GeV, and then gradually fall off. While the lightest heavy neutrinos benchmarks stay above the SM background for most all values of $M(j_1,j_2)$, we observe that the heaviest benchmark mass at $m_N=5\TeV$ only exceeds the background for $M(j_1,j_2)\gtrsim4.5\TeV$. Values of active-sterile mixing below unity will naturally worsen this separation power.

In figure~\ref{fig:h_ptmu2_Preselection_0} we observe that all backgrounds
peak at $p_T^{\mu_2}\sim m_V/2\sim40\GeV-45\GeV$, and quickly dissipate at higher $p_T$. As anticipated, this shows  that backgrounds are driven by resonant weak boson production, though not exclusively. Signal rates become more prominent for $p_T\gtrsim 50\GeV-100\GeV$. For heavy neutrinos masses beyond a few hundred GeV we report high selection efficiencies when requiring $p_T^{\mu_2}$ above this range, but less so for lower masses. In this regime, developing an alternative analysis strategy may increase the sensitivity but goes beyond the scope of this work. 

For $p_T^{\mu_2}\gtrsim300\GeV$, we find that the total background rate is  strongly suppressed. Subsequently, due to its simplicity, we define our signal region by requiring, in addition to pre-selection requirements, that both same-sign leptons carry $p_T$ above $300\GeV$. We summarize this in the bottom of Table~\ref{tab:cut_summary}. For the heavy neutrino masses under consideration, we find that about $\varepsilon\sim15\%$ to $80\%$ of pre-selection signal events survive signal region requirements. For representative masses,  we report in last column of table \ref{tab:signalRates_by_cuts} the signal rate cross section $\sigma^{\rm SR}$ and the corresponding selection efficiency 
$\varepsilon=\sigma^{\rm SR.}/\sigma^{\rm Pre.}$. 

After all selection cuts, we find that the total background rate reaches about \confirm{$\sigma_{b}^{\rm All~cuts}\approx2.35\ab$.} For each background and  their sum, we list in Table~\ref{tab:background} the expected number of background events after full selection for the nominal LHC scenario (LHC) with $\mathcal{L}=300\invfb$ as well as for the high-luminosity scenario (HL-LHC), where our estimate is computed by simply rescaling the luminosity to $\mathcal{L}=3\invab$. At the LHC, less than one background event is expected to pass the selection. 

\subsection{Results}\label{sec:results}

To quantify the expected excess number of $W^\pm W^\pm\to \mu^\pm\mu^\pm$ signal events over the number of SM background events, we follow the recommendations of Ref.~\cite{ATLAS:2020yaz} and employ asymptotic distributions of test statistics. In particular, we define our signal significance $Z$ as~\cite{Cousins:2008zz,Cowan:2010js}:
\begin{align}
Z &= \frac{(n - n_b)}{\vert n - n_b\vert}  \sqrt{2\left[n\log x - \frac{n_b^2}{\delta_b^2}\log y\right]}, \quad \mathrm{with}
\label{eq:significance}
\\
x  &= \frac{n(n_b+\delta_b^2)}{n_b^2+n \delta_b^2}, \quad \mathrm{and} \quad
y  = 1+ \frac{\delta_b^2(n-n_b)}{n_b(n_b + \delta_b^2)}.
\end{align}

 \begin{figure}[!t]
\includegraphics[width=.95\columnwidth]{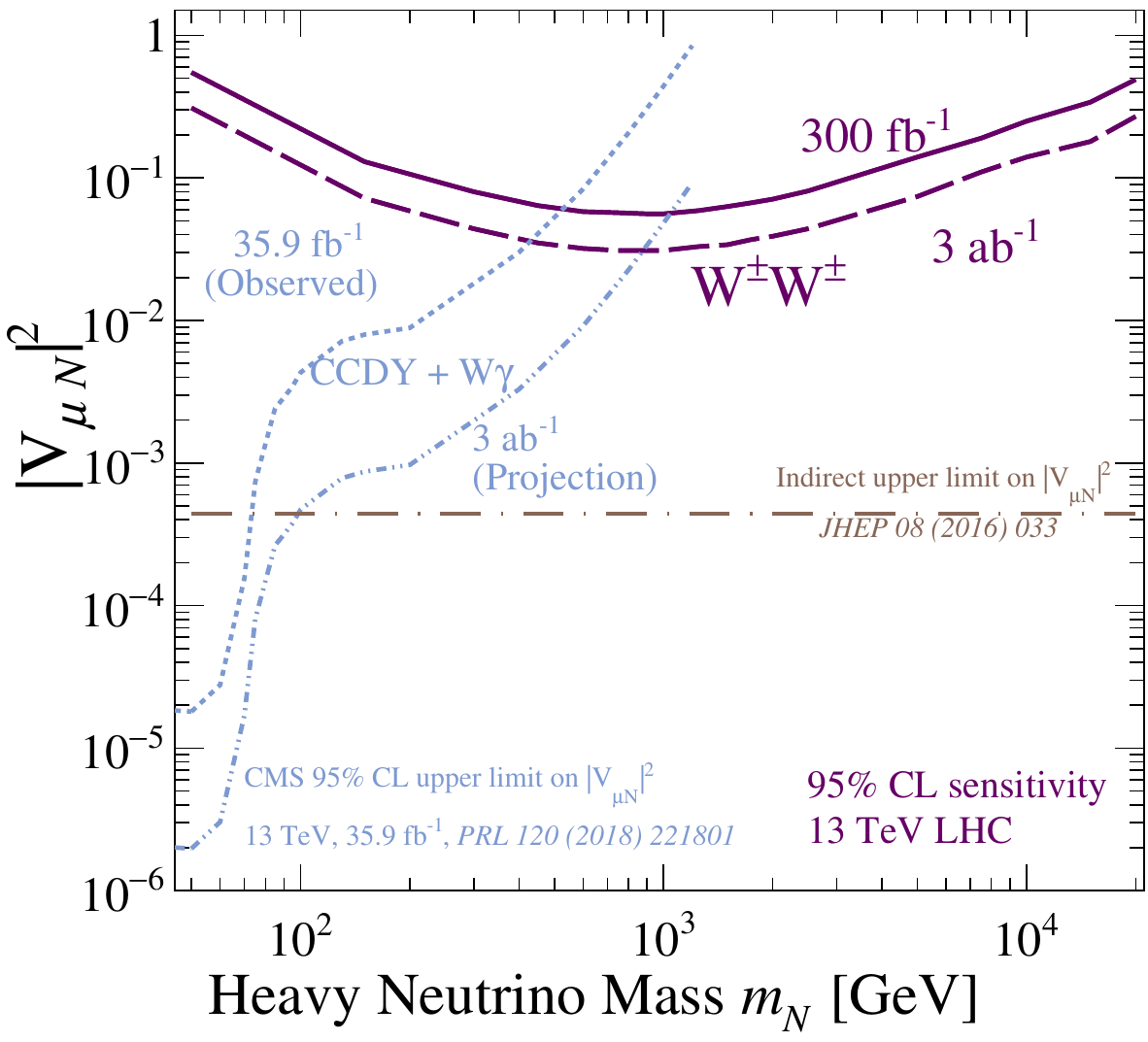}
\caption{Expected 95\% CL sensitivity at the $\sqrt{s}=13\TeV$ LHC (300 fb$^{-1}$) and HL-LHC (3 ab$^{-1}$) on the squared active-sterile mixing element $\vert V_{\mu N}\vert^2$ as a function of heavy neutrino mass following the $W^{\pm}W^{\pm}\to\mu^\pm\mu^\pm$ analysis described in the text. Also shown are the direct limits $(35.9\invfb)$ set by CMS using the CCDY and $W\gamma$ fusion channels~\cite{Sirunyan:2018mtv}, 
an extrapolation of the CMS to the HL-LHC,
and indirect limits~\cite{Fernandez-Martinez:2016lgt}.
}
\label{fig:lhc0vBB_HeavyN_Sensitivity}
\end{figure}

Here, $n=n_s+n_b$ is the total number of observed events, $n_s=\mathcal{L}\times\sigma^{\rm SR}_s$ is the number of signal events expected for  an integrated  luminosity of $\mathcal{L}$ and signal region rate $\sigma^{\rm SR}_s$. The quantity $n_b=\mathcal{L}\times \sigma^{SR}_b$ is the number of background events expected for a signal region rate $\sigma^{\rm SR}_b$, and $\delta_b$ is the uncertainty in $n_b$. Based on experimental measurements of the $W^\pm W^\pm$ scattering process and associated control regions at $\sqrt{s}=13\TeV$~\cite{Aaboud:2019nmv,Sirunyan:2020gyx}, we conservatively estimate our background uncertainty to be 20\%, \ie, we set $\delta_b=0.2$.

As discussed in Ref.~\cite{ATLAS:2020yaz}, the significance estimate $Z$ is consistent with a Poisson-counting likelihood where the background-rate nuisance parameters are constrained by auxiliary Poisson measurements~\cite{Cousins:2008zz}. This constraint is performed, for example, by using control samples enriched with background events. Defining such control samples, which are ultimately employed in the likelihood fits that constrain the normalization of backgrounds in the signal region, is beyond our scope. They are, however, commonly carried out in LHC analyses by choosing control samples in regions of phase space as close as possible to the signal region, but where no signal is expected.

 \begin{table}[!t]
    \centering
    \caption{Expected exclusion (excl.) and discovery (disc.) limits at the LHC (300 fb$^{-1}$) and HL-LHC (3 ab$^{-1}$) on the squared heavy neutrino mixing with the muon $|V_{\mu N}|^2$ following the analysis described in the text, as well as acceptance and efficiency with respect to the generator-level cross section.}
    \setlength\extrarowheight{3pt}
   \begin{tabular}{r|r|r|r|r|r}
   \hline\hline
    \multirow{2}{*}{$m_N$ [GeV]} & 
    \multicolumn{2}{c|}{$\mathcal{L} = 300\mathrm{\,fb}^{-1}$} & 
    \multicolumn{2}{c|}{$\mathcal{L} = 3\mathrm{\,ab}^{-1}$} & \multirow{2}{*}{$\frac{\sigma^\mathrm{SR}}{\sigma^\mathrm{Gen.}}$ [\%]} \rule[-1.0ex]{0pt}{0pt} \\
        & $|V_{\mu N}^\mathrm{excl.}|^2$  & $|V_{\mu N}^\mathrm{disc.}|^2$    
        & $|V_{\mu N}^\mathrm{excl.}|^2$ & $|V_{\mu N}^\mathrm{disc.}|^2$ & \rule[-1.5ex]{0pt}{0pt} \\   \hline
50 & 0.55 & 0.81 & 0.31 & 0.53 & 0.6\\
150 & 0.13 & 0.24 & 0.072 & 0.13 & 3.9\\
300 & 0.080 & 0.15 & 0.044 & 0.077 & 7.8 \\
450 & 0.064 & 0.12 & 0.035 & 0.062 & 12.1\\
600 & 0.058 & 0.10 & 0.032 & 0.056 & 15.6\\
750 & 0.057 & 0.10 & 0.031 & 0.054 & 18.2\\
900 & 0.056 & 0.10 & 0.031 & 0.054 & 21.1\\
1000 & 0.056 & 0.10 & 0.031 & 0.054 & 22.2\\
1250 & 0.059 & 0.11 & 0.033 & 0.057 & 24.2\\
1500 & 0.063 & 0.12 & 0.034 & 0.060 & 26.2\\
1750 & 0.067 & 0.12 & 0.037 & 0.064 & 27.1\\
2000 & 0.071 & 0.13 & 0.039 & 0.068 & 28.4\\
2500 & 0.081 & 0.15 & 0.044 & 0.078 & 29.4\\
5000 & 0.14 & 0.25 & 0.074 & 0.13 & 31.4\\
7500 & 0.19 & 0.36 & 0.11 & 0.19 &  32.2\\
10000 & 0.25 & 0.46 & 0.14 & 0.24 & 32.5\\
15000 & 0.34 & 0.62 & 0.18 & 0.32 & 32.6\\
20000 & 0.49 & 0.81 & 0.27 & 0.47 & 32.6\\
\hline\hline
  \end{tabular}
  \label{tab:limits}
\end{table}

We report in Table~\ref{tab:limits} that $\vert V_{\mu N}\vert^2 \gtrsim 0.06-0.6$ $(0.03-0.3)$ can be probed at 95\% CL for $m_N = 50~{\rm GeV}-20~{\rm TeV}$ with $\mathcal{L}=300\invfb$~(3 ab$^{-1}$). Under the assumption that the mass scale of one or more heavy neutrinos is much heavier than collision scales at $\sqrt{s}=13\TeV$, then in analogy to interpretations of searches for $0\nu\beta\beta$ decay, the LHC expected sensitivity at 95\% CL can be expressed as
\begin{equation}
    \Bigg\vert \sum_{k'=4}^{n_R+3}\frac{V_{\mu N_k}^2}{m_{N_k}}\Bigg\vert \gtrsim
\confirm{2.5~(1.4)\cdot10^{-2}~\TeV^{-1}}.
\end{equation}

Due to higher background rates, we anticipate slightly worse sensitivity for the $e^\pm\mu^\pm jj$ and $e^\pm e^\pm jj$ channels. For  final-states involving hadronic taus, we anticipate even weaker  (but still comparable) sensitivity due to tagging efficiencies. Dedicated studies of these complementary signatures are strongly encouraged.

We find that the proposed analysis has a strong potential to significantly extend the current sensitivity of ATLAS and CMS searches for resonant heavy neutrino masses beyond $m_N \sim 750$ GeV and up to masses at the $\mathcal{O}(10$ TeV$)$ scale, as shown in figure~\ref{fig:lhc0vBB_HeavyN_Sensitivity}. Similar to the existing ATLAS and CMS searches, the proposed analysis however does not reach a sensitivity comparable to the limits set by indirect precision measurements (see section~\ref{sec:theory_constraints}). The analysis does, however, offer a direct test of the $0\nu\beta\beta$ decay mechanism in lepton flavor configurations that are not accessible at nuclear energy scales.

\section{Outlook}\label{sec:outlook}

Discovering the $W^\pm W^\pm\to\ell^\pm_i\ell^\pm_j$ process in LHC collisions would present unambiguous evidence that LN is violated at the TeV scale, and have far-reaching repercussions for both theory and experiment. In light of the encouraging sensitivity reported in section \ref{sec:results},  we now briefly consider the outlook for further improvements to our proposed experimental analysis (section \ref{sec:outlook_ex}), as well as the possible application of our work to other scenarios, including when LN is  conserved (section \ref{sec:outlook_th}).

\subsection{Improving the Experimental Analysis}\label{sec:outlook_ex}

The analysis cuts chosen and outlined in section \ref{sec:event_selection} include only a simple set of selection criteria, which were derived to obtain a good significance for a large range of heavy neutrino masses and with the aim of being robust against the effects of finite detector resolution. Obviously, the selection can be optimized for individual neutrino masses. This is especially true for the lower mass range, where our proposed analysis can add sensitivity to current searches for resonant heavy neutrinos. Such improvements can be roughly grouped into those which further suppress the $W^\pm W^\pm$ or diboson backgrounds. 

An optimized analysis that also takes into account the resolution and acceptance of specific sub-detectors  should considerably improve our sensitivity estimate. An obvious choice would be to use $\met$ in the event selection. An upper cut on $\met$ would especially help  reduce the impact of $W^\pm W^\pm$ production, which has otherwise the same topology as the signal, but was omitted since it is particularly sensitive to detector resolution. As examined in section \ref{sec:wwScattLHC_diff}, the heavy neutrino signal is  characterized by high muon momenta that can reach the TeV scale. Even for a small fractional mis-measurement of muon momentum this can induce a considerable amount of $\met$. Hence correlating the $\met$ and muon momentum (or similarly jet momentum) in the selection would be a way to mitigate some of the resolution effects. 

$WW$ scattering processes, such as the heavy neutrino signal considered in this paper, commonly feature suppressed central hadronic activity. As a consequence, implementing static jet vetoes~\cite{Barger:1990py,Barger:1991ar,Bjorken:1992er,Fletcher:1993ij}, or their dynamic counter part~\cite{Pascoli:2018rsg,Michel:2018hui,Pascoli:2018heg, Fuks:2019iaj,Michel:2020dzo}, can be exploited to further reduce  diboson and top quark processes (or in general all non-VBF backgrounds). One can optimize the corresponding selections based on detector efficiency and resolution for low momentum jets as well as take into account LHC pileup conditions. Since the signal process discussed in this paper is modeled at NLO in QCD with parton shower matching, it can be used to study improvements connected to central hadronic activity. 

Lastly, our projections for the HL-LHC consisted of only a re-scaling of the results obtained for a smaller data set. However, the planned detector upgrades for ATLAS~\cite{ATLASCollaboration:2012ilu} and CMS~\cite{CMSCollaboration:2015zni} will also allow one to refine the analysis's selection criteria. In particular, the extended coverage of the tracking system will be highly relevant for $WW$ scattering processes due to the use of tracking information for jets in the forward region. 

\subsection{Applications to Other Seesaw Searches}\label{sec:outlook_th}

In this study we have focused exclusively on the LN-violating process $p p \to \ell^\pm_i \ell^\pm_j j j$,
when mediated by same-sign $WW$ scattering and the $t$-channel exchange of a heavy Majorana neutrino, as shown in figure \ref{fig:wwScattHeavyN_diagram}. In a complementary fashion, the LN-conserving process $p p \to \ell^\pm_i \ell^\mp_j j j$,
which proceeds through opposite-sign $WW$ scattering, is also possible~\cite{Dicus:1991fk,Datta:1993nm}. One could anticipate that the differences in helicity inversion (see section \ref{sec:wwScattLHC_total}) between the LN-violating $W^\pm W^\pm \to \ell^\pm_i \ell^\pm_j$ sub-process
and the LN-conserving  
$W^+W^-\to \ell^\pm_i \ell^\mp_j$ sub-process
results in substantially different kinematic distributions.
However, for heavy neutrinos in the range of $m_N=750\GeV$ to $5\TeV$, this may not be the case.

As reported in section \ref{sec:wwScattLHC_diff}, we found that the $W^\pm W^\pm \to \ell^\pm_i \ell^\pm_j$ sub-process in  $p p \to \ell^\pm_i \ell^\pm_j j j$ behaves like a factorizable system with kinematics that are nearly independent of the hadronic environment. This means that much of the charged lepton kinematic properties are driven more by momentum conservation in $2\to2$ scattering  than some complex spin correlation. It is arguable that many of the kinematic leverages over backgrounds that we find, {\it e.g.,} large dijet invariant masses and back-to-back charged lepton trajectories, will also hold for the LN-conserving case. While the  $p p \to \ell^\pm_i \ell^\mp_j j j$ collider signature inherently has a much larger background rate than the LN-violating one, we stress that lepton flavor violation is forbidden in the SM. Therefore, requiring that $\ell_i\neq\ell_j$ and that $\met$ is  small, as done for example in Ref.~\cite{Antusch:2018bgr} for low-scale Type I Seesaw models~\cite{Mohapatra:1986aw,Mohapatra:1986bd,Bernabeu:1987gr,Akhmedov:1995ip,Akhmedov:1995vm,Gavela:2009cd}, can significantly reduce SM backgrounds.

As a final remark, we comment on the applicability of our analysis to other neutrino mass models. This include, for example, the Type II Seesaw model  \cite{Magg:1980ut,Schechter:1980gr,Cheng:1980qt,Mohapatra:1980yp,Lazarides:1980nt}, wherein the LN-violating, VBF sub-process $W^\pm W^\pm \to \Delta^{\pm\pm (*)} \to \ell^\pm_i \ell^\pm_j$  can occur through a possibly non-resonant, $s$-channel, doubly charged Higgs boson $\Delta^{\pm\pm}$, as well as  Left-Right (LR) symmetric models~\cite{Pati:1974yy,Mohapatra:1974hk,Mohapatra:1974gc,Senjanovic:1975rk,Senjanovic:1978ev}, wherein the LN-violating, VBF sub-process $W^\pm_R W^\pm_R \to \ell^\pm_i \ell^\pm_j$ can proceed through two  $W_R^\pm$ gauge bosons and a $t$-channel Majorana neutrino. For  LR scenarios this is  interesting in the event that resonant production of $W_R$ is not within the kinematic reach of the LHC~\cite{Dev:2016dja,Ruiz:2017nip}. As both scenarios can mimic our $p p \to \ell^\pm_i \ell^\pm_j j j$ collider signature, its discovery does not automatically prove the existence of RH neutrinos. On the other hand, as both processes occur through the scattering of two color-singlet, massive gauge bosons, most of the color and Lorentz structure remains identical to the original case that we study. Therefore, we anticipate again that the VBF sub-processes approximately factorize, resulting in charged lepton kinematics that resemble those presented in section \ref{sec:wwScattLHC_diff}. As a result, the collider analysis that we propose can readily and justifiably be recast in terms of the Type II and LR symmetric models.

\section{Summary and Conclusions}\label{sec:conclusions}

Motivated by the possible non-conservation of LN in nature, we have investigated the potential to search for heavy Majorana neutrinos in same-sign $W^\pm W^\pm$ scattering at the LHC and the HL-LHC. The experimental signature of two forward jets from VBF, two same-sign leptons, and the lack of substantial missing transverse momenta is interesting in its own right as, to our knowledge, it was not yet explored experimentally at the LHC.

As a benchmark scenario we use the Phenomenological Type I Seesaw model with two key aspects to be probed experimentally: the mass $m_{N}$ of a mostly sterile neutrino $N$  and its mixing with the active neutrinos $\vert V_{\ell {N}}\vert$. Current searches at the LHC target resonant production modes, such as the Drell-Yan and $W\gamma$ fusion  mechanisms, which have the advantage of a factor $\vert V_{\ell N}\vert^2$ less suppression compared to the $W^\pm W^\pm\to\ell^\pm_i\ell^\pm_j$ channel. They however suffer from rapidly falling scattering rates at increasing heavy neutrino masses due to matrix element and phase space suppression. For these reasons LHC searches that employ resonant production modes can only probe masses up to $m_N=3-4\TeV$~\cite{Pascoli:2018heg}.  

To conduct this study, we developed in section \ref{sec:modeling} simulation prescriptions at NLO in QCD with parton shower-matching for both the VBF signal process and backgrounds based on the \textsc{HeavyN} UFO libraries and the \mgamc~simulation suite. We then extensively studied in section \ref{sec:wwScattLHC} the phenomenology  of the $W^\pm W^\pm\to \ell^\pm_i\ell^\pm_j$ process at the amplitude and differential levels. We find that ``bare'' cross section for the full, $2\to4$ signal process at NLO peaks for heavy neutrino masses of around 1 TeV and can reach up to $\sigma/\vert V_{\ell N}\vert^4\sim10\fb$ at $\sqrt{s}=13\TeV$. Apart from the  large rapidity gap between the two leading jets and large dijet invariant mass, the signal also features very high lepton momenta, among other characteristics, which can be exploited for an effective background suppression. 

In section \ref{sec:analysis} we designed our collider analysis, employing the \textsc{Delphes} framework to simulate the response of a typical LHC detector. Our analysis was deliberately kept simple and considers only final states with same-sign muon pairs to obtain reliable and robust results. Accordingly, dedicated  analyses exploiting the suppressed QCD radiation in VBF processes, the angular separation of the same-sign lepton pair,  or the correlation between the measured missing transverse momenta and very high-$p_T$ leptons should improve on our projected sensitivity. 

In section \ref{sec:results} we show that with the LHC Run 2 and expected Run 3 data sets, $\vert V_{\mu N}\vert^2$ can be probed down to $0.06-0.3$ at 95\% CL for heavy neutrino masses in the range $m_N=1-10$ TeV, and that masses at  $m_N=20$ TeV can be probed for $\vert V_{\mu N}\vert^2$ down to $0.5$. At the HL-LHC, this can be improved by a factor of $2$. Comparable results are anticipated for other lepton flavors. We find that the $W^\pm W^\pm$ fusion channel extends significantly the current mass reach based on resonant production modes and adds valuable sensitivity to the masses above a few hundred GeV. Finally in section \ref{sec:outlook}, we give an outlook on areas where our proposed analysis can be improved as well as on complementary applications of our results.

\section*{Acknowledgements}

The authors are grateful to Andreas Hoecker, Jordy de Vries, Juergen Reuter, Alishaan Tamarit for helpful discussions. BF and RR acknowledge the (virtual) hospitality of the DESY ATLAS group. 

JN and KP acknowledge support by the Deutsche Forschungsgemeinschaft (DFG, German Research Foundation) under Germany's Excellence Strategy -- EXC 2121 ``Quantum Universe'' – 390833306.
RR is supported under the UCLouvain fund “MOVE-IN Louvain” and  acknowledge the contribution of the VBSCan COST Action CA16108. This work  has received funding from the European Union's Horizon 2020 research and innovation programme as part of the Marie Skłodowska-Curie Innovative Training Network MCnetITN3 (grant agreement no. 722104), FNRS “Excellence of Science” EOS be.h Project No. 30820817.

Computational resources have been provided by the supercomputing facilities of the Universit\'e catholique de Louvain (CISM/UCL)  and the Consortium des \'Equipements de Calcul Intensif en F\'ed\'eration Wallonie Bruxelles (C\'ECI) funded by the Fond de la Recherche Scientifique de Belgique (F.R.S.-FNRS) under convention 2.5020.11 and by the Walloon Region. 

\appendix

\section{Constraints  on  heavy Majorana neutrinos from $0\nu\beta\beta$ decay searches}\label{app:0vBBLimits}

In this appendix we derive constraints on heavy Majorana neutrinos that arise from direct searches for nuclear $0\nu\beta\beta$ decay as reported  in equation~\eqref{eq:0vBBLimit}. To do this, we assume that the decay is solely mediated by the exchange of heavy states $N_k$ of mass $m_{N_k}$ that couple to SM particles according to the Lagrangian of equation~\eqref{eq:nuLag}. We work in the standard factorization picture~\cite{Doi:1981mi,Doi:1982dn,Primakoff:1959chj,Tomoda:1990rs}. This  stipulates that the transition rate for the decay process of nucleus $(A,Z)$ into nucleus $(A,Z+2)$, 
\begin{equation}
    (A,Z)\to (A,Z+2)+2e^-,
\end{equation}
can be expressed as a product of the {two-body} phase space factor $G_{0\nu}$ for the $(e^-e^-)$-system; a nuclear matrix element (NME)  $\mathcal{A}$; and the propagators for the states $N_k$. Under the assumption that the $0\nu\beta\beta$ decay process is mediated only by the $t$-channel exchange of $W$ bosons and $N_k$, the NME simplifies to $\mathcal{A}\approx\mathcal{A}_{N}$. Explicitly, these are related to the  decay half-life $(T_{1/2}^{0\nu})$ by
\begin{equation}
    \frac{1}{T_{1/2}^{0\nu}} = G_{0\nu}  m_p^2 \vert\mathcal{A}_N\vert^2  \left\vert \sum_k V_{ek}^2 \frac{m_{N_k}}{(t-m_{N_k}^2)} \right\vert^2.
    \label{eq:halflife}
\end{equation}
The proton mass $m_p\approx0.938\GeV$ is introduced to render $\mathcal{A}_N$ dimensionless. Typical momentum transfers in nuclear $0\nu\beta\beta$ decay are of the order  $\sqrt{\vert t \vert} \sim \mathcal{O}(0.1\GeV)$. As we are interested in {EW-scale and TeV-scale} neutrinos, this allows us to  expand equation~\eqref{eq:halflife} and obtain
\begin{align}
    \left\vert \sum_k V_{ek}^2 \frac{m_{N_k}}{(t-m_{N_k}^2)} \right\vert^2
    &=
    \left\vert \sum_k V_{ek}^2 \frac{-1}{m_{N_k}\left(1-\frac{t}{m_{N_k}^2}\right)} \right\vert^2 \\
    &=     \left\vert \sum_k  \frac{V_{ek}^2}{m_{N_k}} \right\vert^2 + \mathcal{O}\left(\frac{t}{m_{N_k}^2}\right).
\end{align}
By  neglecting $\mathcal{O}\left(\vert t\vert/m_{N_k}^2\right)$ contributions we can invert equation~\eqref{eq:halflife} and translate an  experimental lower bound on $T_{1/2}^{0\nu}$ into an upper bound on the mixing-over-mass ratio  of Majorana neutrinos.  In  the following we focus on {the $^{76}$Ge$\to^{76}$Se$+2e^-$ transition}, and consider the recent experimental limits by the GERDA experiment~\cite{Ackermann:2012xja,Agostini:2020xta}.

Following Ref.~\cite{Pascoli:2013fiz}, we use the NMEs of Refs.~\cite{Faessler:2011qw,Meroni:2012qf}. {These employ the so-called Self-consistent Renormalized Quasiparticle Random Phase Approximation (SRQRPA), and make use of two potential models to describe the nucleon-nucleon interactions, namely the Argonne and Charge Dependent Bonn (CD-Bonn) models. The calculations moreover rely on intermediate (Intm.) and large (Large) single-particle spacing, \ie, eigenstate multiplicity, an axial-vector coupling constant $g_A=1.25$, and a nuclear radius $R=1.1~{\rm fm} \times A^{1/3}$}. For $^{76}$Ge, we list  in Table~\ref{tab:0vBBDecay} the  $\mathcal{A}_{N}$ for the four nuclear  potential configurations. The uncertainty in $\mathcal{A}_{N}$ is estimated by considering the envelope spanned by the configurations.

\begin{table}[!t]
 \centering
 \caption{
   Values of the NMEs for the nucleon-nucleon
   interaction models under consideration~\cite{Faessler:2011qw,Meroni:2012qf}
   (second column), and the corresponding exclusion limits at 90\% CL on heavy neutrino masses and mixing extracted from results by the GERDA experiment~\cite{Agostini:2020xta} (third column).
 }
    \setlength\tabcolsep{8pt}
    \renewcommand{\arraystretch}{1.4}
    \begin{tabular}{l l c c}
    \hline\hline
    \multicolumn{2}{c}{NME model}  & $\mathcal{A}_N$ & $\vert \sum_k V_{ek}^2/m_k\vert$\\
    \hline
    Argonne & intm. & 232.8 & $4.12\times10^{-6}$~TeV$^{-1}$      \\
    Argonne & large & 264.9 & $3.62\times10^{-6}$~TeV$^{-1}$       \\
    CD-Bonn & intm. & 351.1 & $2.73\times10^{-6}$~TeV$^{-1}$       \\
    CD-Bonn & large & 411.5 & $2.33\times10^{-6}$~TeV$^{-1}$       \\
    \hline\hline
    \end{tabular}
     \label{tab:0vBBDecay}    
\end{table}

We use the phase space factor $G_{0\nu}=G_{0\nu}^{(0)}g_A^4$, as derived by  Ref.~\cite{Kotila:2012zza}, assuming an axial-vector cutoff  of $g_A=1.25$.
While the polarization-dependent component $G_{0\nu}^{(1)}$ is non-zero, its impact on  the  total $0\nu\beta\beta$ decay  rate vanishes  after phase  space integration. The derivation of the energy-dependent component $G_{0\nu}^{0} \propto 1/R^2$ in Ref.~\cite{Kotila:2012zza} uses a nuclear radius of $R=1.2~{\rm fm} \times A^{1/3}$.  Hence, for consistency with our NMEs, we rescale it by
\begin{equation}
    G_{0\nu}^{(0)}(r_0=1.1~{\rm fm}) =  G_{0\nu}^{(0)}(r_0=1.2~{\rm fm}) \times \left(\frac{1.2}{1.1}\right)^2.
\end{equation}
For $^{76}$Ge we obtain the phase space factor,
\begin{eqnarray}
G_{0\nu}\approx 6.866\times10^{-15}~{\rm yr}^{-1}.
\end{eqnarray}
When added in quadrature, the total estimated uncertainty in this number spans about $\delta G_{0\nu}  \approx  7\%-9\%$~\cite{Kotila:2012zza}. This is considerably smaller than the NME uncertainty, and therefore is neglected in our final constraints on $N_k$.

After an exposure of $\mathcal{E}=127.2$ kg-yr, GERDA  reports a lower limit on the $0\nu\beta\beta$ decay half-life in  $^{76}$Ge of \cite{Agostini:2020xta}
\begin{equation}
    T_{1/2}^{0\nu} > T_{90\%~CL}^{\rm GERDA} = 1.8\times10^{26}~{\rm yr} ~ \text{at 90\%  CL}.
\end{equation}
For the range of NMEs, this translates into the following  upper limit on Majorana neutrino masses and mixing:
\begin{equation}
     \left\vert \sum_{k=4}^{n_R+3}  \frac{V_{ek}^2}{m_k} \right\vert <
     (2.33-4.12)\times10^{-6} \TeV^{-1}.
\end{equation}
For each NME that we consider we report in Table~\ref{tab:0vBBDecay} the corresponding limit on  the mixing-to-mass ratio. For related discussions on $0\nu\beta\beta$ decay, see Refs.~\cite{Mitra:2011qr,Cirigliano:2018yza,Dolinski:2019nrj}.

\section{Tailored phase space cuts on leading leptons in  \mgFull}\label{app:fxfxCuts}

We document here our implementation of tailored phase space cuts in the event generator \mgamc.

As described in section \ref{sec:modeling_bkg_diboson},  our baseline modeling of the diboson spectrum $pp\to 3\ell\nu+X$, provides limited MC statistics when the two same-sign leptons carry $p_T^\ell\gtrsim 100-150\GeV$ but the odd-sign lepton is much softer. To populate this phase space region, we introduce into \mgamc's phase space integration routines tailored generator-level cuts  $(p_T^{\ell-\rm cut})$ on the $p_T$ of the two leading charged leptons, independent of charge. This is in addition to the baseline cuts of equations~\eqref{cuts:gen_3lv_lep} and \eqref{cuts:gen_3lv_FxFx}; no further cut is applied to the trailing charged lepton.

High-statistics samples with $p_T^{\ell-\rm cut}=75\GeV$ and $200\GeV$ are stitched to the baseline FxFx1j sample through cuts on the truth-level $p_T$ of the sub-leading charged lepton. Within statistical uncertainty, we report that the shape and normalization of the high-$p_T$ tails for the leading charged leptons in the stitched sample reproduce those in the baseline FxFx1j sample.  To do this, we make several  modifications\footnote{\label{foot:launchpad}An initial version of these modifications was documented in the URL \href{https://answers.launchpad.net/mg5amcnlo/+question/691233}{answers.launchpad.net/mg5amcnlo/+question/691233}.} to the files \texttt{cuts.f} and \texttt{setcuts.f} in the \texttt{SubProcesses} working directory. Working with version 2.7.1.2 of \mgamc~and for the case of $p_T^{\ell-\rm cut}=75\GeV$ (\texttt{pTlXCut = 75 GeV}), we add to the header  in \texttt{cuts.f} at L72:
\begin{verbatim}
c define user cuts for pTl2
      double precision pTlXCut,pTlXSum
      double precision pTlXMin,pTlXMax
      logical gotLep1
      parameter(pTlXCut = 75.d0)
\end{verbatim}
and at L159 add the lines
\begin{verbatim}
      pTlXSum = 0.d0
      pTlXMax = 0.d0
      pTlXMin = 0.d0
      gotLep1 = .false.
c get pT of hardest and softest charged lepton:
      do i=nincoming+1,nexternal
         if (is_a_lp(i).or.is_a_lm(i)) then
c hypothesize that i hardest and softest
            if(.not.gotLep1) then
               pTlXMax = pt_04(p(0,i))
               pTlXMin = pt_04(p(0,i))
               gotLep1 = .true.
            endif
            pTlXSum = pTlXSum + pt_04(p(0,i))
c update if i is harder or softer
            if(pt_04(p(0,i)).gt.pTlXMax) then
               pTlXMax = pt_04(p(0,i))
            endif
            if(pt_04(p(0,i)).lt.pTlXMin) 
               pTlXMin = pt_04(p(0,i))
            endif
         endif
      enddo
c check if subleading lepton pT is hard enough
      pTlXSum = pTlXSum - pTlXMax - pTlXMin
      if(pTlXSum.lt.pTlXCut.or.
     &   pTlXMax.lt.pTlXCut) then
         passcuts_user=.false.
         return
      endif
\end{verbatim}
In practice, the magnitude of the charged leptons' transverse momenta are first added, then the largest and smallest $p_T$ are subtracted to extract the $p_T$ of the sub-leading charged lepton. If either the leading or sub-leading $p_T$ are below the $p_T^{\ell-\rm cut}$ threshold (\texttt{pTlXCut}), then the phase space point is rejected.

To ameliorate inefficient phase space sampling associated with our cuts, we increment the boundary of the PDF convolution integral $\tau_{\min}=\hat{s} / s$, where $\sqrt{\hat{s}}~ (\sqrt{s})$ is the partonic (hadronic) c.m.~energy, by $1.5\times p_T^{\ell-\rm cut}$. To do this we modify \texttt{setcuts.f} at about L422 with:
\begin{verbatim}
      double precision pTlXCut,cutFact
      parameter (pTlXCut = 75.d0)
      parameter (cutFact =  1.5d0)
\end{verbatim}
and at L421 add the following
\begin{verbatim}
c Increment for pTlXCut on charged leptons
            taumin(iFKS,ichan)=
     & taumin(iFKS,ichan)+pTlXCut*cutFact
            taumin_j(iFKS,ichan)=
     & taumin_j(iFKS,ichan)+pTlXCut*cutFact
            taumin_s(iFKS,ichan)=
     & taumin_s(iFKS,ichan)+pTlXCut*cutFact
\end{verbatim}
This is inserted just after the \texttt{enddo} closure tag at about L421 and just before the line
\begin{verbatim}
            stot = 4d0*ebeam(1)*ebeam(2)
\end{verbatim}

For phase space cuts beyond $p_T^{\ell-\rm cut}\sim 150\GeV$, we observe a severe instability in phase space integration. As documented elsewhere (see footnote \ref{foot:launchpad}), this failure is attributed to inefficient phase space sampling for non-resonant diagrams with massive $\tau$ leptons. Hence, for the $p_T^{\ell-\rm cut}= 200\GeV$ sample, we import into \mgamc~the model file  \texttt{loop\_sm-no\_tau\_mass}, which assumes a massless $\tau$ lepton. For looser $p_T^{\ell-\rm cut}$, we find that this results in sub-percent differences in the cross section normalization from the \texttt{loop\_sm} model file.

\section{Helicity amplitudes for same-sign $W W$ scattering via heavy Majorana neutrinos}\label{app:helicity}

Here we document {our calculation of} helicity amplitudes for same-sign $W W$ scattering to same-sign lepton pairs  when mediated {by} a heavy Majorana neutrino, as shown diagrammatically in figure \ref{fig:wwScattHeavyN_diagram}  and discussed in section \ref{sec:wwScattLHC_total}. To build a succinct picture of the physics, we employ the effective $W$ approximation ~\cite{Dawson:1984gx,Kane:1984bb,Kunszt:1987tk}. In this formalism, which is akin to collinear factorization in perturbative QCD, the $W$ boson is treated as a parton of the proton. This allows us to focus on the $2\to2$ subprocess
\begin{align}
    W^+_\mu(p_1^W,\lambda_1^W) + W^+_\nu(p_2^W,\lambda_2^W) \to 
    \nonumber\\
    \ell^+(p_1^\ell,\lambda_1^\ell) &+ \ell^+(p_2^\ell,\lambda_2^\ell),
\end{align}
where $p$ and $\lambda$ denote the 4-momenta and helicities of external particles. The amplitudes reported here supplement the analytic results reported in section \ref{sec:wwScattLHC}. They are also complementary to the numerical results reported throughout the main text, which evaluate precisely the full $2\to4$ helicity amplitudes for the $pp\to \ell^\pm \ell^\pm j j +X$ process using the \textsc{HeavyN} NLO UFO libraries~\cite{Degrande:2016aje} in conjunction with the \mgamc~\cite{Stelzer:1994ta,Alwall:2014hca} MC event generator (see sections \ref{sec:setup} and \ref{sec:modeling} for related details).

For the above process, we work in the hard-scattering frame, which is equivalent to the $WW$ scattering frame. In this frame, we align coordinate axes such that 
\begin{align}
p_1^W &= \frac{M_{WW}}{2}(1,0,0,+\beta_W), \quad \beta_W = \sqrt{1-4 r_W},\\ 
p_2^W &= \frac{M_{WW}}{2}(1,0,0,-\beta_W), \quad r_W = \frac{m_W^2}{M_{WW}^2},\\
p_1^\ell &= \frac{M_{WW}}{2}(1, \sin\theta_1\cos\phi_1, \sin\theta_1\sin\phi_1, \cos\theta_1),\\
p_2^\ell &= p_1^W + p_2^W - p_1^\ell.
\end{align}
Here $M_{WW}^2  = (p_1^W+p_2^W)^2$ is the invariant mass of the $(WW)$-system and the remaining invariants are {given by}
\begin{align}
t &= (p_1^W-p_1^\ell)^2=m_W^2 - \frac{M_{WW}^2}{2}\left(1-\beta_W\cos\theta_1\right),\\
u &= (p_1^W-p_2^\ell)^2=m_W^2 - \frac{M_{WW}^2}{2}\left(1+\beta_W\cos\theta_1\right).
\end{align}

\begin{table*}[t!]
\renewcommand{\arraystretch}{1.75}
\begin{center}
\caption{
Helicity amplitude components for the 
$W^+_\mu(p_1^W,\lambda_1^W) + W^+_\nu(p_2^W,\lambda_2^W) \to 
    \ell^+(p_1^\ell,\lambda_1^\ell) + \ell^+(p_2^\ell,\lambda_2^\ell)$
 process, according to external helicities (first column) for the $t$-channel (second column) and $u$-channel (third column) heavy neutrino exchange.
}\resizebox{\textwidth}{!}{
\begin{tabular}{rrcc|c|c}
\hline\hline
$\lambda_1^W$ & $\lambda_2^W$  & $\lambda_1^\ell$ & $\lambda_2^\ell$  &
\qquad 
$\varepsilon_\mu(p_1^W,\lambda_1^W)\varepsilon_\nu(p_2^W,\lambda_2^W) {\mathcal{T}}^{\mu\nu}(p_1^\ell,p_2^\ell) ~/~ -i\left(\frac{-ig_W}{\sqrt{2}}\right)^2 
\frac{V_{\ell N} V_{\ell N} m_N }{\left(t - m_N^2\right)}$
\qquad ~ 
&
\qquad
$\varepsilon_\mu(p_1^W,\lambda_1^W)\varepsilon_\nu(p_2^W,\lambda_2^W) {\mathcal{T}}^{\mu\nu}(p_2^\ell,p_1^\ell) ~/~ -i\left(\frac{-ig_W}{\sqrt{2}}\right)^2 
\frac{V_{\ell N} V_{\ell N} m_N }{\left(u - m_N^2\right)}  $
\qquad ~ 
\\ 
\hline
$+1$    & $+1$  & $R$ & $R$ 
& $2e^{-i\phi_1} M_{WW} \cos^2\left(\frac{\theta_1}{2}\right)$
& $2e^{-i\phi_1} M_{WW} \sin^2\left(\frac{\theta_1}{2}\right)$
\\
$+1$    & $-1$    & $R$ & $R$ 
& $0$   
& $0$
\\
$+1$    & $0$    & $R$ & $R$ 
&   $-\frac{1}{2\sqrt{2}r_W} M_{WW} (1+\beta_W)\sin\theta_1$
&   $\frac{1}{2\sqrt{2}r_W} M_{WW} (1+\beta_W)\sin\theta_1$
\\
$-1$    & $+1$    & $R$ & $R$ 
&   $0$   
&   $0$
\\ 
$-1$    & $-1$    & $R$ & $R$ 
&   $2e^{-i\phi_1} M_{WW} \sin^2\left(\frac{\theta_1}{2}\right)$
&   $2e^{-i\phi_1} M_{WW} \cos^2\left(\frac{\theta_1}{2}\right)$
\\ 
$-1$    & $0$    & $R$ & $R$ 
&   $-\frac{1}{2\sqrt{2}r_W} e^{-i2\phi_1} M_{WW} (1-\beta_W)\sin\theta_1$
&   $\frac{1}{2\sqrt{2}r_W} e^{-i2\phi_1} M_{WW} (1-\beta_W)\sin\theta_1$
\\ 
$0$    & $+1$    & $R$ & $R$ 
&   $\frac{1}{2\sqrt{2}r_W}  e^{-i2\phi_1} M_{WW} (1+\beta_W)\sin\theta_1$
&   $-\frac{1}{2\sqrt{2}r_W} e^{-i2\phi_1} M_{WW} (1+\beta_W)\sin\theta_1$
\\ 
$0$    & $-1$    & $R$ & $R$ 
&   $\frac{1}{2\sqrt{2}r_W} M_{WW} (1-\beta_W)\sin\theta_1$
&   $-\frac{1}{2\sqrt{2}r_W} M_{WW} (1-\beta_W)\sin\theta_1$
\\ 
$0$    & $0$    & $R$ & $R$ 
& $-\frac{1}{2 r_W} e^{-i\phi_1} M_{WW}\left(1-2r_W-\beta_W\cos\theta_1\right)$
& $-\frac{1}{2 r_W} e^{-i\phi_1} M_{WW}\left(1-2r_W+\beta_W\cos\theta_1\right)$
\\ \hline\hline
\end{tabular}
} 

\label{tab:scattME}
\end{center}
\end{table*}

Working in the unitary gauge and assuming a clockwise fermion flow of leptons~\cite{Denner:1992vza,Denner:1992me}, the helicity amplitudes in the \textsc{HELAS} basis~\cite{Murayama:1992gi} are given by
\begin{eqnarray}
-i\mathcal{M} &=&
\varepsilon_\mu(p_1^W,\lambda_1^W)\varepsilon_\nu(p_2^W,\lambda_2^W)\mathcal{T}^{\mu\nu}(p_1^\ell,p_2^\ell,\lambda_1^\ell,\lambda_2^\ell)    
\nonumber\\
& & + (t\leftrightarrow u),
\label{appeq:meWWScatt}
\end{eqnarray}
where the $(t\leftrightarrow u)$ term accounts for final-state lepton exchange, and the LN-violating tensor current $\mathcal{T}^{\mu\nu}$ is
\begin{align}
 \mathcal{T}^{\mu\nu}
= & 
-i\left(\frac{-ig_W}{\sqrt{2}}\right)^2 
\frac{V_{\ell N} V_{\ell N}}{\left(t - m_N^2\right)} \times
\\
& 
\left[
\overline{u}(p_1^\ell,\lambda^\ell_1)\gamma^\mu P_R
\left(\not\!p_N + m_N\right) \gamma^\nu P_L  v(p_2^\ell,\lambda_2^\ell)\right]
\nonumber\\
&  =
-i\left(\frac{-ig_W}{\sqrt{2}}\right)^2 
\frac{V_{\ell N} V_{\ell N} }{\left(t - m_N^2\right)} \times m_N \times
\label{appeq:lnvCurrent}
\\
&  
\left[
\overline{u}(p_1^\ell,\lambda_1^\ell)\gamma^\mu
\gamma^\nu P_L  v(p_2^\ell,\lambda_2^\ell)\right].
\nonumber
\end{align}
We assume the exchange of a single sterile neutrino mass eigenstate $N$ with momentum  $p_N = (p_1^W - p_1^\ell)$  and mass $m_N$. For the more general case of multiple heavy neutrinos $N_k$ with masses $m_{N_k}$, one would substitute:
\begin{align}
    \mathcal{T}^{\mu\nu} &\to \sum_{k=3}^{n_R+3}\mathcal{T}^{\mu\nu}_k, \\
    \frac{V_{\ell N} V_{\ell N} m_N }{\left(t - m_N^2\right)}
    &\to 
    \sum_{k=3}^{n_R+3}     \frac{V_{\ell N_k} V_{\ell N_k} m_{N_k} }{\left(t - m_{N_k}^2\right)},
\end{align}
to capture the interference of multiple propagating mass eigenstates. Similarly, for two final-state lepton flavors, one substitutes, $V_{\ell N_k} V_{\ell N_k} \to V_{\ell_1 N_k} V_{\ell_2 N_k}$. Importantly, the spinor and Lorentz index contractions are not modified and therefore are the same for any number of $t$-channel Majorana neutrino {exchange}s.
 
 Explicit evaluation (and inspection) of ${\mathcal{T}}^{\mu\nu}$ shows that the tensor is non-vanishing only when both final-state antileptons carry {right}-handed polarizations, $(\lambda_1^\ell, \lambda_2^\ell) = (R,R)$. Beyond this, the full matrix element also vanishes identically when both incoming $W$ bosons carry opposite transverse polarizations, \ie, when $(\lambda^W_1,\lambda^W_2)=(\pm1,\mp1)$, which follows from an orthogonality of $d$- and $p$-wave states. For both  $t$-channel (second column) and $u$-channel (third column) exchanges of a heavy neutrino, we list in table \ref{tab:scattME} the exact helicity amplitudes as a function of external particle helicity (first column).
 
 We find that $t$- and $u$-channel tensor structures for  the $(\lambda_1^W,\lambda_2^W)=(0,\pm1)$ and $(\pm1,0)$ configurations differ simply by a global factor of $-1$. The  $t$- and $u$-channel structures for  $(\lambda_1^W,\lambda_2^W)=(\pm,\pm1)$  differ by exchanges of sine and cosine functions, whereas for $(\lambda_1^W,\lambda_2^W)=(0,0)$ there is a minor difference in the polar angle dependence. For a  fixed set of helicity polarizations, the matrix element of equation \eqref{appeq:meWWScatt} is obtained by the standard coherent summation of $t$- and $u$-channel terms.

\subsection{Low-mass limit}\label{appsec:wwScattLHC_total_loMass}

When the heavy neutrino and $W$ boson masses are both small compared to the $(WW)$-scattering scale, \ie, $m_N,m_W  \ll M_{WW}$, we can expand each of the squared matrix elements
in powers of the ratios
\begin{equation}
    r_N=\frac{m_N^2}{M_{WW}^2} \quad\text{and}\quad  r_W=\frac{m_W^2}{M_{WW}^2}.
\end{equation}
Doing so reveals that remaining transverse-transverse permutations, \ie,  $(\lambda^W_1,\lambda^W_2)=(\pm1,\pm1)$, as well as LH-longitudinal channels, $(\lambda^W_1,\lambda^W_2)=(-1,0)$  and $(0,-1)$, either vanish or are sub-leading. In the latter cases, we see the emergence of a helicity suppression that can compete or overcome longitudinal polarization enhancements, which  scale as $\varepsilon^\mu(k,0)\sim k^\mu/m_W+\mathcal{O}(m_W/k^0)$. 

To lowest order in $r_N$ and  $r_W$, the RH-longitudinal helicity configurations, $(\lambda^W_1,\lambda^W_2)=(+1,0)$  and $(0,+1)$, are also sub-leading, but at a parametrically milder degree than the previous combinations. This follows from the alignment of the $(WW)$-system's angular momentum with that of the dilepton system and a single longitudinal polarization enhancement. Explicitly, we obtain
\begin{align}
&    \vert\mathcal{M}(W^+W^+\to \ell^+\ell^+)\vert^2_{(\lambda^W_1,\lambda^W_2)=(+1,0),(0,+1)} 
\nonumber\\ 
& = 2g_W^4 \vert V_{\ell N}\vert^4  \left(\frac{r_N}{r_W}\right) \cot^2\theta_1 + \mathcal{O}(r_N^2,~r_W^0).
\end{align}
This expression does not account for the $(1/2!)$ multiplicity factor for identical final-state particles. 

We find that the leading polarization configuration in this kinematic limit is the longitudinal-longitudinal channel, $(\lambda^W_1,\lambda^W_2)=(0,0)$. We attribute its survival in the expansion  to the double longitudinal enhancement and is 
\begin{align}
       \vert\mathcal{M}(W^+W^+&\to \ell^+\ell^+)\vert^2_{(\lambda^W_1,\lambda^W_2)=(0,0)} 
\nonumber\\ 
&  = g_W^4  \vert V_{\ell N}\vert^4 \left(\frac{r_N}{r_W^2}\right) + \mathcal{O}(r_N^2,~r_W^{-1}).
\end{align}

To build the partonic, $2\to2$ cross section $\hat{\sigma}$ we employ the standard relationship between scattering rates and matrix elements. This is given by the  phase space integral
\begin{align}
   \hat{\sigma}(W^+W^+\to \ell^+\ell^+) = \int {\rm d}PS_2 \frac{{\rm d}\hat{\sigma}}{{\rm d}PS_2},
   \label{eq:xsecTotDef}
\end{align}
where the 2-body phase space volume measure is
\begin{align}
{\rm d}PS_2(p_1^W+p_2^W; p_1^\ell,p_2^\ell)=\frac{{\rm d}\cos\theta_1 {\rm d}\phi_1}{2(4\pi)^2}\beta_W,      
\end{align}
and the differential scattering rate is
\begin{align}
   \frac{{\rm d}\hat{\sigma}}{{\rm d}PS_2} &= 
   \frac{1}{2M_{WW}^2 \beta_W}
   \frac{1}{\mathcal{S}}
   \sum_{\{\lambda\}} \vert \mathcal{M}(\{\lambda\})\vert^2.
    \label{eq:xsecDifDef}
\end{align}
The symmetry factor $\mathcal{S}=3^2\cdot2!$ accounts for spin-averaging over initial-state $W$ polarizations  and identical, final-state particles, while the incoherent summation runs over all external helicities $\{\lambda_1^W,\lambda_2^W,\lambda^\ell_1,\lambda^\ell_2\}$. The velocity factor $\beta_W=\sqrt{1-4r_W}$ accounts for the masses of incoming beam particles.

After phase integration over the azimuthal direction, the leading contribution to the polar distribution of $\ell_1^+$ in $W^+W^+\to\ell^+\ell^+$ scattering is given analytically by
\begin{align}
   \frac{{\rm d}\hat{\sigma}}{{\rm d}\cos\theta_1} = 
   \frac{g_W^4}{2^6 3^2 \pi m_W^4}
   \vert V_{\ell N}\vert^4  m_N^2
   + \mathcal{O}\left(r_N^2, r_W^{-1}\right).
\end{align}
After integration over the polar angle, the total rate is
\begin{align}
   \hat{\sigma} 
   = 
   \frac{g_W^4}{2^5 3^2 \pi m_W^4}
   \vert V_{\ell N}\vert^4  m_N^2
   + \mathcal{O}\left(r_N^2, r_W^{-1}\right).
\end{align}

For this limit, we report agreement between this expression and numerical evaluations of the same $2\to2$ process using {the \textsc{HeavyN} model with \mgamc}. For multiple heavy neutrinos coupling to potentially different charged lepton flavors, the above generalizes to
\begin{align}
  \hat{\sigma} (W^+W^+&\to\ell^+_1\ell^+_2)
   = \frac{g_W^4(2-\delta_{\ell_1\ell_2})}{2^5 3^2 \pi m_W^4} \\
   &
   \times \Bigg\vert \sum_{k=4}^{n_R+3} V_{\ell_1 N_k} m_{N_k} V_{\ell_2 N_k} \Bigg\vert^2
   + \mathcal{O}\left(r_N^2, r_W^{-1}\right).
   \nonumber
\end{align}

\subsection{High-mass limit}\label{appsec:wwScattLHC_total_hiMass}

When the heavy neutrino mass is large compared to the $W^\pm W^\pm\to\ell^+\ell^+$ scattering scale, \ie, when $m_N  \gg M_{WW}, m_W$, one can work in the decoupling limit~\cite{Appelquist:1974tg} and treat the exchange of $N$ as a point-like, contact interaction. Formally, this entails expanding the heavy neutrino propagator such that
\begin{align}
    \frac{1}{t-m_N^2} & = \frac{-1}{m_N^2} + \mathcal{O}\left(\frac{\vert t\vert^2}{m_N^4}\right), \\
    \frac{1}{u-m_N^2} & = \frac{-1}{m_N^2} + \mathcal{O}\left(\frac{\vert u\vert^2}{m_N^4}\right). 
\end{align}

Inserting this expansion into the amplitudes listed in table \ref{tab:scattME} reveals a strong destructive interference among most of the helicity permutations. In particular, the only non-vanishing channels correspond to those with incoming $W$ bosons carrying identical polarizations, \ie~$(\lambda^W_1,\lambda^W_2)=(0,0)$ and $(\pm1,\pm1)$. Explicitly, the matrix elements for these configurations are given by
\begin{align}
&\mathcal{M}(W^+W^+\to \ell^+\ell^+)\Big\vert_{(\lambda^W_1,\lambda^W_2)=(\pm1,\pm1)} 
\\
&  = -i e^{-i\phi_1} g_W^2\frac{V_{\ell N}^2}{m_N}  M_{WW}
+ \mathcal{O}\left(\frac{\vert t\vert^2}{m_N^4}, \frac{\vert u\vert^2}{m_N^4}\right), 
\nonumber
\end{align}
\begin{align}
&\mathcal{M}(W^+W^+\to \ell^+\ell^+)\Big\vert_{(\lambda^W_1,\lambda^W_2)=(0,0)} 
\\ 
&  = i e^{-i\phi_1} g_W^2 \frac{V_{\ell N}^2}{m_N} M_{WW}\frac{(1-2r_W)}{2r_W}
+ \mathcal{O}\left(\frac{\vert t\vert^2}{m_N^4}, \frac{\vert u\vert^2}{m_N^4}\right). \nonumber
\end{align}
In comparing the two expressions one {can see} the impact of {the} longitudinal polarization enhancements, which are responsible for the relative factor of $(M_{WW}/m_W)^2$.

    After squaring and integrating over the azimuthal angle, we obtain the leading contributions in the decoupling limit to the polar distribution of $\ell_1^+$ in $W^+W^+\to\ell^+\ell^+$ scattering. For each polarization  channel, this is:
\begin{align}
   \frac{{\rm d}\hat{\sigma}}{{\rm d}\cos\theta_1}&\Big\vert_{(\lambda^W_1,\lambda^W_2)=(\pm1,\pm1)} 
\\
&=   \frac{g_W^4}{2^6 \pi }
   \frac{\vert V_{\ell N}\vert^4}{m_N^2}
  + \mathcal{O}\left(r_N^{-1}\right),
   \nonumber   
\end{align}
\begin{align}
   \frac{{\rm d}\hat{\sigma}}{{\rm d}\cos\theta_1}&\Big\vert_{(\lambda^W_1,\lambda^W_2)=(0,0)}
\\
&=   \frac{g_W^4}{2^8 \pi}
   \frac{\vert V_{\ell N}\vert^4}{m_N^2}\frac{(1-2r_W)^2}{r_W^2}
   + \mathcal{O}\left(r_N^{-1}\right).
   \nonumber
\end{align}
We report good agreement between these expressions and numerical evaluations of helicity-polarized cross sections in this kinematic limit using the \textsc{HeavyN} model~\cite{Degrande:2016aje} with \mgamc~in conjunction with the formalism of Ref.~\cite{BuarqueFranzosi:2019boy}.

After integrating over the polar angle and averaging over all $W$ boson helicities, the total scattering rate is  
\begin{align}
   \hat{\sigma}
   = 
 \frac{g_W^4}{2^7 3^2 \pi}
   \frac{\vert V_{\ell N}\vert^4}{m_N^2}\frac{(1-4r_W+12r_W^2)}{r_W^2}
   + \mathcal{O}\left(r_N^{-1}\right).
\end{align}
For $n_R$ heavy neutrinos coupling to potentially different charged lepton flavors, the above generalizes to
\begin{align}
  \hat{\sigma} (W^+W^+&\to\ell^+_1\ell^+_2)
   =     \frac{g_W^4 (2-\delta_{\ell_1\ell_2})}{2^7 3^2 \pi r_W^2}\\
&    
  \times\Bigg\vert \sum_{k=4}^{n_R+3} \frac{V_{\ell_1 N_k}  V_{\ell_2 N_k}}{m_{N_k}} \Bigg\vert^2
   + \mathcal{O}\left(r_N^{-1},  r_W^{-1}\right).
   \nonumber
\end{align}

\bibliography{wwScattHeavyN_refs}

\end{document}